\documentclass[iicol]{sn-jnl}
\usepackage{subcaption}
\usepackage{graphicx}
\usepackage{algorithm}
\usepackage{algpseudocode}
\usepackage{amsmath}
\usepackage{amsfonts}
\usepackage{xcolor}
\usepackage{blindtext}
\usepackage{upgreek}

\usepackage{natbib}
\setcitestyle{authoryear,open={(},close={)}}

\graphicspath{ {./Images/} }

\jyear{2022}%

\theoremstyle{thmstyleone}%
\theoremstyle{thmstyletwo}%

\theoremstyle{thmstylethree}%

\begin{document}

\title[\hspace{8.5cm} A Comprehensive Review of Digital Twin - Part 1: Modeling and Twinning Enabling Technologies]{A Comprehensive Review of Digital Twin - Part 1: Modeling and Twinning Enabling Technologies}


\author[1]{\fnm{Adam} \sur{Thelen}}\email{acthelen@iastate.edu}

\author[2]{\fnm{Xiaoge} \sur{Zhang}}\email{xiaoge.zhang@polyu.edu.hk}

\author[3]{\fnm{Olga} \sur{Fink}}\email{olga.fink@epfl.ch}

\author[4]{\fnm{Yan} \sur{Lu}}\email{yan.lu@nist.gov}

\author[5]{\fnm{Sayan} \sur{Ghosh}}\email{sayan.ghosh1@ge.com}

\author[6]{\fnm{Byeng D.} \sur{Youn}}\email{bdyoun@snu.ac.kr}

\author[7]{\fnm{Michael D.} \sur{Todd}}\email{mdtodd@eng.ucsd.edu}

\author[8]{\fnm{Sankaran} \sur{Mahadevan}}\email{sankaran.mahadevan@vanderbilt.edu}

\author*[1]{\fnm{Chao} \sur{Hu}}\email{chaohu@iastate.edu}

\author*[9]{\fnm{Zhen} \sur{Hu}}\email{zhennhu@umich.edu}

\affil[1]{\orgdiv{Department of Mechanical Engineering}, \orgname{Iowa State University}, \orgaddress{\city{Ames}, \postcode{50011}, \state{IA}, \country{USA}}}

\affil[2]{\orgdiv{Department of Industrial and Systems Engineering}, \orgname{The Hong Kong Polytechnic University}, \orgaddress{\city{Kowloon}, \country{Hong Kong}}}

\affil[3]{\orgdiv{Intelligent Maintenance and Operations Systems}, \orgname{EPFL}, \orgaddress{\city{Lausanne}, \postcode{12309}, \state{NY}, \country{Switzerland}}}

\affil[4]{\orgdiv{Information Modeling and Testing Group}, \orgname{National Institute of Standards and Technology}, \orgaddress{\city{Gaithersburg}, \postcode{20877}, \state{MD}, \country{USA}}}

\affil[5]{\orgdiv{Probabilistic Design and Optimization group}, \orgname{GE Research}, \orgaddress{\city{Niskayuna}, \postcode{12309}, \state{NY}, \country{USA}}}

\affil[6]{\orgdiv{Department of Mechanical Engineering}, \orgname{Seoul National University}, \orgaddress{\city{Gwanak-gu}, \postcode{151-742}, \state{Seoul}, \country{Republic of Korea}}}

\affil[7]{\orgdiv{Department of Structural Engineering}, \orgname{University of California, San Diego}, \orgaddress{\city{La Jolla}, \postcode{92093}, \state{CA}, \country{USA}}}

\affil[8]{\orgdiv{Department of Civil and Environmental Engineering}, \orgname{Vanderbilt University}, \orgaddress{\city{Nashville}, \postcode{37235}, \state{TN}, \country{USA}}}

\affil[9]{\orgdiv{Department of Industrial and Manufacturing Systems Engineering}, \orgname{University of Michigan-Dearborn}, \orgaddress{\city{Dearborn}, \postcode{48128}, \state{MI}, \country{USA}}}

\abstract{
As an emerging technology in the era of Industry 4.0, digital twin is gaining unprecedented attention because of its promise to further optimize process design, quality control, health monitoring, decision and policy making, and more, by comprehensively modeling the physical world as a group of interconnected digital models. In a two-part series of papers, we examine the fundamental role of different modeling techniques, twinning enabling technologies, and uncertainty quantification and optimization methods commonly used in digital twins. This first paper presents a thorough literature review of digital twin trends across many disciplines currently pursuing this area of research. Then, digital twin modeling and twinning enabling technologies are further analyzed by classifying them into two main categories: physical-to-virtual, and virtual-to-physical, based on the direction in which data flows. Finally, this paper provides perspectives on the trajectory of digital twin technology over the next decade, and introduces a few emerging areas of research which will likely be of great use in future digital twin research. In part two of this review, the role of uncertainty quantification and optimization are discussed, a battery digital twin is demonstrated, and more perspectives on the future of digital twin are shared.
}

\keywords{Digital twin; Optimization; Machine learning; Enabling technology; Perspective; Industry 4.0, Review}



\maketitle

\section{Introduction}\label{sec1}
\label{Sec1}
%



This paper is the first in a series of two that analyze the roles of modeling and twinning enabling technologies, uncertainty quantification (UQ), and optimization in digital twins. Modeling and twinning enabling technologies are fundamental methods used to bridge the information gap between a physical system and its digital counterpart. In this paper, we review and analyze many methods and modeling techniques currently used in digital twins, and classify them into groups based on the direction in which data flows. 


We begin by reviewing an existing definition of digital twin. A digital twin, as originally proposed by~\cite{grieves2014digital}, is a virtual representation of a complex physical asset in the digital space for the purpose of closely characterizing the operations of the original physical process or system. The precise representation of the physical system in cyberspace is enabled by continuous data synchronization and information exchange between the digital twin and the physical counterpart. Having a digital replica of the physical entity of interest leads to enormous benefits spanning its entire life cycle, including the development phase (e.g., product design, resource planning, manufacturing process design), manufacturing phase (e.g., production process planning, manufacturing process control, maintenance of manufacturing equipment), service phase (e.g., performance and health monitoring, maintenance and control of fielded products,
path planning), and disposal phase (e.g., end-of-life reuse, remanufacturing, and recycling). With the ability to conduct what-if analyses in the digital space, the benefits of digital twin technology are already materializing in a wide range of applications ranging from high-value manufacturing industry \citep{ayerbe2021digitalization} to personalized medicine \citep{corral2020digital}, oil refinery management \citep{min2019machine}, risk identification, and city planning \citep{lu2020developing}. 

The digital twin concept centers around “individualized” digital models that capture the unique characteristics of individual products or process units. These models allow decision making to be optimized for each product or process unit, rather than based on the average characteristics of the entire population. This emerging technology poses new and challenging optimization problems at the forefront of model-based design, smart manufacturing, Industrial Internet of Things (IIoT), machine learning (ML), and predictive maintenance. The industry-scale adoption of the digital twin concept entails creating novel optimization solutions that use data collected from sensors and inspections (physical to digital) to provide decision makers with actionable information (digital to physical), thereby closing the digitalization loop. Major benefits include the ability to optimize control/maintenance actions to individual units and the potential to optimize the design of next-generation products.

The promise of digital twin concept and the need to guide the effective development of digital twin technology have motivated researchers to put together many review papers on digital twin. For example, motivated by the need to consolidate the many definitions and characteristics of digital twin, several review papers have been reported in recent years, such as \cite{barricelli2019survey, jones2020characterising, vanderhorn2021digital, liu2021review}. These papers summarized various definitions of digital twin and suggested generalized definitions and characteristics to distinguish digital twin from other models. \cite{lim2020state} presented a comprehensive review of digital twin by considering its benefits to engineering product life cycle management and business innovation. \cite{qi2021enabling} surveyed enabling technologies and tools for digital twins and proposed a five-dimensional digital twin model. \cite{fuller2020digital} analyzed enabling technologies for artificial intelligence, Internet of Things (IoT), and digital twins. \cite{rasheed2020digital} reviewed the latest methodologies and techniques related to the construction of digital twins from the modeling perspective.
 
There are also numerous application-specific literature review papers. For instance, \cite{jiang2021digital} reviewed the implementation of digital twins in civil engineering and discussed several problems that need to be addressed. \cite{xie2021digital} discussed applications of digital twins at different states of a cutting tool's life cycle. \cite{errandonea2020digital} reviewed the applications of digital twins for maintenance. \cite{kritzinger2018digital} provided a categorical review of digital twins in manufacturing and grouped applications of digital twins into different categories according to the level of integration. Focusing on a similar topic, \cite{cimino2019review} reviewed the applications of digital twins in manufacturing. Along the same line,  \cite{ayerbe2021digitalization} discussed recent advances in the digitalization of lithium-ion battery cell manufacturing processes and proposed a three-layer structure of a digital twin of large-scale battery manufacturing. \cite{ozturk2021digital} conducted a literature survey over the applications of digital twins in the architectural, engineering, construction, operation, and facility management industries. \cite{wagg2020digital} summarized the state-of-the-art and discussed future research directions of digital twin, focusing on applications in engineering dynamics. A survey of digital twins for verification and validation of industrial automation systems is available in \cite{locklin2020digital}. \cite{tao2018digital} presented a comprehensive review of digital twins in industry. \cite{he2021digital} reviewed applications of digital twins for intelligent and sustainable manufacturing. \cite{boje2020towards} discussed the future research directions of digital twin in construction engineering. \cite{uhlemann2017digital} discussed the importance of digital twins in realizing cyber-physical production systems for Industry 4.0.

While current review papers cover a good amount of literature on different aspects of digital twin, such as definitions, enabling technologies, and applications in different domains, not all important aspects of digital twin have been covered in the current literature. Digital twin technology is an emerging technology that exploits the synergy between several key enabling technologies, including multiphysics simulations, sensing, ML, UQ, data analytics, to name a few. Modeling and twinning methods are key to integrating different technologies in a digital twin. However, a comprehensive review of the modeling and twinning methods and the associated UQ and optimization methods enabling digital twins is still missing. Such a review is essential because it may guide effective, industry-scale implementations of digital twins for different types of physical systems. Based on a literature review of over 230 research papers on digital twins, this paper aims to fill this void by (1) categorizing the modeling and twinning methods according to their roles in digital twins, and (2) providing a detailed discussion on the importance of UQ and optimization in the modeling and twinning methods of digital twins. The challenges and future research directions on digital twin will also be discussed based on the literature survey and lessons learned from case studies. 

\begin{figure}[!h]
  \centering
    {\includegraphics[scale=0.6]{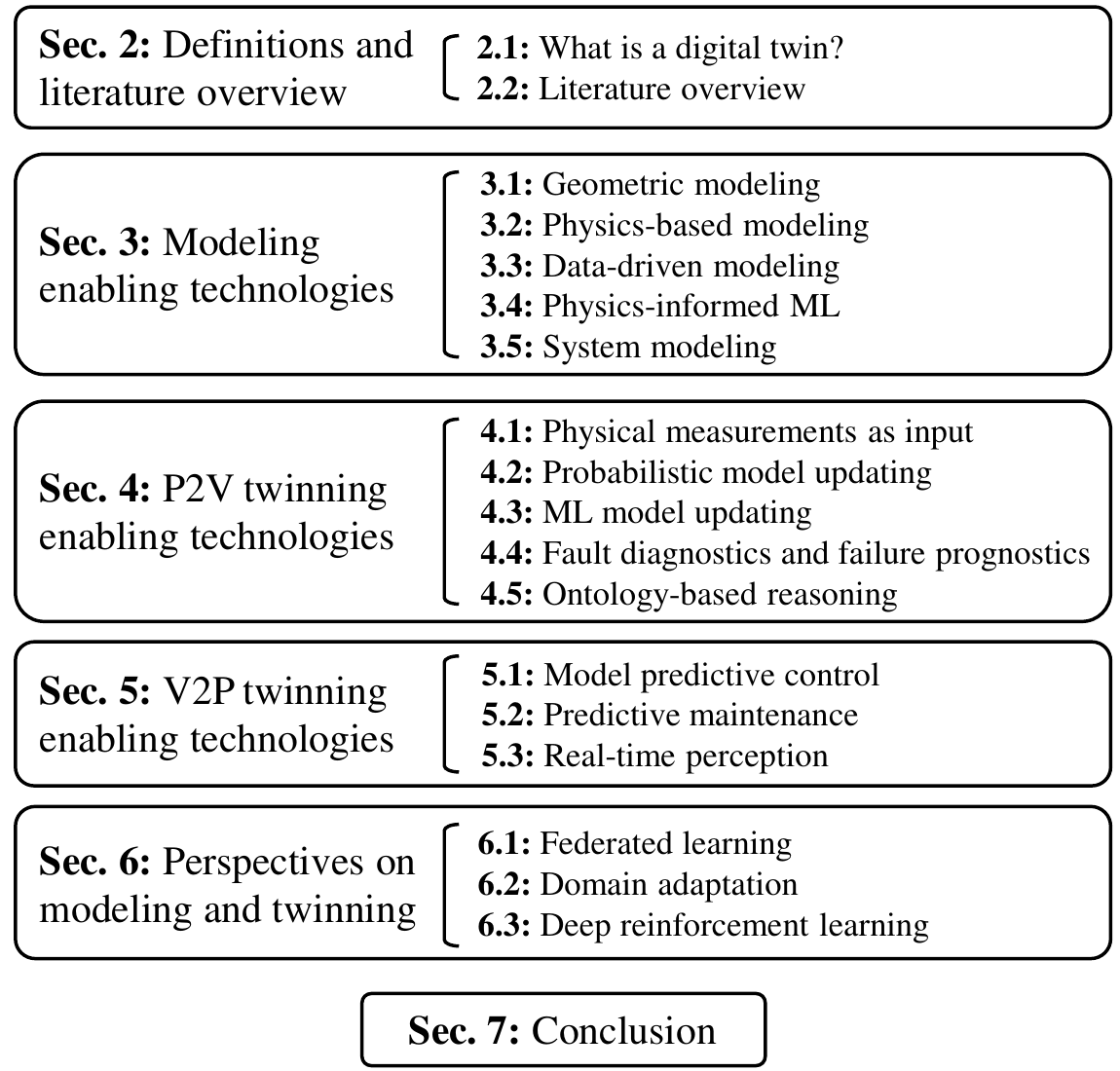}}
  \caption{Overview of topics covered in Part 1}
  \label{fig:outline}
\end{figure}

In consideration of the large number of subjects covered, we split the review paper into two parts to make it easier to read. The first paper is an introduction to various definitions of digital twin in the literature and our proposed five-dimensional definition of digital twin. In addition, the first paper also covers commonly used modeling and twinning technologies to map a physical system to its digital counterpart (P2V) and the actions/decisions sent back by the digital twin that can be taken in the physical system (V2P). Figure \ref{fig:outline} depicts the overall structure of Part 1 of the review paper.

Part 2 of the review paper~\citep{SMO2} centers around describing the roles of UQ and optimization in digital twins so as to support a better characterization of physical systems in the digital space. This second part of the review also presents a case study on battery health management to showcase the specific implementation of a battery digital twin and the power of UQ and optimization in the digital twin in an end-to-end fashion. Additionally, Part 2 reviews the applications of digital
twins at an industrial scale and available open-source tools and datasets related to digital twins, followed by a discussion on
challenges and future research directions.

The remainder of this paper is organized as follows. Sec. \ref{sec2} first discusses a variety of digital twin definitions reported in the literature and then conducts a comprehensive analysis on the trend of digital twin research. Sec. \ref{sec3} categorizes the modeling enabling technologies in digital twins. Sec.~\ref{sec:physical2virtual} summarizes commonly used approaches for physical-to-virtual twinning. Sec.~\ref{sec:V2P} elaborates the actions informed by virtual-to-physical connection that can be carried out at various stages of a physical system. Sec. \ref{sec:perspectives} presents perspectives on modeling and twinning in digital twins. Sec. \ref{sec9} concludes the first part of the two-part review paper.



\section{Definitions of digital twin and literature overview}
\label{sec2}

\subsection{What is a digital twin?}
According to the literature and other documents, the concept of \lq\lq digital twin" was first implemented in the NASA Apollo 13 program in the 1960s. A digital twin model of Apollo 13 was created on earth to allow engineers on the ground to test possible solutions for a rescue mission in space \citep{barricelli2019survey,nguyen2021digital}. Michael Grieves made the first documented definition of digital twin in his presentation in the context of product life cycle management in 2003 and later in a white paper \citep{grieves2014digital}. Since Grieves's presentation on digital twin in 2003, various digital twin concepts have been proposed. In a 2012 conference paper, Edward H. Glaessgen at NASA's Langley Research Center and David S. Stargel at the U.S. Air Force Office of Scientific Research gave the first specific definition of digital twin in the aerospace domain as:

\emph{\lq\lq A Digital Twin is an integrated multiphysics, multiscale, probabilistic simulation of an as-built vehicle or system that uses the best available physical models, sensor updates, fleet history, etc., to mirror the life of its corresponding flying twin \citep{glaessgen2012digital}."} 

Since the first definition in the aerospace domain, digital twin has received increasing attention in the past decade in various application domains, such as manufacturing, automobile, energy, and civil engineering. Due to the broad applications of digital twin in different domains, numerous application-specific digital twin definitions have been proposed to capture the nature of digital twin in these domains. Aiming to reduce the risk of diluting the digital twin concept by a large number of definitions, efforts have been made in recent years to consolidate digital twin definitions and terminologies in several review papers \citep{barricelli2019survey, jones2020characterising, vanderhorn2021digital, liu2021review}. In particular, \cite{vanderhorn2021digital} proposed  a consolidated and generalized definition of digital twin as \emph{\lq\lq a virtual representation of a physical system (and its associated environment and processes) that is updated through the exchange of information between the physical and virtual systems"}. Amongst the many definitions of digital twin, the one specified in \cite{glaessgen2012digital} is considered one of the most broadly accepted definitions, even though the definition may vary from application to application.

\begin{figure}[!h]
  \centering
    {\includegraphics[scale=0.68]{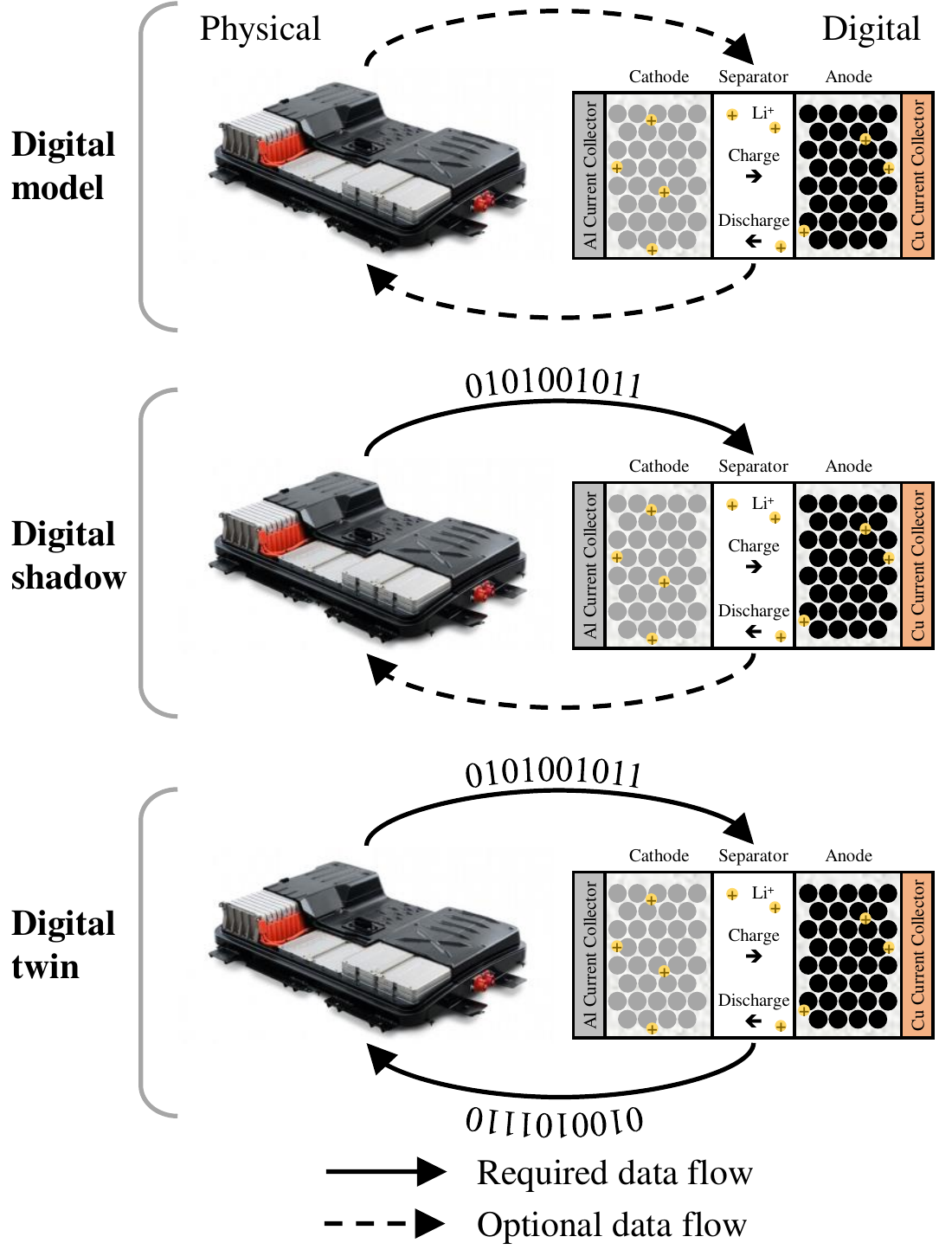}}
  \caption{Digital model, digital shadow, and digital twin concept for a lithium-ion battery cell within a larger pack}
  \label{fig:digital shadow}
\end{figure}

The distinction between a digital model, a digital shadow, and a digital twin is unclear in many cases. Further, having a large number of vague and inconsistent definitions of digital twin circulating in the literature may adversely affect industry interest in the adoption of this technology, creating a barrier to unleashing the maximum potential of the digital twin technology \citep{wright2020tell}. In what follows, we attempt to distinguish these three terms. It is argued that, for a digital model, data flow between the physical space and virtual space is optional, or at the very most, achieved manually as shown in Fig. \ref{fig:digital shadow}. For a digital shadow, data flow is unidirectional from physical to digital. But for digital twin, the data flow has to be bidirectional (see Fig. \ref{fig:digital shadow}) \citep{kritzinger2018digital}. When digital twin is used in control-related applications, the bidirectional data flow needs to be automatic, often enabled by monitoring and control software. For the application of digital twin to support decision making, such as predictive maintenance, which will be reviewed in Sec. \ref{sec:predictive maintenance}, however, the data flow from virtual to physical involves humans in the loop who carry out maintenance actions and, therefore, is not fully automatic but should be handled on time to avoid unexpected breakdown. Moreover, \cite{zhang2021building} pointed out that since a physical entity usually has many different aspects that can be modeled, it is always necessary to clarify for what aspect the digital twin is constructed. Take a simple mechanical shaft as an example. The shaft may fail due to fatigue, extreme torque, etc. Since it is difficult for the digital space to mirror all aspects of the physical space, a digital twin is always constructed for specific aspect(s) relevant to the engineering problem the digital twin is used to solve, and these aspect(s) need to be specified before the construction of the digital twin.

\begin{figure*}[!ht]
  \centering
    {\includegraphics[scale=0.75]{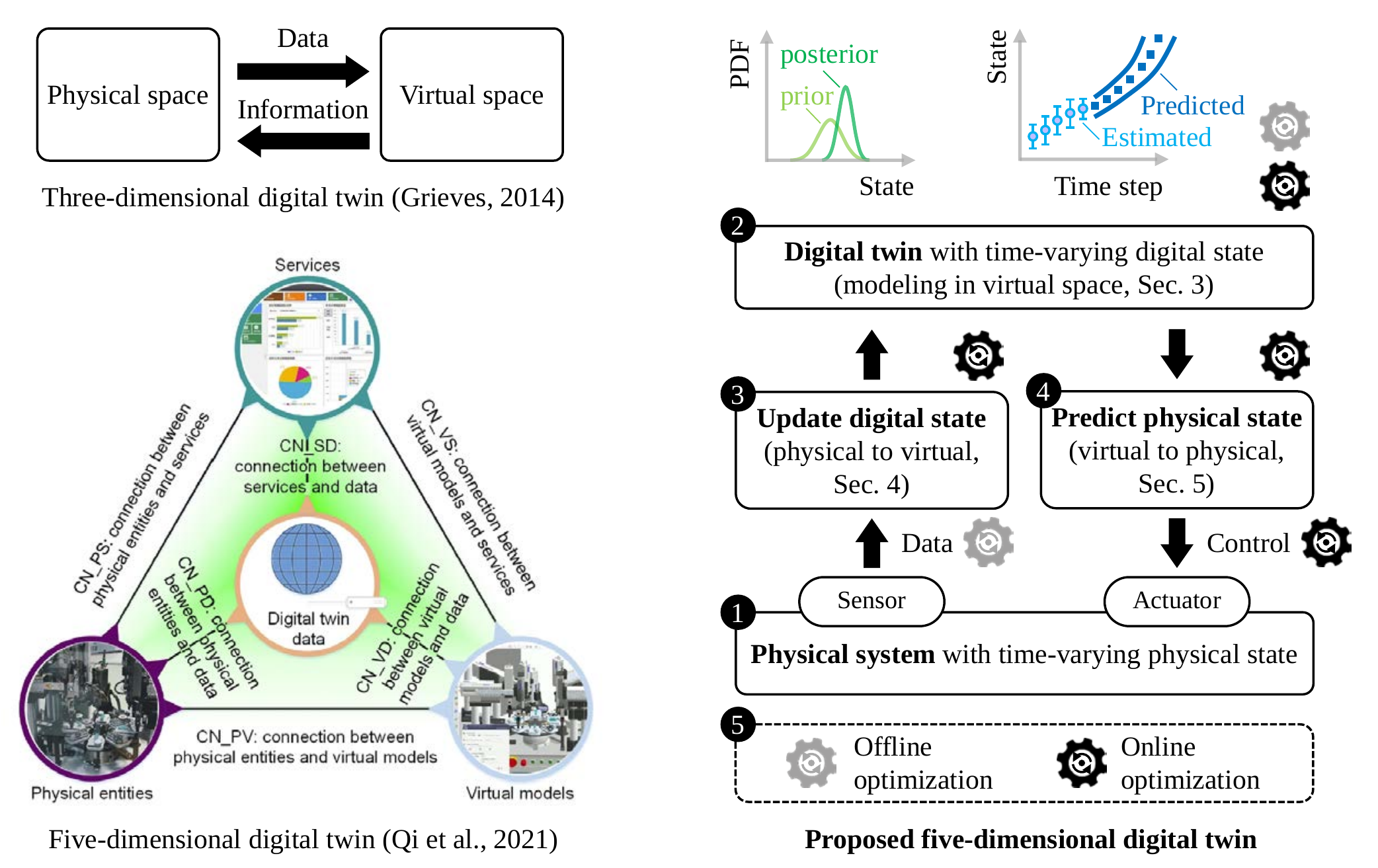}}
  \caption{Comparison of different digital twin (DT) models and the proposed model}
  \label{fig:digital twin}
\end{figure*}

In addition, various digital twin models have been proposed in the past few decades. Two highly cited digital twin models in the literature are the three-dimensional digital twin model proposed by \cite{grieves2014digital} and the five-dimensional digital twin model suggested by \cite{tao2018digital}. As shown in Fig. \ref{fig:digital twin}, the three-dimensional digital twin model consists of a physical space, a virtual space, and connections between them. \cite{tao2018digital} expands the three-dimensional digital twin model into a five-dimensional digital twin model for manufacturing, which includes a real machine, a virtual machine, a services component, digital twin data, and the connections as depicted in Fig. \ref{fig:digital twin}. 

Based on the definition of digital twin given in \citep*{glaessgen2012digital} as mentioned above, we suggest a new five-dimensional digital twin model as illustrated in Fig. \ref{fig:digital twin}. The proposed five-dimensional digital twin model can be considered an extended version of the three-dimensional model and is defined as follows
\begin{equation}
\label{eq:DT}
{\rm{DT}} = \mathbb{F}({\rm{PS}},\;{\rm{DS}},\;{\rm{P2V}},\;{\rm{V2P}},\;{\rm{OPT}})
\end{equation}

As indicated in the equation above, the proposed five-dimensional digital twin model consists of \emph{a physical system (PS)}, \emph{a digital system (DS)}, \emph{an updating engine (P2V)}, \emph{a prediction engine (V2P)}, and \emph{an optimization dimension (OPT)}, and $\mathbb{F}(\cdot)$ integrates all five dimensions together to be an effective digital twin. The physical system (PS) with sensing capability gathers data from multiple sources using various sensing and data acquisition techniques. The updating engine (P2V) updates the state of the digital model based on the sensor data. The updated digital models (DS) are then employed to predict the future state of the physical system using the prediction engine (V2P), thereby enabling predictive decision making feeding back to the physical system through actuators (control, maintenance, path planning, etc.). Finally, an optimization dimension (OPT, i.e., the fifth dimension) supports the functionalities of the other four dimensions in the digital twin by optimizing the data collection, modeling, state estimation, decision making, etc. The optimization dimension consists of two elements, namely offline optimization and online optimization. These two optimization elements will be explained in detail in Part 2 of the review paper~\citep{SMO2}. The seamless integration of the five dimensions (PS, DS, P2V, V2P, and OPT) enables digital twins to fulfill the need for a real-time mirror of the life cycle of a physical system, thereby supporting optimal decision making. 

To highlight how a five-dimensional digital twin might operate in practice, we can look at the example of a digital twin for a lithium-ion battery cell within a battery pack, shown in Fig. \ref{fig:digital shadow}. When designing a lithium-ion battery pack, engineers are most concerned with managing the rate and amount of degradation of each cell in the pack driven by the cell's usage (e.g., charging and discharging rates, depth of discharge, and temperature). In this scenario, a digital twin model of each cell within a pack can be constructed to track and forecast the capacity degradation trajectory of the cell and ultimately drive decision making around when the pack needs to be retired from service. Sensor measurements of the cell's state of health (PS) can be used to update (P2V, OPT) the digital model (DS) of the cell. Then, the digital model (DS) can be used to predict the cell's future state of health and degradation trajectory (V2P). Engineers can then use the predicted future degradation trajectory to inform decision making about the optimal time when the pack may need to be retired from service (V2P, OPT). The complete flow of data through all five digital twin dimensions outlined in Fig. \ref{fig:digital twin} is what ultimately characterizes this type of workflow as a true digital twin. This digital twin application for forecasting the future health of a lithium-ion cell and using the forecast health for predictive maintenance is extensively investigated as a case study in Part 2 of this review.

With a focus on the modeling and twinning technologies (i.e., the updating and prediction engines in Fig. \ref{fig:digital twin}), in the subsequent sections, we first perform an extensive literature review of the state-of-the-art and then classify the enabling technologies into distinct categories. The role of the optimization dimension in digital twin and future research directions will also be discussed.

\begin{figure*}[!ht]
  \centering
    {\includegraphics[scale=0.75]{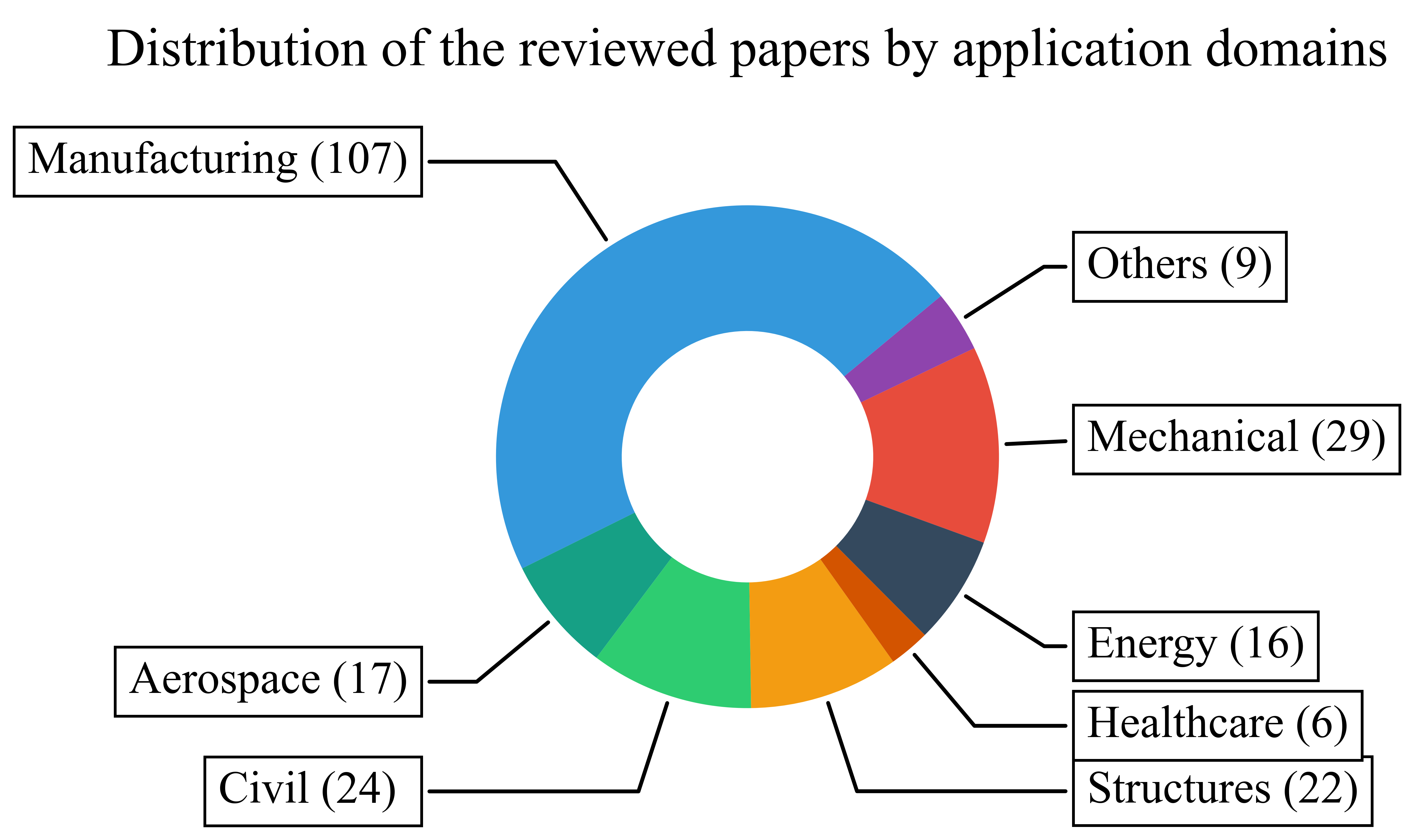}}
  \caption{Distribution of reviewed papers by domain (\lq\lq Civil" includes civil engineering, building engineering, smart city, and transportation; \lq\lq Mechanical" includes mechanical engineering, automotive, robotics, and marine; \lq\lq Others" includes supply chain, optical communication system, and agricultural engineering.}
  \label{fig:review_pie}
\end{figure*}

\subsection{Literature overview} 
Our literature review starts with a Google scholar search using the keyword \lq\lq digital twin". In addition to the review papers summarized in Sec. \ref{Sec1}, 230 research papers, excluding papers from predatory journals, were reviewed. The majority of these 230 research papers were published between 2017 and 2022, with some published between 2011 and 2012. They are classified into different groups by application domain as shown in Fig. \ref{fig:review_pie}. This figure indicates that manufacturing, followed by mechanical, civil, and structures, accounts for most of the digital twin papers. 107 out of 230 papers come from the manufacturing domain, which implies a leading role of manufacturing in the digital twin field.

The following information is then extracted from the literature survey:
\begin{itemize}
    \item \textbf{Top-cited papers}, which are papers with an average number of citations per year greater than 25 (the average number of citations per year to a paper is its total number of citations divided by the number of years elapsed since the publication of the paper).
    \item \textbf{Major journals}, which are the journals where research papers on digital twin are frequently published based on the 230 research papers reviewed.
    \item \textbf{Publicly available tools/datasets}: tools and datasets related to digital twin from the 230 papers and associated websites.
    \item \textbf{Modeling methods}: methods used in the papers to create the virtual space.
    \item \textbf{Twinning methods}: methods used to establish physical-to-virtual and virtual-to-physical connections.
    \item \textbf{Optimization}: roles of optimization in the digital twin models presented in the papers.
\end{itemize}

Next, we will summarize the major findings related to the first two items listed above. The methods of modeling, twinning (i.e., physical-to-virtual and virtual-to-physical), and optimization will be summarized and classified into different categories in Sec. \ref{sec3}. The publicly available tools/datasets will be provided in Sec.~4.2 of Part 2 of the review paper.


\subsubsection{Top-cited papers and trends in different domains}
From the 230 research papers, we first identify papers whose average number of citations per year is greater than 25. It is found that many papers in the manufacturing domain satisfy this criterion. It is not entirely surprising since digital twin has been a hot topic in smart manufacturing. We, therefore, list the top-cited papers in the manufacturing domain separately from those in the other domains. 

\begin{table*}[!ht]
    \centering
    \caption{Top ten top-cited papers in manufacturing domain (Based on yearly average number of Google scholar citations as of April 2022)}
    \begin{tabular}{ p{7cm}|p{3cm}|p{1.5cm}|p{2.5cm}}
     \hline \hline
     \textbf{Title} & \textbf{Author and Publication year} & \textbf{Citations} &  \textbf{Application}\\
     \hline
     Digital twin-driven product design, manufacturing and service with big data   & \cite{tao2018digital}    & 1435 &   Design and Manufacturing\\
    \hline
     Shaping the digital twin for design and production engineering   & \cite{schleich2017shaping}    & 719 &   Production\\
         \hline
     Digital twin and big data towards smart manufacturing and industry 4.0: 360 degree comparison   & \cite{qi2018digitala}    & 789 &   Smart manufacturing\\
         \hline
     Digital twin shop-floor: a new shop-floor paradigm towards smart manufacturing   & \cite{tao2017digital}    & 585 &   Shop floor\\
         \hline
     Digital twin-based smart production management and control framework for the complex product assembly shop-floor   & \cite{zhuang2018digital}    & 313 &   Shop floor\\
         \hline
     Toward a Digital Twin for real-time geometry assurance in individualized production   & \cite{soderberg2017toward}    & 372 &   Geometry assurance\\
         \hline
     Digital twin-driven manufacturing cyber-physical system for parallel controlling of smart workshop   & \cite{leng2019digital}    & 235 &   Shop floor\\
         \hline
     Defining a digital twin-based cyber-physical production system for autonomous manufacturing in smart shop floors   & \cite{ding2019defining}    & 225 &   Shop floor\\
         \hline
     Digital twin-based designing of the configuration, motion, control, and optimization model of a flow-type smart manufacturing system   & \cite{liu2021digital}    & 100 &   Configuration optimization\\
         \hline
     Digital twin service towards smart manufacturing   & \cite{qi2018digital}    & 231 &   Service in Manufacturing\\
     \hline \hline
    \end{tabular}
    \label{tab:top-cisted papers_Manuf}
\end{table*}

Table \ref{tab:top-cisted papers_Manuf} summarizes the top ten most-cited papers in the manufacturing domain from the 230 reviewed papers, based on the average number of citations per year. Qi and Tao are the the most highly-cited researchers in this field at this time \citep{tao2018digital, qi2018digitala, tao2017digital, qi2018digital}. Most of the top-cited papers in the manufacturing domain focus on digital twins of manufacturing shop-floors, such as assembly, production management and control, and configuration optimization. There is one paper that focuses on geometry assurance in individualized production. Table \ref{tab:top-cisted papers} lists the top-cited papers from the civil, aerospace, energy, and healthcare domains, which reflect the trends of digital twin outside the manufacturing sector. As shown in this table, aerospace, civil, and mechanical are leading the trend of digital twin outside the manufacturing domain. Five out of the 12 top-cited papers concentrate on fault diagnostics and prognostics of physical assets using digital twin \citep{xu2019digital,li2017dynamic,glaessgen2012digital,jain2019digital,tuegel2011reengineering}. The others focus on optimization, management, and service. 

\begin{figure}[!h]
  \centering
    {\includegraphics[scale=0.55]{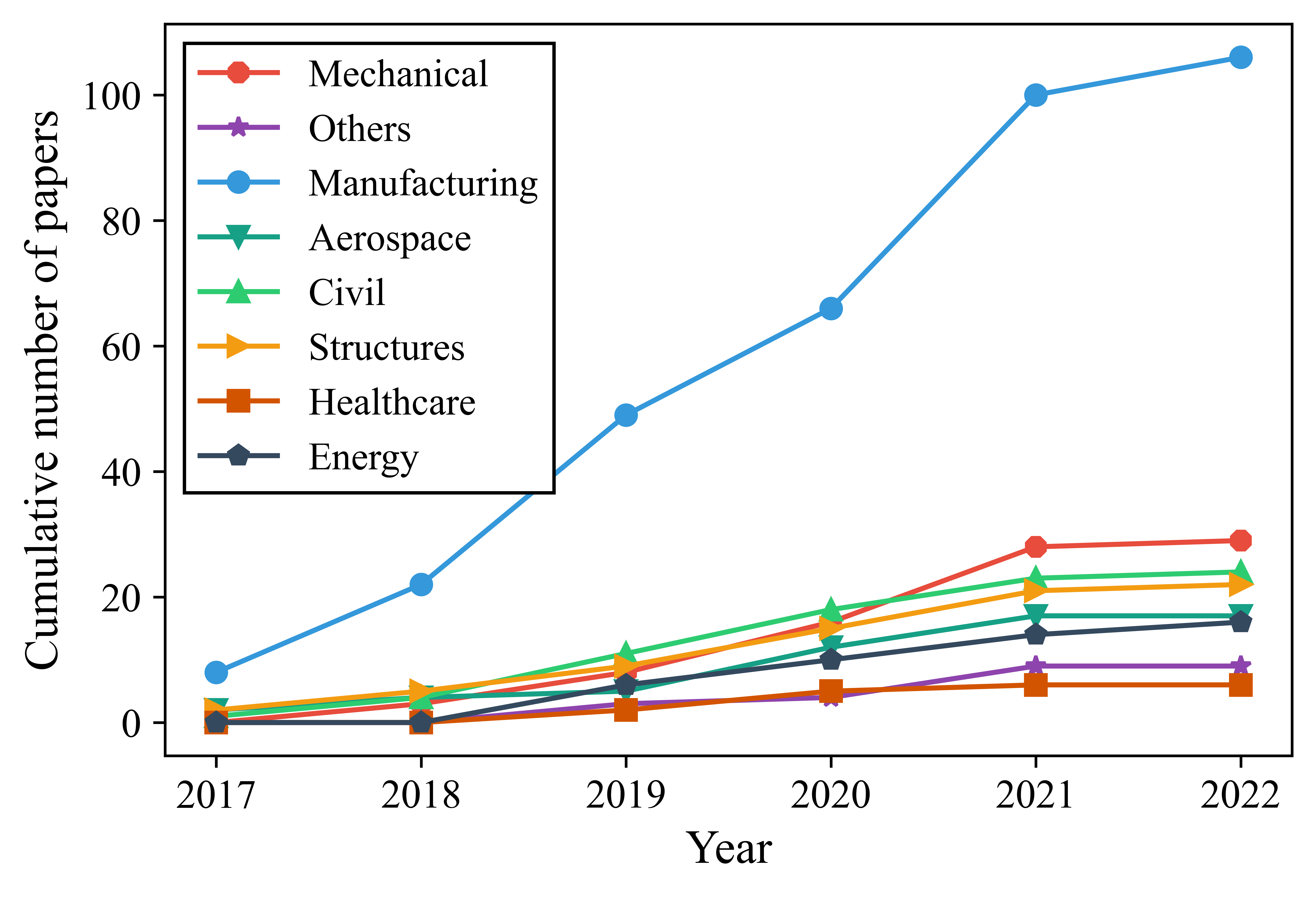}}
  \caption{Cumulative number of papers of different application domains}
  \label{fig:accumu}
\end{figure}

Following that, Fig. \ref{fig:accumu} shows the cumulative number of papers over the past five years in each different application domain. It is clear that the number of papers in the manufacturing domain is much higher than the other domains. This agrees with the earlier analysis of the most highly cited papers which found that a majority of them were in the manufacturing domain. Note that the trends abruptly flatten from 2021 to 2022 because this review paper was written in early 2022 and did not account for papers published later in the year. It is expected that the increasing trend continue into 2022.

\begin{figure}[!h]
  \centering
    {\includegraphics[scale=0.55]{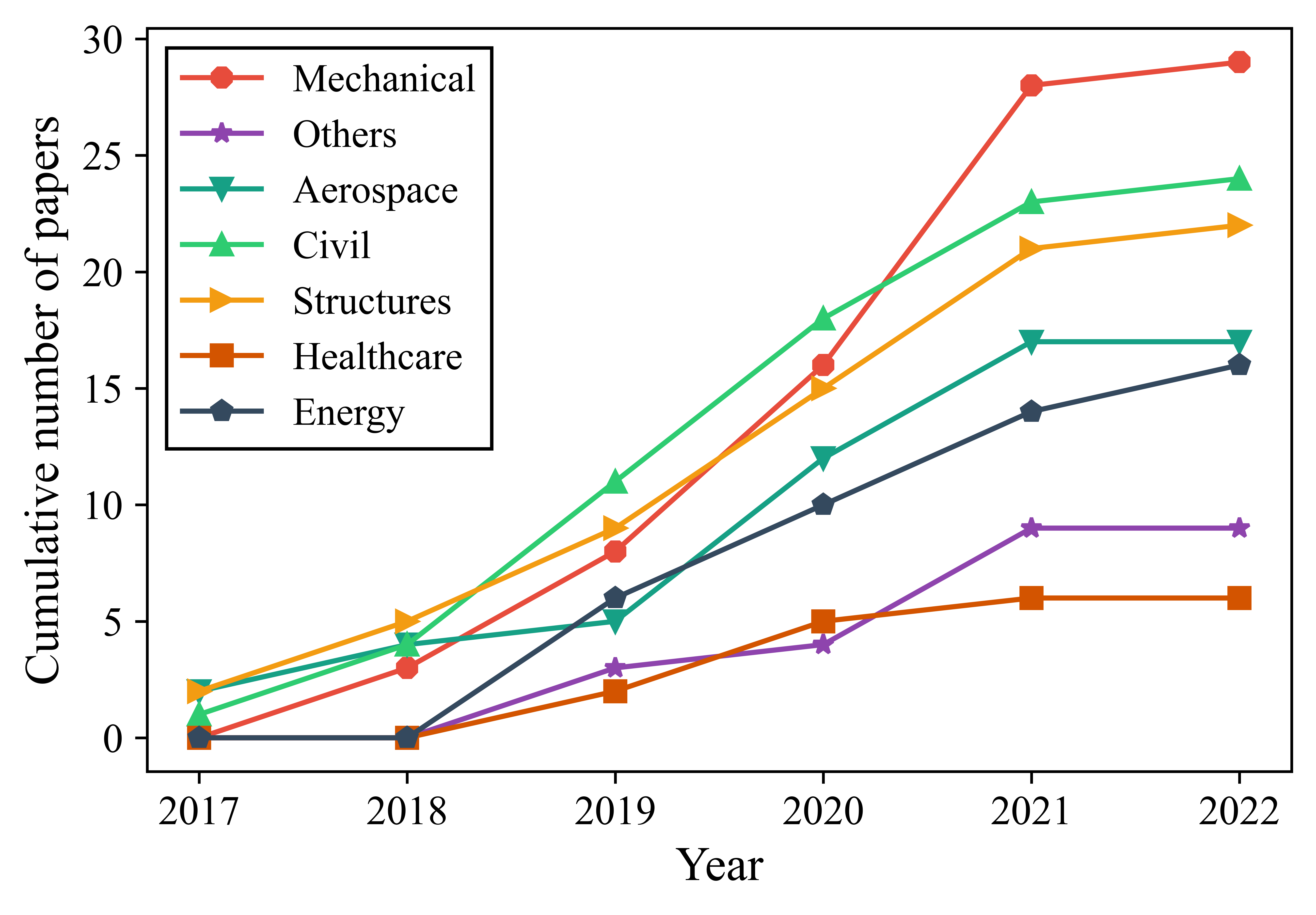}}
  \caption{Cumulative number of papers of different application domains excluding manufacturing}
  \label{fig:accumu_no_Manuf}
\end{figure}

To better analyze the trends in the other domains, we remove the manufacturing data, and re-scale the plot. Fig. \ref{fig:accumu_no_Manuf} shows the cumulative numbers of papers in the past five years, excluding manufacturing. In recent years, the cumulative number of papers in the other domains, especially civil, mechanical, and energy, have been growing rapidly.

\begin{table*}[!ht]
    \centering
    \caption{Top-cited papers in each application domain excluding manufacturing (based on yearly average number of Google Scholar citations as of April 2022)}
    \begin{tabular}{ p{7cm}|p{3cm}|p{1.5cm}|p{2.5cm}}
     \hline \hline
     \textbf{Title} & \textbf{Publication year} & \textbf{Citations} &  \textbf{Application domain}\\
    \hline
     A digital-twin-assisted fault diagnosis using deep transfer learning   & \cite{xu2019digital}    & 154 &   Mechanical (Automotive)\\
         \hline
     The digital twin paradigm for future NASA and U.S. Air Force vehicles   & \cite{glaessgen2012digital}    & 1226 &   Aerospace\\
         \hline
    Reengineering aircraft structural life prediction using a digital twin   & \cite{tuegel2011reengineering}    & 639 &   Aerospace\\
         \hline
     Dynamic Bayesian network for aircraft wing health monitoring digital twin   & \cite{li2017dynamic}    & 171 &    Aerospace\\
         \hline
     Machine learning based digital twin framework for production optimization in petrochemical industry   & \cite{min2019machine}    & 141 &   Energy (Oil industry)\\
         \hline
     A digital twin approach for fault diagnosis in distributed photovoltaic systems   & \cite{jain2019digital}    & 106 &   Energy\\
         \hline
     A BIM-data mining integrated digital twin framework for advanced project management   & \cite{pan2021bim}    & 60 &   Civil (Construction)\\
         \hline
     Developing a digital twin at building and city levels: Case study of West Cambridge campus   & \cite{lu2020developing}    & 82 &    Civil (Smart city)\\
         \hline
     Digital twin--Proof of concept   & \cite{haag2018digital}    & 288 &   Mechanical (Structural Health Monitoring)\\
              \hline
     An application framework of digital twin and its case study   & \cite{zheng2019application}    & 206 &   Others (Service)\\
              \hline
     How to tell the difference between a model and a digital twin   & \cite{wright2020tell}    & 104 &   Others (Service)\\
              \hline
     A novel cloud-based framework for the elderly healthcare services using digital twin   & \cite{liu2019novel}    & 171 &   Healthcare\\
     \hline \hline
    \end{tabular}
    \label{tab:top-cisted papers}
\end{table*}

\subsubsection{Major journals}

Table \ref{tab:major journals} presents major journals where the 230 reviewed papers were published. Only journals with more than three published papers are listed. Shown in the table, manufacturing-related journals, namely \emph{Journal of Manufacturing Systems}, \emph{International Journal of Advanced Manufacturing Technology}, and \emph{Robotics and Computer Integrated Manufacturing}, are leading journals. Other leading journals include \emph{Journal of Ambient Intelligence and Humanized Computing}, \emph{IEEE Access}, \emph{Engineering Fracture Mechanics}, and \emph{International Journal of Production Research}.

\begin{table}[!ht]
    \caption{Major journals on digital twin based on the reviewed papers}
    \centering
    \begin{tabular}{ p{5cm}|p{2cm} }
     \hline \hline
     \textbf{Journal Name} & \textbf{No of Papers}\\
     \hline
    {Journal of Manufacturing Systems}                           & 25                        \\
    {IEEE Access}                                                & 10                        \\
    {International Journal of Advanced Manufacturing Technology} & 9                         \\
    {Robotics and Computer Integrated Manufacturing}             & 6                         \\
    {International Journal of Computer Integrated Manufacturing} & 6                         \\
    {Procedia CIRP}                                              & 6                         \\
    {CIRP Annuals - Manufacturing Technology}                    & 5                         \\
    {IFAC Papers Online}                                         & 5                         \\
    {Journal of Intelligent Manufacturing}                       & 4                         \\
    {Journal of Ambient Intelligence and Humanized Computing}    & 4                         \\
    {Engineering Fracture Mechanics}                             & 4                         \\
    {International Journal of Production Research}               & 3                         \\
    {Advanced Engineering Informatics}                           & 3                         \\
    {Automation in Construction}                                 & 3                         \\
    {Computer and Structures}                                    & 3                         \\ 
\hline \hline
\end{tabular}
    \label{tab:major journals}
\end{table}

\section{Modeling enabling technologies}
\label{sec3}
Modeling plays an essential role in digital twins by creating a digital replica of the physical entity of interest. According to the purpose and complexity of different modeling methods, we classify them into the following five categories.

\subsection{Geometric modeling}
\label{sec:Geo_modeling}

\paragraph{(a) Solid modeling}
Computer-aided design (CAD) is a virtual solid modeling technique where 3D objects are created from a number of 2D drawings using computer software. Many individual components can be combined in an assembly to create a 3D solid model of a larger system. Solid models help designers and engineers see how a product will look and function in the real world. There are a number of free and paid-for 3D solid modeling software on the market. Each software package differs in its intended audience and scope of available features. For engineering, popular software includes Autodesk Inventor and AutoCAD, PTC Creo, SolidWorks, and Catia \citep{cai2020using}. Free software packages include Unity3D, Unreal Engine, and Blender \citep{leng2020digital, matulis2021robot, liu2021genetic}.

A simple but useful application of solid modeling for digital twin is to create models of all the equipment for a production line. Then, the modeled equipment can be moved around, checked for timing, or simulated consistent with a range of different goals. For example, \cite{liu2021intelligent} created a virtual workshop with various machining tools so that the order in which different parts were machined could be optimized. Similar work by \cite{yi2021digital} investigated ideas for using 3D models of the equipment in an assembly line to optimize the layout for reduced travel distance between steps. \cite{vathoopan2018modular} proposed creating a solid model of a robotic arm to detect and alarm when the robot's joints may have moved further than physically possible as predicted by the solid model. \cite{ayani2018digital} created a complete solid model of an old analog-controlled production machine so that its movements, timings, and controls could be uploaded to the cloud and further studied. \cite{huang2022building} modeled a CNC tool changing system and all the tools inside of it for the purpose of improving tool change speeds. Creating solid models for optimization problems which focus on the physical size and location of manufacturing equipment and tools is relatively easy because minor details like material properties and movable joints can be ignored. If, however, the solid model will be used for simulation of stresses, movements, or interference using physics-based simulations (see Sec. \ref{sec:physics_based_modeling}), the modeling process requires all details about the equipment be defined, increasing the modeling time substantially.

Another application of solid modeling in digital twin is simulating and studying human-environment interaction. Solid modeling enables researchers to investigate how humans fit into the digital twin environment. \cite{vatankhah2021digital} investigated concepts of human-robot interaction and ways to improve energy savings. \cite{pairet2019digital} proposed modeling an entire offshore oil drilling platform to study human-robot interaction in the dense 3D space. Similarly, \cite{lohtander2018micro} created a 3D model of a human assembly area to study ergonomics and optimize the placement of part bins. This area of research is of great importance, especially in a manufacturing setting, where humans and robots coexist in a constrained environment. It is envisioned that solid modeling of environments will play a large role in designing smarter robots which can better interact safely in proximity to humans.

Other applications of solid modeling to digital twins include tolerancing, geometry variance control, and defect control in manufacturing settings. Digital twin models of a production process can be used to improve quality control in various ways. \cite{liu2022digital} investigated integrating solid modeling and surface roughness prediction for adaptive manufacturing control. \cite{schleich2017shaping} proposed computer aided tolerancing and geometry variation control via a solid modeling approach. \cite{zambal2018digital} discussed how one might make a solid model system of a carbon fiber production process so that deviations in the equipment or manufacturing process could be simulated. Through simulation, the effects of equipment deviations on the final finished carbon fiber product can be quantified and understood.

Solid modeling has many promising uses in digital twin applications. With the ability to model the geometry of any object, solid modeling enables additional optimization dimensions not captured through sensor measurements and other techniques. One of the key problems using solid models in digital twins is the excessive amount of time required to create models which can be used in accurate physics-based simulations, such as finite element analysis (FEA) or motion simulations in Sec. \ref{sec:physics_based_modeling}. In many cases, the solid models of the parts and machines researchers aim to use in digital twin simulations have already been created by the original equipment manufacturers (OEMs). Similar to ideas discussed in \cite{grieves2014digital} white paper, it is entirely conceivable that future high value assets (e.g., aircraft, spacecraft, industrial equipment, and even entire production factories) could be delivered to the customer accompanied by a complete digital solid model for use in a digital twin framework. In turn, the extensive time commitment the OEM has sunk into creating the initial 3D solid model of the product could now be recouped by supporting the end user with a long-term reliability and predictive maintenance contract. With lasting communication between the OEM and the end user through the online solid model, safety issues can be managed quicker, replacement parts can be identified more easily, and future upgrades can be tested and checked virtually for compatibility with a given customer's asset before committing to manufacturing the parts. Ultimately, the sharing of solid models between OEMs and end users is necessary to accelerate the adoption of digital twin technology.

\paragraph{(b) Laser scanning}
Laser scanning, also commonly referred to as LiDAR (light detection and ranging), is a geometry modeling technique which uses laser light to repeatedly sample the distance to different points on an object's surface. The recorded measurements are virtually assembled into a 3D point cloud of (X, Y, Z) coordinate data to describe the surface morphology of the scanned object in great detail. Unlike solid modeling techniques which can model internal components, material changes, and geometries, laser scanning can only model surface-level features. However, laser scanning makes up for its inability to model internal geometries by generally being much faster at modeling objects. When it is paired with the appropriate software, laser scanning can create a point cloud of data points on a computer in near real-time.

Laser scanning has been used in many applications for digital twin. In the manufacturing domain, laser scanning has been proposed for quickly checking the geometry of traditionally manufactured or 3D printed parts \citep{almalki2022digital, warmefjord2017inspection, white2021digital}. Also in manufacturing, \cite{droder2018machine} showed a proof of concept for a digital twin considering human-robot interaction. The researchers used a Microsoft Kinect 2.0 LiDAR sensor to detect humans in a robotic arm cell. Based on the proximity of the human to the robotic arm, the robotic arm's path and speed were automatically optimized in real-time to create a safe environment for the human while not ceasing operation entirely. In contrast to other systems which use laser trip-wires that completely stop the robot when a human enters the area, the LiDAR-based digital twin system significantly improved manufacturing performance while maintaining a safe working environment. In a different application of laser scanning and robots, \cite{wang2020digital} laterally mounted an industrial camera on a robotic welder to take images of the process for improved real-time quality control via augmented reality data streaming. LiDAR and other laser scanning techniques will be extremely useful for the development of more complex manufacturing sensor networks connected to a digital twin. Real-time data streaming to and from the physical and virtual dimensions will enable more productive manufacturing lines without compromising worker safety.

The ability of laser scanning to quickly map and model large structures, landscapes, and buildings makes it particularly useful for digital twin modeling in the civil engineering domain \citep{lu2020digital,pan2021bim}. One of the most common uses of LiDAR and laser scanning for digital twin is for scanning the surface of large buildings \citep{deng2021systematic}, bridges \citep{shim2019development}, or other structures \citep{lim2020digital}. Specifically in \cite{shim2019development}, the authors discussed how LiDAR laser scanning could be used to periodically model the surface features of different bridge components for preventative maintenance. In this system, the raw scanned point cloud data is fed in batches to the cloud where structure recognition deep learning (DL) algorithms sort the scanned images according to the type of structure it is. Then, within the digital twin network, the images are mapped into their respective ontological structure file (see Section \ref{sec:sys_modeling} (b)) where they are further analyzed for cracks, pitting, shifting, or other undesirable degradation indicators. Any positively identified defects would then be loaded into a global solid model simulation for physics-based stress analysis. In the next section, we discuss such physics-based simulations, and their role in digital twins. Altogether, applications of LiDAR and laser scanning for digital twin modeling will become extremely important in the coming years as cities struggle to maintain the ever increasing number of bridges and other critical infrastructure.

\paragraph{(c) VR, AR, and MR technologies}
In addition to solid modeling and laser scanning, other techniques, such as virtual reality (VR), augmented reality (AR), mixed reality (MR), have also been explored as viable methods of constructing digital replicas of various physical entities. These visualization and interaction techniques differ in how they display relevant information. The most simple method, AR, overlays relevant contextual data to the user by projecting over assets in the physical world. Typically, the projections are done using light projection onto surfaces, or in some cases, through a wearable headset with transparent lenses. MR technologies further extend the capabilities of AR by allowing interactions between the digital objects and the physical world. This technology is closely followed by VR, which is a completely immersive digital environment where interactions happen exclusively in the virtual world. These data visualization and interaction technologies are promising enablers of digital twin modeling because of their ability to connect users, assets, and data streaming in real-time. VR, AR, and MR technology research is an active field. For example, \cite{tadeja2020aerovr} developed an immersive aerospace design environment built upon VR to aid the design process of aerodynamic components by interactively visualizing performance and geometry. Other work by ~\cite{cai2020using} leveraged AR to link the layout information between a physical and digital asset in a reconfigurable additive manufacturing system made of robotic arms for tool path planning and simulation. ~\cite{choi2022integrated} integrated MR in a digital twin to achieve safety-aware human-robot collaboration, where MR was combined with safety-related monitoring to track the shared workplace and real-time safety distance calculation. Research into VR, AR, and MR technology remains a hot and active topic, as the role of these powerful data visualization and interaction techniques is not yet fully understood. In the future, it is expected that these technologies will enable high-fidelity interactive digital twin models that can be used to visualize, control, and optimize various processes and designs in real-time.

\subsection{Physics-based modeling} 
\label{sec:physics_based_modeling}

Scientists observe the basic principles of how our world works and translate them into physical laws. These laws define the rules for the motion of matter and other physical phenomena through space and time. It is well-known that partial differential equations (PDEs) can mathematically describe certain physical laws. Two well-known examples of PDEs are the Navier–Stokes equations describing fluid motion \cite{chorin1990mathematical} and the heat equation describing heat diffusion \cite{widder1976heat}. Combining PDEs into a mathematical model and solving this model allows engineers and scientists to simulate and observe the physical phenomena governing a system or process. In computer-aided engineering (CAE), engineers and researchers use software packages to formulate mathematical models consisting of PDEs that describe the underlying physics, initial conditions, and boundary conditions. After defining the models, the software solves the PDEs to simulate the physical effects so that engineers and researchers can better understand the physical interactions. Simulating the solution approximates the input-output relationship of the physical system, where the input consists of the physical system properties, parameterized in the mathematical model, as well as the operating and environmental conditions, and the output is the performance of interest. Physics-based modeling software is diverse and complex in the range of applications it can be used for. Below we outline some of the most common types of physics engineers and researchers aim to simulate.

\begin{itemize}
\item \textbf{Types of physics}: Physics-based modeling and simulation for use in digital twins covers a wide scope. Below, we highlight and summarize some of the most common types of physics and physical modeling techniques used for digital twin.
\begin{enumerate}
    \item Solid body structural analysis of stress and strain is typically conducted using finite element analysis (FEA) software where the objects are modeled as a tessellation mesh of triangles, and the forces are calculated at each node in the mesh. FEA is flexible in that the resolution of the mesh can be tuned to balance the trade-off between simulation accuracy and computational costs. FEA has so far been an important modeling technique for digital twin \citep{li2017dynamic,angjeliu2020development,millwater2019probabilistic,moi2020digital,karve2020digital,ye2020digital,liu2021digital,bellalouna2021case,wang2022structural}. Specific examples of FEA for digital twin include stress analysis of a ground vehicle suspension \citep{hu2011adaptive}, wear analysis of cutting tool in manufacturing \citep{zhang2021digital}, strain analysis \citep{revetria2019real}, and dynamic structural analysis of an aircraft wing \citep{seshadri2017structural}. As an initial digital twin proof of concept, \cite{haag2018digital} demonstrated simultaneous simulation and measurement of forces in a beam bending test bench.
    
    \item Thermal and fluid flow analyses are typically conducted using computational fluid dynamics (CFD) software \citep{zhou2021digital}. Fluid physics are especially relevant in aircraft and spacecraft simulation, where digital twin was first conceived \citep{grieves2014digital}. In many cases, fluids are closely tied to heat transfer, as flowing fluids are exceptionally effective at exchanging heat. \cite{fan2013parametric} performed thermal analysis for a battery cooling system of an electric vehicle. In the future, thermal and fluid flow analyses will prove exceptionally useful in digital twins.
    
    \item Kinematic and dynamic analyses of mechanisms using multi-body dynamics (MBD) models are another type of physics commonly modeled. \cite{ha2018computational} analyzed the motion of robots using MBD; \cite{xia2021digital} used MBD to simulate and test the interference and collisions of the components in their digital twin model.
    
    \item Multiphysics simulations simultaneously simulate multiple coupled physical phenomena by sharing information between individual physics-based simulations \citep{delaurentis2000uncertainty, hu2018adaptive}. Examples include coupled FEA and CFD simulations in fluid-structure interaction for aircraft wings \citep{tezduyar2001fluid} and wind turbines \citep{hsu2012fluid}, and coupled electrochemical-mechanical simulations for modeling Li-ion battery cells \citep{allu2014new,wang2022digital}. Another example of coupled multiphysics interaction is the hypersonic vehicle shown in Fig. \ref{fig:multiphysics}. This multiphysics simulation features coupled aerodynamic and aeroelastic analyses to predict the complex vehicle response under hypersonic airflow conditions \citep{culler2010studies}.

\end{enumerate}

\begin{figure}[!h]
  \centering
    {\includegraphics[scale=0.52]{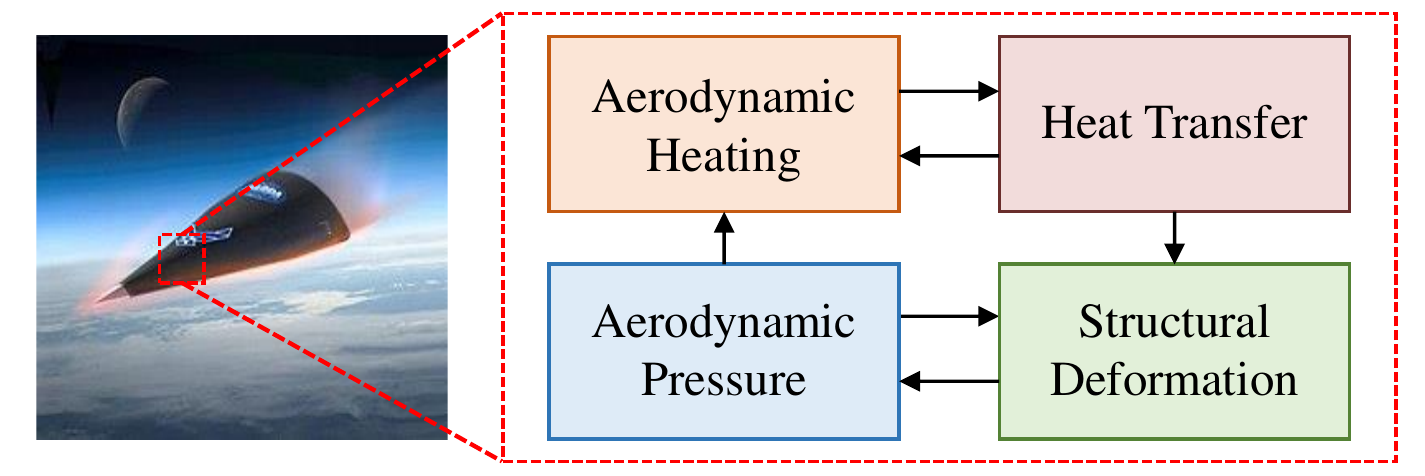}}
  \caption{Coupled multiphysics simulation of a hypersonic vehicle panel \citep{hu2018adaptive}}
  \label{fig:multiphysics}
\end{figure}

\item \textbf{Levels of fidelity}: When building a physics-based model, it is important to consider the minimum level of detail (fidelity) required to approximate the physical input-output relationship in order to optimize the balance between computational costs and model accuracy. Take for example a fluid dynamics problem which can be solved using different CFD models. With the highest fidelity, direct numerical simulation can provide an accurate high-fidelity solution to model turbulence flow. With slightly lower fidelity, large eddy simulation and the Reynolds-averaged Navier–Stokes equations can also be used to solve the same problems more quickly, albeit with a slightly lower accuracy. Another example of multi-fidelity modeling arises in the battery modeling community. There is a well-established hierarchy of electrochemical-thermal models that consists of (from high- to low-fidelity) the 3D Doyle–Fuller–Newman (DFN) model, 2+1D DFN model, the DFN model (or pseudo-two-dimensional), and the single particle model \cite{marquis2020suite}. The field of multi-fidelity modeling is rich with literature, and interested readers should read several of the dedicated review papers \citep{fernandez2016review, peherstorfer2018survey, giselle2019issues}. One final note regarding model fidelity is that the data-driven models presented in Sec. \ref{sec:data_driven_modeling} fall into the broad category of low-fidelity models, and great effort is being done to find ways to combine physics-based and data-driven models for improved fidelity (see Sec. \ref{sec:PI_ML})
\end{itemize}

When defining a relevant physics-based modeling problem, the engineering problem at hand determines which types of physics should be modeled. Many commercial CAE software tools streamline the execution of three basic steps: (1) loading in 3D geometry data from a geometric model created in CAD software or via laser scanning as reviewed in Sec. \ref{sec:Geo_modeling}, (2) defining the initial conditions and determining the geometric and physical bounds of the simulation, and (3) solving the model to approximate the performance of interest. Examples of software tools used to solve physics-based simulations are ANSYS Mechanical, Abaqus FEA, and Altair HyperMesh for FEA, ANSYS Fluent, Autodesk CFD, and SimScale for CFD, MATLAB \& Simulink (Simscape Multibody), MSC ADAMS, and RecurDyn for MBD, and COMSOL Multiphysics and MSC Nastran for coupled multiphysics simulation. A few noteworthy open-source packages relevant to digital twin are OpenFOAM and Stanford University's Unstructured (SU2) package for CFD, MBDyn for MBD, and for multiphysics simulation, Advanced Simulation Library (ASL) and Chrono open-source options.


It is important to strike a balance between accuracy and computational cost when choosing the level of fidelity for a physics-based simulation. No matter how high the fidelity of a physics-based model might be, it is still only an approximation of reality because of missing physics, flawed assumptions, finite mesh discretization, etc. If the resulting model bias is large (likely for a low-fidelity model), it is important to quantify the bias offline and compensate for it online (see Sec. 2.1.2 in Part 2 of the review paper for a detailed discussion on the offline calibration of dynamic system models). Furthermore, it may be counter-intuitive to say that high-fidelity models are not always preferred even when computational costs incurred in DS, P2V, and V2P are not a concern. Practitioners need to keep in mind that high-fidelity models contain larger numbers of parameters that need to be calibrated offline, and during online deployment, a subset of the parameters need to be further tuned. Offline validation and online fine tuning requires the use of previously collected data and measurements. However, most of the time, very limited data is available for both offline model parameter estimation and online fine tuning. Practioners who deploy a high-fidelity model but have inadequate data to properly tune the parameters may find that model underperforms, and exhibits poor accuracy. As a result, practical implementations of physics-based models are (1) a lower-fidelity model that offers the desired accuracy at a lower cost and with a smaller number of parameters for offline calibration and (2) a small and critical subset of the model parameters are identified for online updating, reducing the amount of data required (see also a discussion in Sec. \ref{sec:probabilistic_updating} (a)).

\subsection{Data-driven modeling}
\label{sec:data_driven_modeling}

Data-driven models are needed for modeling physics in a digital twin under either of two main scenarios: 
\begin{enumerate}
 \item the underlying physics is too complicated or is not fully understood, and as a result, building an accurate physics-based model is impossible; or
 \item the physics is well understood and can be modeled using available software, but the simulation is too computationally expensive or time consuming to be useful in a digital twin, especially when many model runs are required (such as with UQ tasks).
\end{enumerate}

\begin{figure*}[!ht]
  \centering
    {\includegraphics[scale=0.60]{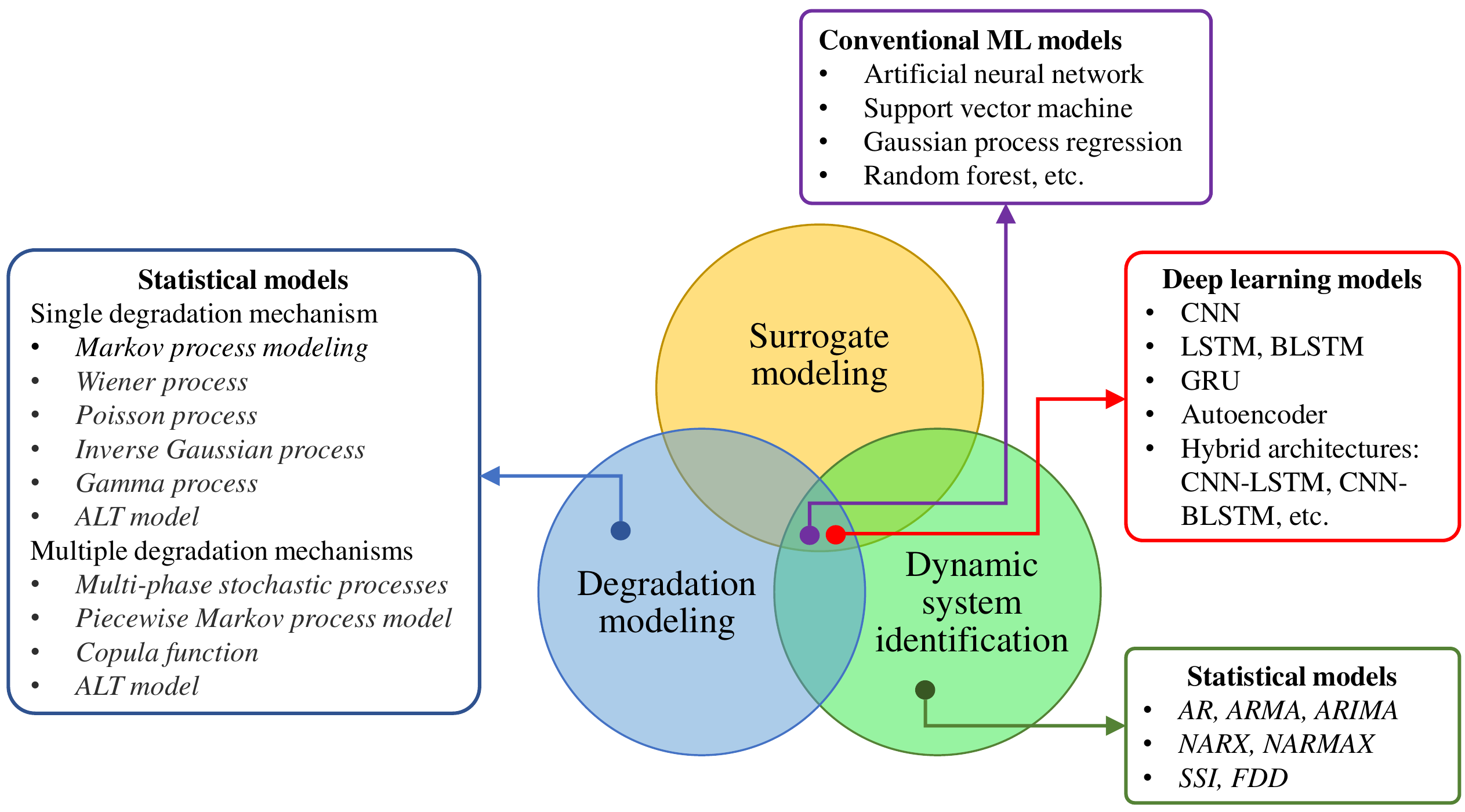}}
  \caption{Overview of data-driven models, where \emph{italic} font represents statistical-modeling methods; and \emph{non-italic} font indicates ML methods.}
  \label{fig:data_model}
\end{figure*}

In either scenario, data-driven models must be constructed to represent the underlying physical input/output behavior, enabling efficient prediction of digital states for real-time control and optimization in a digital twin. Shown in Fig. \ref{fig:data_model}, data-driven models can be grouped into three classes according to their application domain, namely (1) data-driven models for \emph{degradation modeling}, (2) data-driven \emph{surrogate models} which can entirely replace the physics-based models discussed in Sec. \ref{sec:physics_based_modeling}, and (3) data-driven models for \emph{dynamic system identification} using sensor measurements from a physical system (PS in Eq. (\ref{eq:DT})). Moreover, it is worth mentioning that system identification is fundamentally very similar to surrogate modeling. System identification is generally the determination of a model's characteristic parameters (regardless of model class) that minimize some predictive error metric, and one may note that essentially a surrogate model is trying to mimic a system's behavior also with a goal towards some minimized prediction error metric. The main difference between the two is that surrogate modeling is typically constructed and optimized using computer simulation data for any type of physical simulation at any time scale, while system identification mainly uses online sensor monitoring data and/or offline experimental data collected to tune the specific model's parameter(s). The close similarity is why many of the surrogate modeling methods and system identification methods can be used interchangeably. This relationship is further explained in Fig. \ref{fig:data_model}.

Data-driven models for use in digital twins can be classified into two categories according to the fundamental theory and construction (e.g. numerical or analytical), namely \emph{statistical models} and \emph{ML models}. Statistical models and ML models have been used for both prediction and inference \citep{ij2018statistics}. \lq\lq \emph{Statistical methods have a long-standing focus on inference, which is achieved through the creation and fitting of a project-specific probability model, while ML concentrates on prediction by using general-purpose learning algorithms to find patterns in often rich and unwieldy data}" \citep{ij2018statistics,bzdok2017machine}. In what follows, we discuss the latest data-driven modeling trends and provide insights on how each method can or has been used in a digital twin framework.

\subsubsection{\textbf{Statistical models}} Statistical models play a vital role in establishing both P2V and V2P connections in digital twins. For P2V connection, statistical models have been used in the context of dynamic system identification to build state-space models based on measurement data. For V2P connection, statistical models have been extensively used to build degradation models for failure prognostics and predictive maintenance of physical systems.

\begin{itemize}
\item \emph{Statistical models for dynamic system identification (P2V).} System identification is the process of learning a mathematical representation of a dynamic system based on pairs of measured input/output data  \citep{ljung1998system}. System identification has been extensively studied using statistical methods in the past decade, especially for linear dynamic systems.

\hspace{0.5cm} Many approaches have been developed for linear system identification in both the time domain and frequency domain \citep{ljung1981frequency}. In the time domain, simple models, like auto-regressive moving average models (ARMA), have been studied for the identification of dynamic systems with stationary responses \citep{rojo2004support,martinez2006support}. To overcome the stationary limitation of ARMA, auto-regressive \emph{integrated} moving average (ARIMA) models have been developed for non-stationary dynamic systems \citep{xuemei2010hybrid}. We highlight the \emph{I} in ARIMA as it denotes the model's ability to model trending data as a series of differences, making it a much more versatile statistical model. Other widely used approaches for time-domain system identification include stochastic subspace identification (SSI) methods \citep{peeters1999reference} and eigensystem realization algorithms \citep{juang1985eigensystem}. In the frequency domain, commonly used modeling methods include frequency domain decomposition (FDD) \citep{brincker2001modal} and the least squares complex frequency domain method \citep{guillaume2003poly}. Among the mentioned linear system time-domain and frequency-domain methods, SSI, a parametric identification method based on stochastic state-space models, is one of the most widely used approaches. Many variants of the SSI method have been proposed in the past decade, such as reference-based SSI \citep{peeters1999reference} and covariance-driven SSI \citep{dohler2011pre}. A detailed explanation of SSI is available in \citep{peeters1999reference,brincker2006understanding}.

\hspace{0.5cm} Numerous efforts have also been made to develop statistical methods for nonlinear system identification \citep{billings2013nonlinear,piroddi2003identification,worden2018confidence}. The most widely used methods include the nonlinear autoregressive exogenous (NARX) model \citep{piroddi2003identification}, and the  nonlinear autoregressive moving average model with exogenous inputs (NARMAX) model \citep{billings2013nonlinear}. In a NARX model, the value of the current output $y_k$ is represented as a nonlinear function of past values of the output (i.e., ${y_{k - 1}},\;{y_{k - 2,}},\; \cdots ,{y_{k - p}}$) and current and past values of exogenous inputs (i.e., ${u_k},\;{u_{k - 1}},\; \cdots ,\;{u_{k - q}}$) as follows
\begin{equation}
\label{eq:NARX_model}
\begin{split}
 & {y_k} = g({y_{k - 1}},\;{y_{k - 2,}},\; \cdots ,{y_{k - p}},\;  \cr 
& \;\;\;\;\;\;\;\;{u_k},\;{u_{k - 1}},\; \cdots ,\;{u_{k - q}}) + {\varepsilon _k},\cr
\end{split}
\end{equation}
where $g(\cdot)$ is a nonlinear function, such as a nonlinear polynomial function, $p$ and $q$ are respectively the number of lags of the output and the exogenous inputs included in $g(\cdot)$, and ${\varepsilon _k}$ is a Gaussian random variable with zero mean and unknown standard deviation estimated based on observations along with parameters of $g(\cdot)$. In a NARMAX model, the noise sequences (i.e., ${\varepsilon _k}, {\varepsilon _{k-1}}, \cdots , {\varepsilon _{k-e}}$) are also included in $g(\cdot)$ in addition to past values of exogenous inputs and the response \citep{billings2013nonlinear}. When an ML method is employed to construct the nonlinear function $g(\cdot)$, the nonlinear system identification methods fall into the intersection of statistical methods and ML methods \citep{worden2018confidence}.  

\hspace{0.4cm} There are also several review papers summarizing various system identification approaches. Namely, \cite{aastrom1971system} conducted a detailed review of various system identification methods of dynamic systems, \cite{ljung2010perspectives} shared his perspectives on system identification, \cite{sirca2012system} reviewed approaches for system identification in structural engineering, and \cite{kerschen2006past} conducted a comprehensive review on developments in system identification of nonlinear dynamical structures and illustrated them using numerical and experimental examples.

\hspace{0.5cm} The dynamic system models identified using the system identification methods reviewed above can enable real-time state estimation and control, a highly desirable ability of a digital twin. For instance, \cite{he2019data} used a discrete-time state-space model identified from data to model various component-level dynamics in constructing a virtual system in their digital twin model. Similarly, \cite{xia2021intelligent} used the MATLAB system identification toolbox (see Table 1 in Part 2 of the review paper) to build digital models for their digital twin.

\item \emph{Statistical methods for degradation modeling in predictive maintenance (V2P).} One of the most important elements in digital twins is degradation modeling. Degradation modeling is vital for predicting the end of life of a physical system to enable predictive maintenance (i.e., V2P). In some cases, degradation mechanisms of the physical system involve complicated physics which may not be fully understood. In these scenarios, a physics-based degradation model cannot be constructed, and an empirical statistical model must be used instead. Empirical modeling leverages historical data collected from failed units to construct a degradation model. Shown in Fig. \ref{fig:data_model}, many statistical models have been proposed for degradation modeling of both single and multiple failure mechanisms. For example, numerous stochastic process models, including Markov models \citep{giorgio2011age,lee2010online,vega2021novel}, semi-Markov models \citep{kharoufeh2010semi,liu2018diagnosis,compare2016semi}, Wiener process, Poisson process, inverse Gaussian process, Gamma process, and accelerated life testing (ALT) models \citep{hu2018sequential}, have been developed for degradation modeling of single degradation mechanisms \citep{ye2015stochastic,gorjian2010review}. For degradation modeling with multiple degradation mechanisms, multi-phase stochastic process models \citep{liao2021remaining}, piece-wise Markov process models \citep{vega2020optimal}, copula functions \citep{peng2016bivariate}, and system-level ALT models \citep{moustafa2021system}, have been proposed. 

\hspace{0.5cm} While statistical degradation models allow for effective prediction of the end of life of physical systems, there are several challenges that need to be addressed in order for some of the models to be used in a digital twin. For example, a Markov degradation model usually describes the degradation as a set of highly abstract discrete states, such as ratings given in letters (e.g., A, B, C, D). It is often difficult to directly use online sensor data, collected from the physical system continuously and in real-time, to update the highly abstract discrete states. This poses a real challenge in seamlessly establishing the P2V connection in a digital twin system (see Fig. \ref{fig:digital twin}). To tackle this challenge, researchers have proposed approaches integrating Bayesian model updating methods reviewed in Sec. \ref{sec:probabilistic_updating} with Markov chain model for model updating and failure prognostics in \cite{chiachio2020markov, vega2021novel}. In addition, the uncertainty of a degradation model needs to be properly quantified in order for such models to be fully integrated into a maintenance decision making framework. Accurately quantifying the uncertainty of prognostic models is essential to the effective integration of these models into digital twins for predictive maintenance, especially for high-value physical assets and in safety-critical applications, as will be discussed in Sec. \ref{sec:predictive maintenance}. A detailed discussion on the role of UQ in digital twins will be discussed in Section 2.1 of Part 2 of the review paper.

\end{itemize}

\subsubsection{\textbf{ML models}}
\label{sec:ML_models}
With the increase in IIoT connected devices, big datasets have become commonplace, and advanced data analysis methods have been developed to take advantage of the collected knowledge. ML models, which can learn from a dataset and be used to predict on new, unseen data, have become a popular method for modeling physical systems in a digital twin \citep{min2019machine,ibrahim2020machine, liu2020digital,priyanka2022digital,elayan2021digital,ladj2021knowledge,hu2021toward}. As shown in Fig. \ref{fig:data_model}, ML methods can be further categorized into conventional ML methods and newer, emerging DL methods.

\emph{Conventional ML methods}, as indicated in Fig. \ref{fig:data_model}, include feed-forward neural networks \citep{chen1992neural}, support vector machines \citep{zhang2006robust}, random forests \citep{chai2019enhanced}, and Gaussian process regression \citep{worden2018confidence}. All of these methods have been used in digital twins for (1) degradation modeling, (2) surrogate modeling to replace  first-principles models, and (3) dynamic system identification from data. For example,  for surrogate modeling, \cite{chakraborty2021machine,molinaro2021embedding} used Gaussian process regression models to replace computational simulations in digital twins, and thereby enable real-time prediction; \cite{zhu2021digital} trained ML models to build a digital workpiece model for a manufacturing system digital twin; ~\cite{zhou2022real} investigated using multiple ML approaches to characterize the relationship between crack size/shape and crack-front stress intensity factors in a helicopter. For the identification of dynamic systems, particularly nonlinear systems, various ML-based approaches have been proposed by following the NARX format given in Eq. (\ref{eq:NARX_model}). When a Gaussian process regression model is used as $g(\cdot)$ in Eq. (\ref{eq:NARX_model}), it is called a GP-NARX model \citep{doerr2018probabilistic}. When feed-forward neural networks are used, it is called a NARX-net \citep{lin1996learning}.  

As a subarea of ML, \emph{DL} is a relatively new modeling technique which uses large multi-connected networks with many parameters to approximate physical process or functions ~\citep{lecun2015deep}. Deep learning models are easily reconfigured to have any number of parameters or architecture, and for this reason, they have a significant advantage over traditional ML models when dealing with massive high-dimensional heterogeneous datasets ~\citep{tang2020deep,rassolkin2019digital,hu2022new}. As shown in Fig. \ref{fig:data_model} and similar to conventional ML models, DL models have been applied in various digital twin applications for degradation modeling, surrogate modeling, and system identification. In particular, the application of DL models for \emph{degradation modeling} to establish a V2P connection in a digital twin has gained much attention recently. Applications of ML and DL models for degradation modeling in failure prognostics and predictive maintenance will be reviewed and discussed in detail in Secs. \ref{sec:failure_prognostics} and \ref{sec:predictive maintenance}. Here, we briefly summarize the application of DL models for surrogate modeling and dynamic system identification.

\begin{itemize}
\item In recent years, a large number of DL methods have been proposed for \emph{surrogate modeling}. DL models can replace many expensive physics-based simulations by approximating the underlying physics using the highly-interconnected model parameters. For example, ~\citet{guo2022deep} developed a surrogate model for thermal signature prediction in laser metal deposition using DL models. \cite{tang2020deep} proposed a DL-based surrogate modeling method to predict dynamic subsurface flow in channelized geological models. Similarly, \cite{weber2020deep} developed a DL surrogate model for earth system simulation models. In deep neural network-based surrogate modeling methods, it is worth highlighting a particular type of neural network -- long short-term memory (LSTM) networks, which are a type of recurrent neural network. Unlike regular feed-forward neural networks, LSTM has a feedback loop inside each cell which allows information to persist in time throughout the network \citep{hochreiter1997long}. As a result, it can capture long-term dependencies in time series data. Because of this appealing feature, LSTM networks have been extensively leveraged to build surrogate models for sequence-to-sequence prediction with different number of inputs and outputs (e.g., one to many, many to one, and many to many), such as in aerodynamics \citep{zhang2021aerodynamic}, elasto-plastic finite element simulation \citep{im2021surrogate}, and microstructure evolution applications \citep{montes2021accelerating}, among others. Moreover, numerous hybrid DL architectures, such as convolutional neural network (CNN)-LSTM \citep{zhao2022surrogate} and CNN-Bidirectional LSTM \citep{zhang2021novel}, have been proposed in recent years for surrogate modeling of various physical systems.

\item For \emph{system identification}, DL models are becoming more popular due to their advantage over conventional ML and statistics-based models in handling high-dimensional and heterogeneous data sources, such as videos and images. As mentioned above, the main difference between system identification and surrogate modeling is that system identification is mainly done using sensor data collected during online operation, whereas surrogate modeling may be done with either physics-based simulation data or sensor measurement data.

\hspace{0.4cm} There are two main approaches for system identification using DL. In the first approach, DL models are used to learn the \emph{input-output} relationship directly in a nonlinear autoregressive exogenous scheme. In this way, the DL model learns from data to act as a black-box state-space model of the system. The learned states and state transition functions are encoded in the DL model's parameters and are typically not easy to visualize because the model's contain many thousands of parameters. ~\citet{zhang2020bayesian} used an LSTM DL model to learn the relationship between the measured aircraft velocity and flight trajectory to facilitate the safety assessment of en-route flight. ~\citet{ma2021self} used a probabilistic LSTM to learn the relationship between geometric sensor data and thermal elongation.~\cite{hu2022grasps} developed a grasps generation-and-selection CNN to facilitate performance assessment of robotic grasps in the context of digital twin, where grasping state (i.e., position, rotation angle, and gripper width) was captured and characterized in the form of red–green-blue-depth images. \cite{wu2019deep} presented a deep convolution neural network-based method for system identification of structural dynamic systems. They showed that CNN-based system identification methods are more robust than conventional multilayer perception neural networks-based methods when the signals are contaminated by large noise.

\hspace{0.4cm} In the second approach to system identification using DL, the state-space model of a system is learned in a more easily interpreted, low-dimensional form. The most popular DL model for this is the autoencoder \citep{masti2021learning,li2021hierarchical,otto2019linearly}. Shown in Fig. \ref{fig:autoencoder}, an autoencoder is given the name \emph{encoder} because it takes high dimensional inputs $I_{k-1}$ and encodes them as a low-dimensional state vector $x_k$. The low-dimensional state vector can easily be viewed and visualized since it is of much lower dimensionality than the system inputs and outputs. To estimate the system outputs $O_k$, the state vector $x_k$ is decoded in the decoder. In a recurrent fashion, the encoded state vector $x_k$ is combined with control system inputs $u_k$ using another small neural network which learns the state transition function $x_{k+1}=f(x_k, u_k)$ to update the state estimate,  $x_{k+1}$, at the next time step. The specifically designed high- and low-dimensional parts of recurrent autoencoder networks offer practitoners the ability to work with high-dimensional inputs and outputs while still maintaining a better level of clarity and visualization by observing the low-dimensional states $x$. Autoencoders and similar DL models will play a vital role in understanding the many complex input/output relationships in future digital twin frameworks.

\begin{figure}[!ht]
  \centering
    {\includegraphics[scale=0.5]{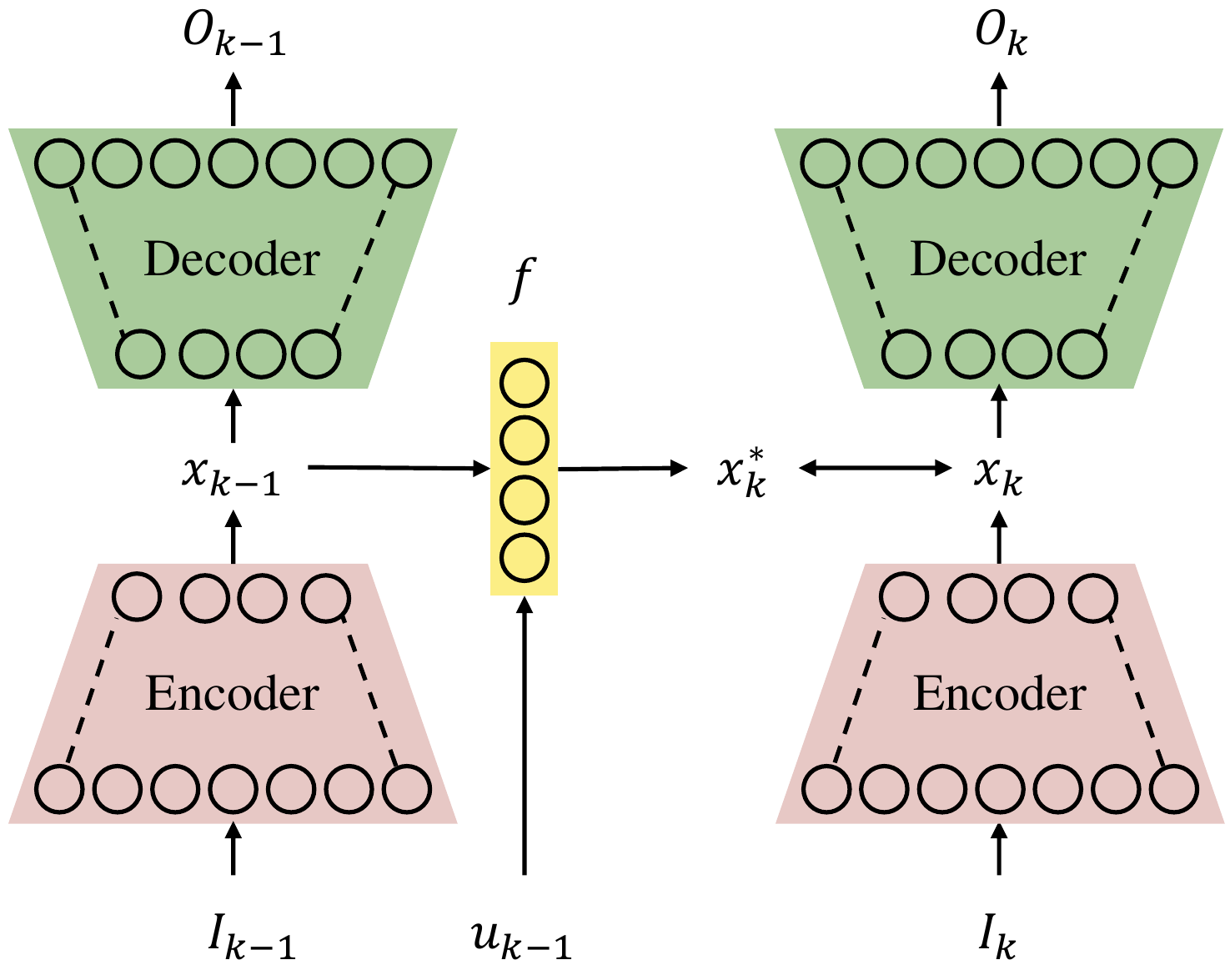}}
  \caption{A recurrent autoencoder-based DL model for system identification of a nonlinear dynamic system \citep{masti2021learning}.}
  \label{fig:autoencoder}
\end{figure}
 
\end{itemize} 

We conclude our discussion on data-driven modeling by listing a few topics that are crucial for the successful deployment of data-driven models in a digital twin.

\begin{enumerate}
\item \textbf{Model selection}: Even though both statistical models and ML models are getting unprecedented attention in recent years, they have their own advantages and disadvantages. For example, statistical models are highly explainable while complex DL models are difficult to interpret and are often thus used in a black-box way. There are so many data-driven models available, however, selection of an appropriate data-driven model should always follow the “Occam’s Razor” principle. For example, if a conventional ML model, such as a Gaussian process regression model, can be employed to build an accurate data-driven model, there is no need to construct a very complicated DL model for the same task. The selection of an appropriate data-driven model for a specific application of digital twin is an important optimization problem. 
\item \textbf{UQ of ML models}: George E.P. Box famously said “All models are wrong, but some are useful,” as recorded in~\cite{box1976science}. This is also true for all data-driven models including statistical models, conventional ML models, and DL models. It is therefore important to quantify the uncertainty in data-driven models, especially ML models. UQ of ML models allows us to quantify the confidence level of ML predictions. It plays a vital role in determining when the model prediction will go wrong and how bad the prediction is if it is wrong. A detailed discussion of such an important topic is given in Section 2.1.1 in Part 2 of the review paper.

\item \textbf{Data collection}: Data is essential for the learning of data-driven ML models. The quantity and quality of available data significantly affect the performance of data-driven models, and eventually the effectiveness of digital twins. In engineering practice, high-fidelity physics-based simulations may be computationally  very expensive and physical experiments may be very time- and resource-consuming. In those situations, it is difficult to collect large volumes of high quality training data to build data-driven models. Ways to tackle this challenge include (1) physics-informed ML, reviewed in Sec. \ref{sec:PI_ML}; (2) active learning where data are collected adaptively to refine data-driven models in a data-efficient manner \citep{cohn1996active}, with \cite{bichon2008efficient} being the first application of active learning to reliability analysis under uncertainty; and (3) online model updating, as will be reviewed in Sec. \ref{sec:ML_updating}, where an ML model is first trained using a small amount of data offline and then is updated using data collected online to improve the prediction accuracy. 
\end{enumerate}

\subsection{Physics-informed ML}
\label{sec:PI_ML}
\begin{figure*}[!ht]
  \centering
    {\includegraphics[scale=0.65]{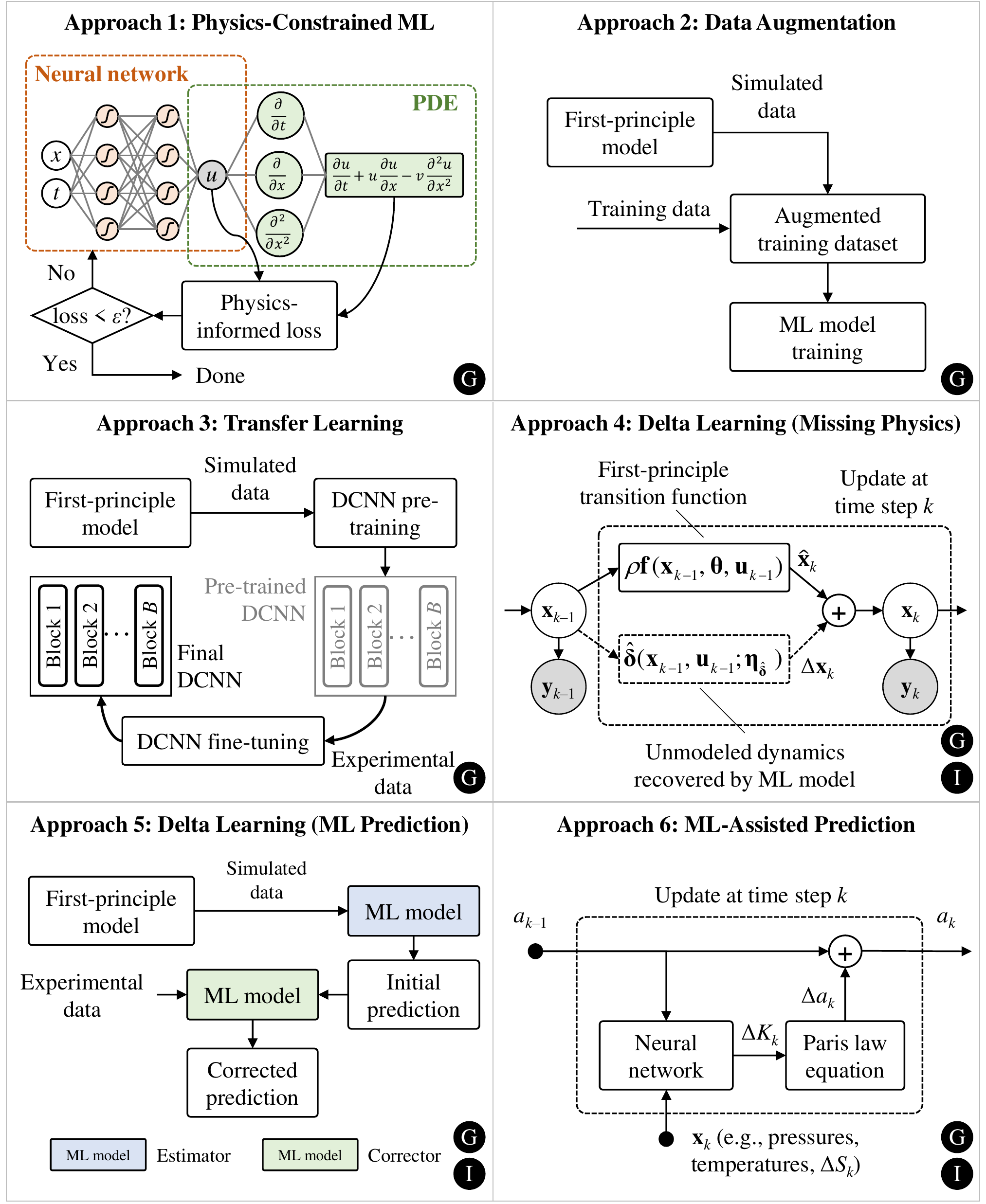}}
  \caption{Six approaches to construct a hybrid of a physics-based and data-driven ML model: (1) Modifying the loss function of an ML model by adding a physics-informed penalty term \citep{karniadakis2021physics}; (2) Generating synthetic data using first-principle simulations \citep{ritto2021digital}; (3) Pre-training an ML model on physics-based synthetic data and then fine-tuning it on experimental data \citep{shen2020deep,kapusuzoglu2020physics,huang2022transfer}; (4) Correcting a first-principle model by learning unmodeled physics from data \citep{jiang2022model}; (5) Correcting an ML model, trained with physics-based synthetic data, by learning its prediction residual from experimental data \citep{thelen2022integrating}; and (6) Learning to predict the input of a first-principle model \citep{yucesan2022hybrid}.  The icon with ``G" inside a black circle indicates improvement in generalization over pure ML; the icon with ``I" inside indicates improvement in interpretability.}
  \label{fig:hybrid_model}
\end{figure*}

For some aspects of the physics in digital twins, both data from first-principle (or physics-based) simulations and data collected from a physical system may be available. When both physics and data are present, methods presented in Sec. \ref{sec:physics_based_modeling} and Sec. \ref{sec:data_driven_modeling} can be integrated, resulting in hybrid physics-ML approaches for modeling physical systems. For instance, \cite{luo2020hybrid} fused a first-principle model with operational data to build a multi-domain model in a digital twin. Similar approaches have been attempted, sometimes labeled as physics-informed ML, for physical system modeling in digital twins \citep{liu2018role,yu2020hybrid,ritto2021digital,broo2022design}. Incorporating physical knowledge into data-driven ML models can offer several attractive benefits, including better generalization of ML models to unseen ``out-of-distribution'' test data, as will be discussed in detail in \ref{sec:ML_updating}, the ability to explain the underlying physics of the predictions from the ML models, and speeding up the training process because some physical laws can restrict the feasible region of solutions and limit the values the model parameters can take. According to the way that a first-principle model is integrated with a data-driven ML model, existing hybrid approaches for physical system modeling can be broadly classified into six approaches, shown in Fig. \ref{fig:hybrid_model} and explained in what follows.
\begin{itemize}
    \item \textbf{Approach 1: Physics-Informed Loss Function} This approach modifies the loss function of a data-driven ML model by adding a physics-informed loss term that penalizes model predictions not compliant with first principles (or physics), thus constraining the training of the ML model toward solutions that comply with physics \citep{raissi2019physics,jagtap2020adaptive,lu2021deepxde}. This approach is the most common way to build hybrid physics-ML models and has found success in many digital twin applications such as solving PDEs and surrogate modeling \citep{raissi2019physics, zhu2019physics, haghighat2021physics, gao2021phygeonet, zhou2022generic}, inverse modeling \citep{raissi2019deep, jagtap2020conservative}, fault diagnostics \citep{shen2021physics}, and degradation modeling \citep{yan2022integration}. This method of physics-informed ML is different than the other approaches in that it does not necessarily require any training data from a physics-based simulation, a significant advantage over other methods.
    
    \item \textbf{Approach 2: Data Augmentation} This second approach first runs first-principle simulations to generate data at various states an operating conditions of a physical system. Examples of states include  healthy, degraded/damaged, and failed for predictive maintenance applications. It then combines the generated data (or synthetic data) with actual data (e.g., experimental data) to create an augmented training dataset for training data-driven ML models. An example that falls into this category is the digital twin model presented by \cite{ritto2021digital}. The authors generated different damage scenarios using first-principle simulations to augment a training dataset for an ML classifier used for damage detection of a bar structure. Another example is the application of physics-informed ML to degradation diagnostics of lithium-ion batteries, where a simple physics-based half-cell model was used to generate high-degradation simulated data to allow training ML models on small, early-life experimental data \citep{thelen2021physics}. Using physics-based synthetic data to train an ML model partially constrains the ML model’s output with the physics carried by the first-principle model, thereby improving the generalization performance of the ML model. Additionally, this data augmentation approach can reduce the data size required for training large-scale DL models. A necessary condition for the success of this approach is that the first-principle model captures physics highly relevant to the problem at hand. However, the model does not need to be highly accurate. 
    
    \item \textbf{Approach 3: Transfer Learning} This physics-informed ML approach is inspired by the concept of transfer learning, which has been explored in various digital twin applications, including multi-fidelity modeling, such as \cite{huang2022transfer} for melt pool modeling in additive manufacturing, process quality monitoring, such as \cite{kapusuzoglu2020physics} (the third type of strategy) for in-line quality monitoring of additive manufacturing processes, and data-driven prognostics such as \cite{zhang2018transfer} for turbofan engine prognostics, \cite{shen2020deep} for battery capacity estimation, and \cite{wang2021deformable} for bearing fault diagnosis. It first pre-trains an ML model using a large quantity of data from first-principle simulations. It then fine-tunes the model using a small quantity of real data (e.g., data from physical system experiments). Similar to the data augmentation approach, this transfer learning approach can improve the generalization of ML models and alleviate the data size requirement. Its success also highly depends on how relevant the physics captured by the first-principle model is to the classification/regression problem that needs to be solved. 
    
    \item \textbf{Approach 4: Delta Learning (Missing Physics)} The basic idea of this delta learning approach is to add a data-driven ML model, co-existing with a first-principle simulation model, that learns to recover the unmodeled physics. This hybrid modeling approach is commonly used in multi-fidelity modeling and quantification of simulation model discrepancy. For example, \cite{jiang2022model} added a polynomial chaos expansion model on top of a simplified physics-based degradation model for failure prognostics of a structural system called miter gates. The role of the polynomial chaos expansion model was to represent the discrepancy of the simplified physics-based degradation model due to unmodeled dynamics. \cite{yucesan2020physics} built a data-driven LSTM model to augment a physics-based fatigue model, which created a hybrid model for fatigue prognostics of wind turbines.
    
    \item \textbf{Approach 5: Delta Learning (ML Prediction)} In this approach, an ML model learns residuals on top of initial predictions by another ML model, trained using data generated by a physics-based model. The final predictions are the sum of the initial predictions and residuals. The initial predictions can also be made directly by a physics-based model \citep{zeng2020tossingbot}. The logic behind this hybrid physics-ML approach is two-fold: (1) the first-principle model use baseline physics to produce initial, population-based predictions that may not be accurate for individual system units but can generalize well to different operating conditions; and (2) adding the data-driven residuals to compensate for missing physics and unit-to-unit variability. The resulting performance gains are improved generalization over pure data-driven ML and improved prediction accuracy over purely physics-based modeling. Just like the above physics-informed ML approaches, this delta learning approach may not work well if the first-principle model does not capture relevant physics. Issues may also arise when the first-principle model cannot capture the effects of testing or operating conditions on experimental data.
    
    \item \textbf{Approach 6: ML-Assisted Prediction} This sixth approach to physics-informed ML uses a data-driven ML model to predict the input $\mathbf{x}$ or parameters $\boldsymbol{\uptheta}$ of a first-principle model. When implementing this approach in practice, the first step is identifying the input or a subset of the parameters as variables, $\mathbf{{x}}_{\mathrm{FPM}}$ or $\boldsymbol{{\uptheta}}_{\mathrm{FPM}}$, to be predicted by the ML model. Then, the ML model estimates or, in some cases, forecasts $\mathbf{{x}}_{\mathrm{FPM}}$ or $\boldsymbol{{\uptheta}}_{\mathrm{FPM}}$. Finally, the first-principle model uses the ML estimates or forecasts to predict the response of interest $y$ at current (estimation) or future (forecasting) time steps. A clear benefit of this approach is the physical interpretability, coming from (1) using physically meaningful variables as intermediate $y$, predicted by an ML model, and (2) using a first-principle model to predict the final $y$. Another advantage is the improved generalization performance, attributable to the use of the first-principle model. A typical challenge with this approach is the low identifiability of the first-principle model’s input or parameters from observational data. In extreme cases, large errors in the ML model’s estimates or forecasts propagate through the first-principle model, causing the final prediction performance to be worse than pure ML. Examples of Approach 6 can be found in grease degradation forecasting (ML) for bearing fatigue life prediction (first-principle modeling) \citep{yucesan2022hybrid} and degradation parameter forecasting (ML) for battery capacity fade prediction (first-principle modeling) \cite{ramadesigan2011parameter, downey2019physics, lui2021physics}. 
\end{itemize}

We note that the above list is not exhaustive but rather an incomplete collection of representative approaches to physics-informed ML. For example, an approach that could, in general, be categorized as Approach 6 is that the output and parameters of a first-principle model are used to augment the input of an ML model, as demonstrated for in-line quality monitoring of additive manufacturing processes in \citep{kapusuzoglu2020physics} (the second type of strategy), and turbofan engine diagnostics in \cite{chao2019hybrid} and prognostics in \cite{chao2022fusing}. The only difference is that the information flow is in the opposite direction (i.e., the first-principle model now feeds an additional input to the ML model). A seventh approach could be defining physics-based kernel functions for ML models, such as the FEA-Net \citep{yao2020fea} and physics-informed kernel functions for Gaussian process regression \citep{yang2018physics}. Interested readers can find more comprehensive and detailed reviews in review papers more dedicated to the topic of physics-informed ML, such as \cite{karniadakis2021physics} for computational physics, \cite{aykol2021perspective} for battery lifetime prediction, and \cite{willard2020integrating} for a comprehensive collection of engineering applications. Despite tremendous advances in combining physics and data for physical system modeling, much more research is needed to address practical challenges impeding industry-scale adoption of hybrid physics-ML modeling approaches. Among those challenges are (1) a lack of large, realistic datasets relevant to different digital twin applications for training, validating, and testing hybrid models and benchmarking their generalization and interpretation performance (see also the discussion in Section 4.2 of Part 2 of the review paper), (2) a shortage of digital twin benchmark problems that allow researchers and practitioners to compare prediction accuracy and computational efficiency of different hybrid approaches and determine their suitability for real-time physical system modeling, and (3) a lack of academia-industry collaborative effort in building open-access, well-maintained platforms that can streamline data sharing, model building, and benchmarking, such as many existing efforts to streamline ML workflows.


\subsection{System modeling}
\label{sec:sys_modeling}
A physical system usually consists of many subsystems and components that interact with each other in a variety of means. It is, therefore, important to model the high-level interactions and the overall interconnection structure among different components in the physical system. This section briefly reviews two widely used techniques for modeling interactions in physical systems.

\paragraph{(a) Unified Modeling Language (UML) and Systems Modeling Language (SysML)}
Towards systems modeling, the Unified Modeling Language (UML) is a general-purpose language that is commonly used for software development in model-based systems engineering (MBSE) \citep{ramos2011model}. In brief, UML is a graphical modeling language that provides a standardized way to specify the structure of a software system and visualize the interactions of different models in the system \citep{glatt2021modeling,weilkiens2011systems}. In engineered systems, UML has been increasingly utilized to characterize the design of a system with regard to its high-level structure, behavior, and interactions. To this end, different diagrams are created to reflect the model in line with the intended purpose from a certain perspective, and these diagrams can be grouped into two categories:
\begin{enumerate}
    \item \textbf{Structure diagram}: It describes the static structure of elements in a system, which is often adopted to document the architecture of software systems. Structure diagram includes class diagram, component diagram, object diagram, deployment diagram, etc. Take the class diagram as an example, it is often used to describe a set of classes, interfaces, and collaborations and their relationships in object-oriented software systems. For example, both classes students and professors belong to a super class person, and the relationship between professors and students is that professors supervise students.
    
    \item \textbf{Behavior diagram}: Differing from the structure diagram, behavior diagram emphasizes on characterizing and visualizing the dynamic aspects of the system. It is composed of activity diagram, use case diagram, sequence diagram, state chart. Take the same students versus professors as an example, in the case of course taking, the sequence diagram illustrates the very basic steps in this task ordered by the timeline, starting from the very beginning course registration to homework assignment/submission and homework grading and report. In short, the sequence diagram summarizes the dynamic interactions between different users (i.e., students, professors, administrators) and applications (i.e., course registration, homework submission, homework grading and report) over time. 
\end{enumerate}

The promise of UML in modeling complex system interactions has been demonstrated in multiple studies. For example,~\citet{luo2019digital} leveraged UML to characterize the connections among multiple physical subsystems (i.e., mechanical, electrical, hydraulic) in the digital twin of a CNC machine tool. Note that UML is intensively used for interaction modeling, focusing heavily on the software side. 

Differing from UML, the Systems Modeling Language (SysML) is independent of the specific discipline (e.g., software, hardware, personnel, or facilities), and it is a general-purpose architecture modeling language in MBSE through extending and omitting certain UML elements~\citep{weilkiens2011systems}. As driven by user requirements, SysML consists of four key pillars: structure (e.g., system hierarchies), behavior (e.g., state machine), properties (e.g., attributes of time variables), and requirements (e.g., requirements' hierarchies and traceability). Using SysML facilitates the design, specification, analysis, verification, and validation of engineering products and processes across a wide range of systems and systems-of-systems. Thanks to SysML's rich features in cross-system extensibility and traceability, there is an increasing number of studies investigating how SysML is adopted in digital twins~\citep{sun2020digital}. For example,~\cite{bachelor2019model} leveraged the traceability of the allocation of requirements to system function definitions in SysML, and used a digital twin model of an ice protection system to deal with changes in system design through data traceability across the abstracted lifecycle. \cite{zhang2022multi} used SysML as a model fusion tool to establish the relationship among various domain models in a shop-floor digital twin. A similar approach is also developed by \cite{liu2021construction} to model the interaction of different models in Unity3D for their shop-floor digital twin model.

In digital twins, as software elements (i.e., simulation models, database, machine learning models, decision support modules) play increasingly important roles, UML and SysML have been frequently used to manage the growing software complexity, where a unified interface is provided to describe the structure, behavior, and interactions among software modules. Such architecture-level view of all elements and their interdependencies in the overall system brings multiple benefits, such as ensuring the planned functionality and the envisioned benefits, preventing delays of the actual implementation due to planning errors, easing project transfer due to employee leave. The state-of-the-art literature has witnessed prominent adoptions of UML/SysML in the manufacturing industry. A representative usage of UML is showcased by~\citet{glatt2021modeling}, where they considered the digital twin of cyber-physical production systems. As the digital system contained a significant number of elements and complex relationships between those elements,~\citet{glatt2021modeling} adopted a UML-based approach to model system structure and component interactions associated with different system functions (i.e., monitoring, diagnostics, prediction). The unified representation of class relationships and system workflow substantially eases the understanding of the object-oriented software systems in digital twins. 


\paragraph{(b) Ontology}
\begin{figure*}[!ht]
    \centering
    \includegraphics[scale=0.50]{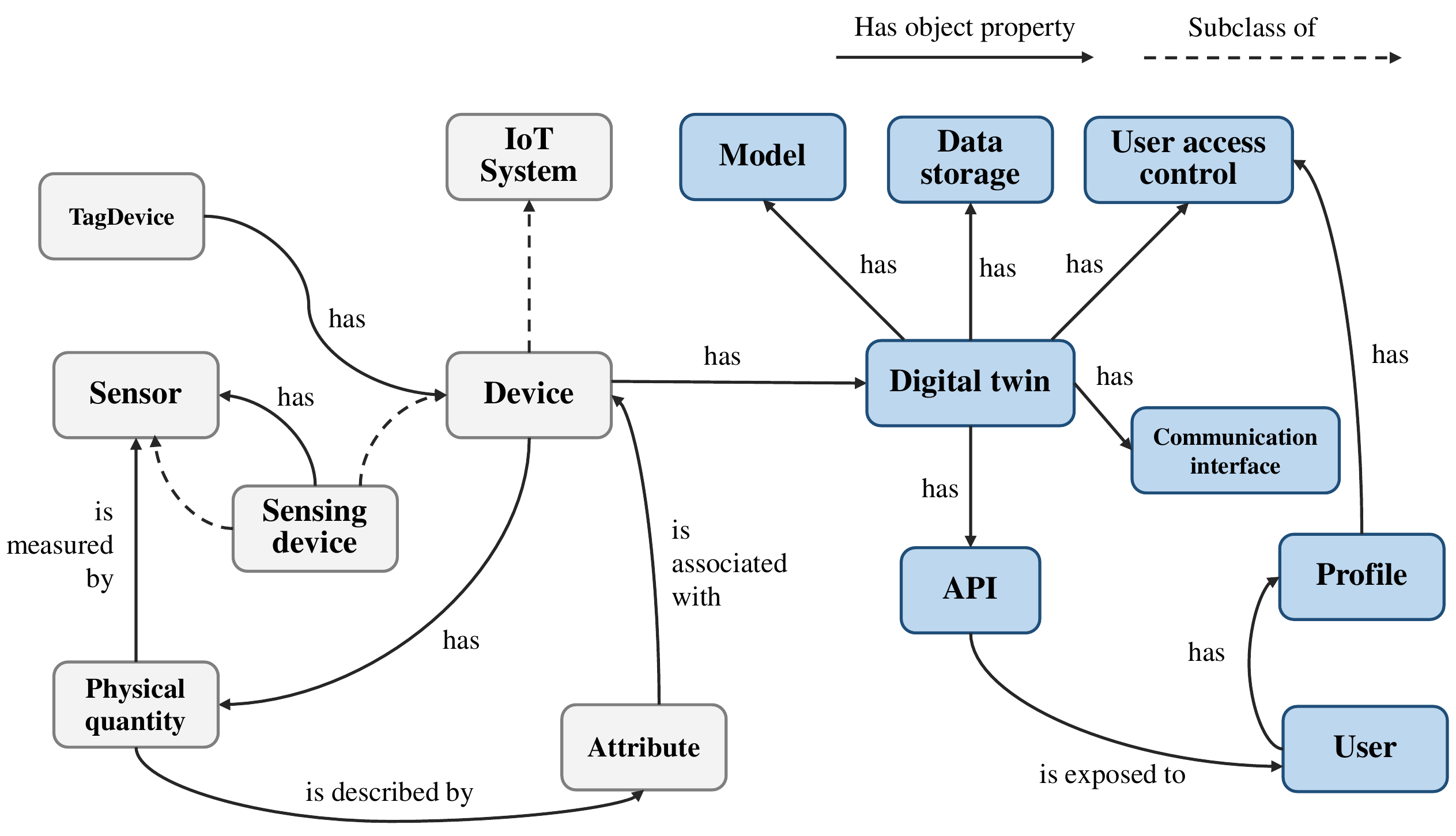}
    \caption{A simple ontology model for digital twins. Adopted from~\cite{steinmetz2018internet}.}
    \label{fig:ontology_example}
\end{figure*}


In computer science and information science, ontology is a prevailing knowledge engineering and representation paradigm to describe the properties of a subject area and define how they are related to each other with explicit descriptions and specifications. Towards this end, ontology encompasses a set of domain-specific concepts, entities, and categories as well as their definitions, attributes, and interrelationships to represent the subject matter of concern. Suppose we are interested in building an ontology map for an IoT system, and we have the basic information related to each device and its digital counterpart. Then a tentative ontology map can be constructed like Fig. \ref{fig:ontology_example}. In this figure, rectangles denote the distinct types of entities (i.e., device, sensor, and attribute), and arrows represent the relationships between entities (e.g., each sensor has a sensing device). 

Note that ontology maps are generalized data models in the sense that they only account for general types of entities and their abstractions with certain properties in common. But they do not include any specific information about an entity in the domain. For example, rather than focus on the characteristics of a specific device, an ontology map concentrates on the general concept of devices by capturing common characteristics that most devices have. The construction of such a unifying conceptual framework significantly fosters the communication and sharing of the structure of information specific to each domain that is understandable by both humans and machines~\citep{castells2006adaptation,compton2012ssn}. The advantage of ontology in enabling the sharing and reuse of domain knowledge has been one of the key drivers in the recent surge in ontology applications, which has been broadly used in machine learning and information science as a popular means of knowledge representation~\citep{ji2021survey}. For example, ~\cite{akroyd2021universal} demonstrated a digital twin built with a general-purpose dynamic knowledge graph in the World Avatar project, where ontology-based knowledge representation was repeatedly exploited to facilitate a unifiable interface for data queries.

Knowledge graphs, being a popular way to instantiate ontology, are commonly used to connect concepts in ontologies with specific data for knowledge engineering and information retrieval~\citep{wang2017knowledge,lin2015learning}. In a knowledge graph, nodes and edges represent the instances of concepts and the relationships between instantiated entities, respectively. Attributable to knowledge graphs' power in querying, inferring, and generating knowledge in the semantic space, they have been adopted as a key enabling technology to achieve semantic interoperability of heterogeneous data and information in engineered systems ~\citep{noy2019industry,psarommatis2021generic}. For example,~\cite{banerjee2017generating} exploited the reasoning power in a knowledge graph and developed a query language to extract and infer knowledge from large-scale production line data to support manufacturing process management in digital twins.

In the context of digital twin, ontology and knowledge graphs offer a common language to uniquely refer to each possible object (i.e., component or sensor) in the physical asset. Doing this greatly enhances our flexibility in describing the evolution of digital twin when it is subject to systems/subsystems/components changes (e.g., retrofitting) or modifications in the asset management process (e.g., monitoring) over the entire life-cycle. For example,~\cite{erkoyuncu2020design} demonstrated the benefits of an ontology-based framework explicitly, where ontology was used to enable the co-evolution of the digital twin with its physical asset. 



\section{Physical-to-virtual (P2V) twinning enabling technologies}
\label{sec:physical2virtual}
\begin{figure*}[!ht]
    \centering
    \includegraphics[scale=0.56]{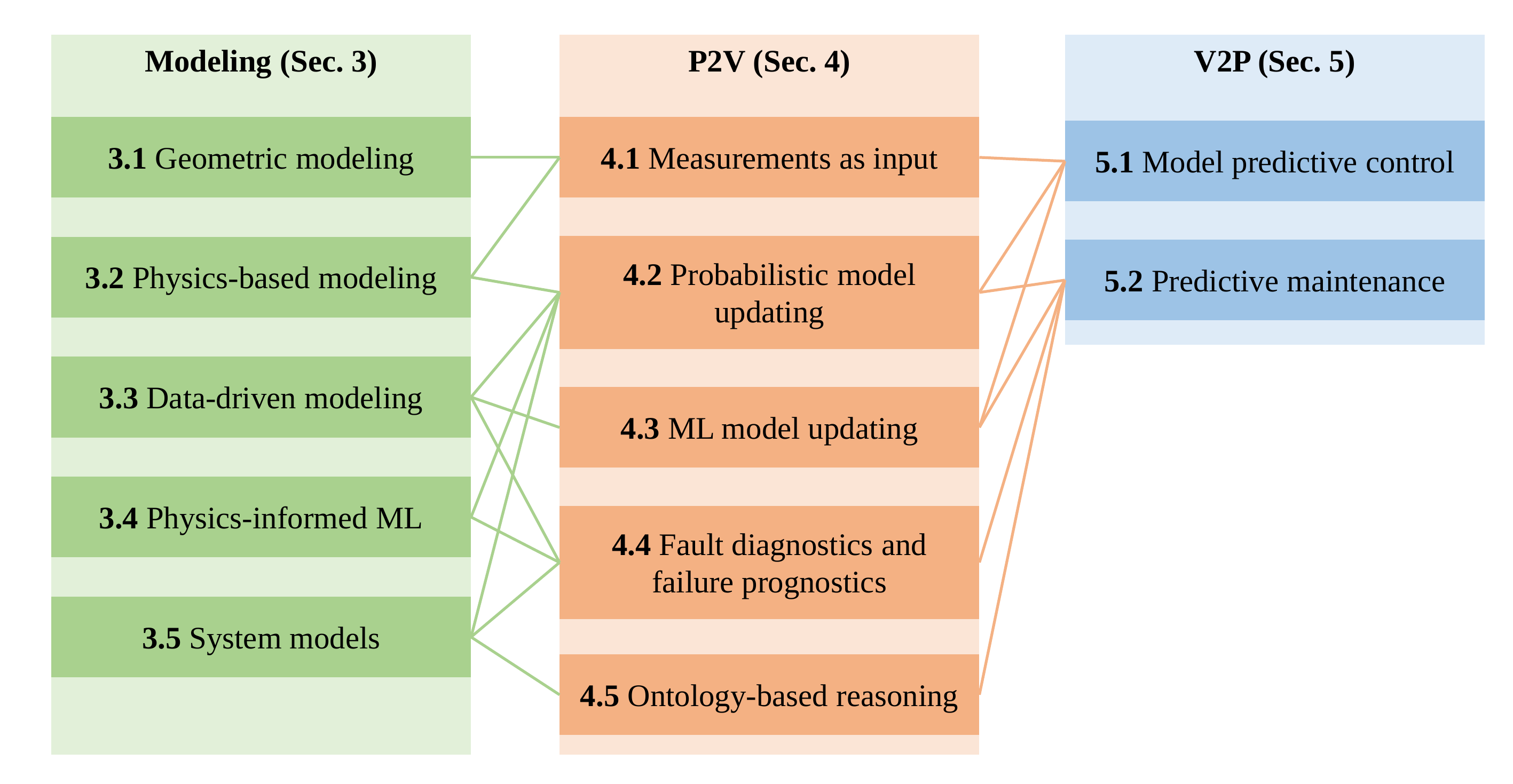}
    \caption{Relationship between modeling methods, P2V methods, and V2P methods.}
    \label{fig:connection}
\end{figure*}

In this section, we summarize five categories of widely employed physical-to-virtual twinning methods. The P2V twinning methods vary depending on the modeling approach used. Fig. \ref{fig:connection} presents the relationship between the different modeling methods reviewed in Sec. \ref{sec3}, the P2V methods reviewed in this section, and the V2P methods that will be discussed in Sec. \ref{sec:V2P}. Shown in the Fig. \ref{fig:connection}, some of the P2V twinning methods (e.g., probabilistic model updating) are widely used across many different modeling methods, while some other P2V methods (e.g., ontology-based reasoning) are designed for specific modeling techniques. Next, we will explain the five P2V twinning techniques in detail.

\subsection{Physical measurements as input to the virtual space}
\label{sec:measurements_as_inputs}
A straightforward way of connecting the physical space with the virtual space (P2V) is to use physical measurements as inputs \citep{lydon2019coupled}. Doing so enables the digital model in the virtual space to predict the response of physical system in real-time, thus supporting timely risk assessment and decision making (V2P). The current approaches for this type of P2V connection can be categorized into two main groups as follows:
\begin{enumerate}
\item \emph{Using physical measurements to update the movements or position of digital models.} This group of P2V updating strategy is mainly for the modeling methods described in Sec. \ref{sec:Geo_modeling}. For instance, in the manufacturing domain, real-time position data of various components in a manufacturing system collected by the IoT system have been used to generate behavior models in a digital twin \citep{ricondo2021digital}, detect disturbances by comparing the position measurements with prediction \citep{glatt2021modeling}, drive the movement of digital models on the virtual shop floor and thereby optimize manufacturing system configurations or shop scheduling \citep{leng2020digital,liu2021construction, fang2019digital,guo2021digital,zhang2020digital,zhang2021digital}. In the smart city domain, data on mobility (i.e., movement data of human, vehicles, and other objects), infrastructure, and building conditions are collected in real-time, and are used as inputs for simulations in the virtual space for digital twin-based city development and planning \citep{white2021digital,schrotter2020digital}.
\item \emph{Using physical system monitoring data as input for physics-based analysis or design.} This group of updating strategy is mainly for the modeling methods summarized in Sec. \ref{sec:physics_based_modeling}. For example, in the aerospace domain, \cite{millwater2019probabilistic} and \cite{guivarch2019creation} employed the physical flight condition data collected from sensors as inputs of a digital model for risk assessment of aircraft structures. In the design domain, \cite{bellalouna2021case} utilized operational data collected from sensors and transmitted the data to a digital environment in the PTC Creo platform for physics-based topology optimization, demonstrating a digital twin concept. In the manufacturing domain, \cite{liu2021digital} performed physics-based hardware-in-the-loop simulation using real-time data from a physical system to replace physical tests, thereby reducing the verification time and cost. 
\end{enumerate}

When discussing P2V connections using sensor measurements, it is imperative to also consider the available methods of data transfer, as they should be carefully selected for the application at hand. Digital twin simulations can require substantial speeds and rates of data transfer in order to enable real-time optimization and control (see Section \ref{sec:real-time-ML}). In general, data transfer methods can be categorized into two classes: wired and wireless. Wired connections include Ethernet cables (twisted pair cables), coaxial cables, and optical fiber cables, where each cable is rated for different data transfer distances and rates. Wired data transfer is generally preferred, as it is easier to prevent data loss and is generally quicker as the connections are direct from source to receiver. Wireless communication methods include Bluetooth, Wi-Fi, ultra-wide band, and near-field communication (NFC) \citep{hu2021digital}. More recently, the 5G wireless communication standard has come to market offering higher data throughput with lower latency. It is envisioned that 5G technology will enable deployment of more complex and expansive sensor networks to collect and transmit data in real-time for use in digital twins \citep{cheng2018industrial}. These connected sensor and machinery networks, known more commonly as the industrial internet of things (IIoT), will be responsible for the data collection and transmission that will enable digital twins. Ultimately, each of the aforementioned data transmission techniques will play a critical role in enabling real-time simulation, process control, and optimization for digital twins.

This straightforward P2V connection is only applicable to cases where digital states (e.g., position, mobility, traffic flow) can be directly updated using monitoring data. However, for many engineering problems, digital states cannot be directly updated and are affected by various uncertainty sources. In those cases, more advanced P2V twinning methods are required and will be discussed in the subsequent sections.

\subsection{Probabilistic model updating}
\label{sec:probabilistic_updating}

\subsubsection{Introduction to digital state}
\label{sec:intro_digital_state}
As shown in Fig. \ref{fig:digital twin}, updating a digital system with a time-varying state is an essential characteristic of a digital twin. This state can be called a digital state, defined as a set of digital state variables that characterize the digital model(s) in a digital twin. Examples of digital states enabling predictive maintenance include the health of rotating equipment (e.g., motors, pumps, and compressors) \citep{wang2019digital} and machine tools \citep{luo2020hybrid} on a production floor, the material properties and structural health parameters of an unmanned aerial vehicle (UAV) \citep{kapteyn2020data}, and the state of charge (SOC) and state of health parameters (e.g., capacity and internal resistance) of a lithium-ion battery cell \citep{plett2004extended,hu2012multiscale,li2020digital}. A digital state often cannot be directly measured (or observed) but can be estimated through noisy measurements that depend on the digital state. The digital state of a physical system in operation changes over time and can be estimated as new information about the physical system becomes available. The time evolution of the digital state can be modeled as a dynamic system perturbed by a certain process noise.

\begin{figure*}[!ht]
  \centering
    {\includegraphics[scale=0.68]{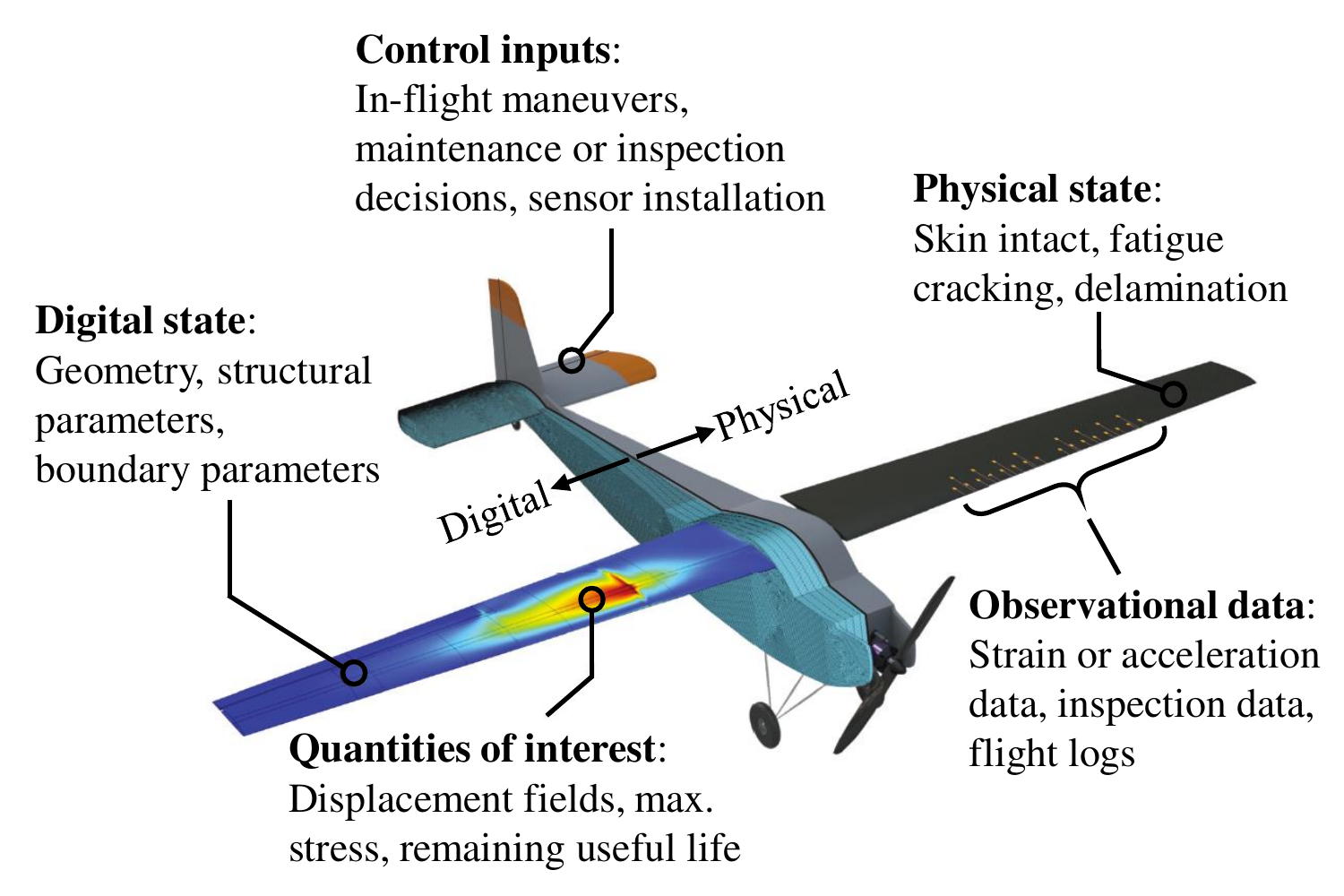}}
  \caption{An abstraction of a digital twin enabling a self-aware UAV, shown as a physical asset-digital twin system. In this example, the digital state could be a small subset of the physical state, allowing for tractable digital-state estimation. Reproduced with permission \citep{kapteyn2021probabilistic}. Copyright 2021, Springer Nature.}
  \label{fig:UAV}
\end{figure*}

Before we talk about the specifics of probabilistic model updating, let us look at a few examples of digital states. A recent paper demonstrated a structural digital twin of a UAV and provided a distinction between a physical state and a digital state \citep{kapteyn2021probabilistic}. A key argument in this distinction is that the physical state space should allow capturing variation in the physical system's state (e.g., the structural state of the UAV) and, therefore, could have high complexity and dimensions, while the digital state space should be designed to be simple enough to make the online estimation of the digital state feasible and tractable. As a result, digital models composing a digital twin should provide a digital state space that strikes a balance between complexity (variation in the physical asset captured by the digital state space is sufficient for performing meaningful diagnostics, prognostics, and decision making) and tractability (the number of digital state variables is sufficiently small to allow for good observability for a fixed number of sensor types and quantity of sensors). Fig. \ref{fig:UAV} illustrates the physical and digital states of the structure in a self-aware UAV application. The digital state encompasses geometric parameters, structural parameters (e.g., material properties and structural health parameters), and boundary conditions. Most of these digital state parameters are calibrated offline, and only a small subset of the structural parameters is dynamically updated online with sensor data, likely to ensure good model identifiability \citep{arendt2012quantification}. It is worth mentioning that the observational data (control inputs) flow from the physical (digital) to digital (physical) space, which shares similarities with the data and control elements in our proposed digital twin model in Fig. \ref{fig:digital twin}. 

Another example of a digital state is in an industrial application of digital twin for smart manufacturing \citep{wang2019digital}. Although not clearly defined, the digital state of a rotor system seems to refer to parameters of a geometric model,  parameters of a rotor dynamics model, and material properties. In the online phase, two parameters of rotor dynamics, namely the critical speed and vibration amplitude under unbalance, are dynamically estimated based on vibration data collected from the rotor system. Similar to \cite{kapteyn2021probabilistic}, a very small number of digital state parameters are updated online, suggesting the importance of identifying a small yet critical set of parameters for online updating. 

The third and last example comes from a review paper on battery digital twin \citep{li2020digital}. Similar to the digital state definitions in \cite{wang2019digital} and \cite{kapteyn2021probabilistic}, the digital state of a lithium-ion battery cell/module/pack can be defined as parameters fully characterizing a digital model of the cell/module/pack. Of particular interest to battery state estimation is the real-time SOC and state of health, the estimation of which are essential functions of a battery management system. Since each battery system in the field is equipped with a battery management system actively estimating and controlling battery SOC and state of health, one could thus argue that the digital twin concept is implemented on every battery system. 

\subsubsection{State estimation and Bayesian filters}
\label{state est and bayesian filters}

\begin{figure*}[!ht]
  \centering
    {\includegraphics[scale=0.78]{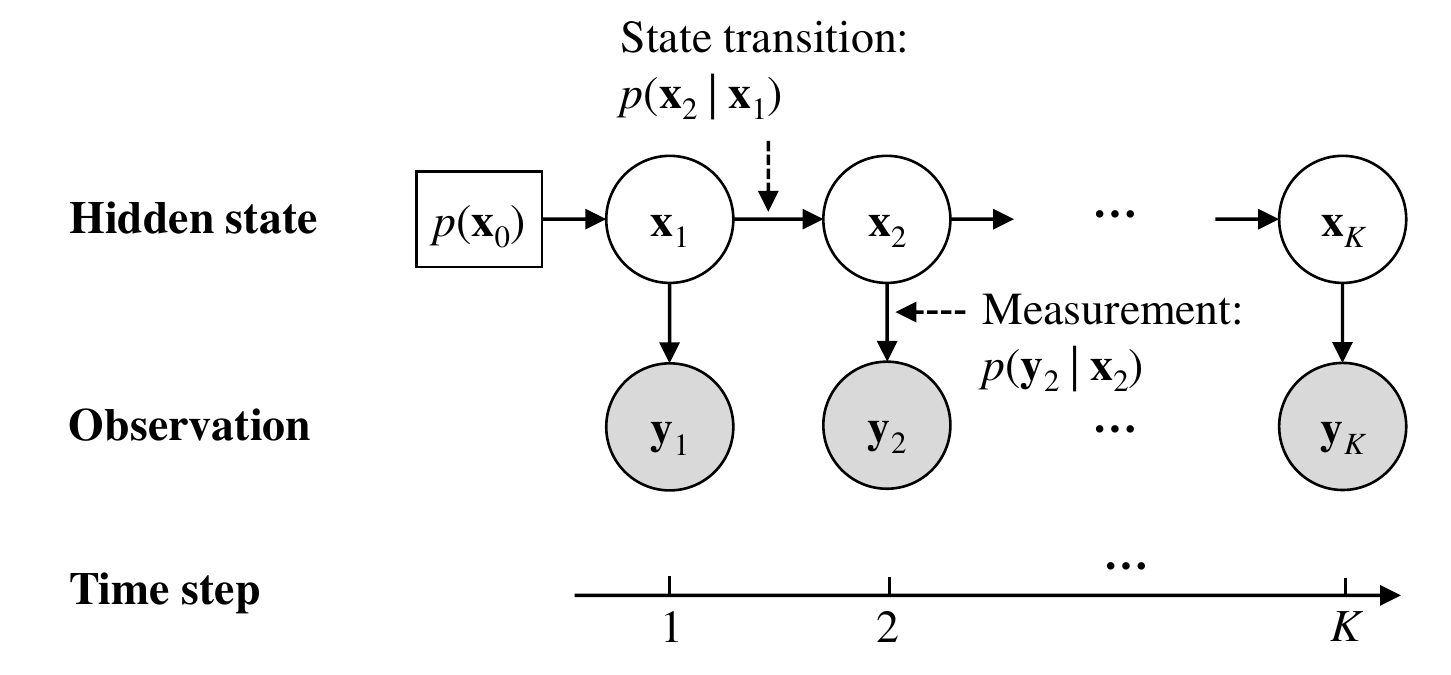}}
  \caption{An illustrative schematic of a hidden Markov model. This model consists of: (1) the hidden state ${\bf{x}}_k$ that evolves according to the state transition equation in Eq. (\ref{eq:state-space1}), and (2) the measurement ${\bf{y}}_k$ directly related to the hidden state by the measurement equation in  Eq. (\ref{eq:state-space1}). A prior probability density $p({{\bf{x}}_0})$, usually obtained from an offline calibration of the digital model (see Sec. 2.2.2 of Part 2 of the review paper), specifies the initial probability distribution of the hidden state at the $0^{th}$ time step (i.e., $k = 0$)}
  \label{fig:Hidden_MKC}
\end{figure*}

Let us consider a sequence of unknown digital state vectors, ${{\bf{x}}_{0:k}} = \{ {{\bf{x}}_0},\;{{\bf{x}}_1},\; \cdots ,\;{{\bf{x}}_k}\} $ , and a sequence of noisy measurement vectors, ${{\bf{y}}_{1:k}} = \{ {{\bf{y}}_1},\; \cdots ,\;{{\bf{y}}_k}\} $. We can formulate a discrete-time state-space model that takes the following form at time step $k$ (i.e., time $t_k$)

\begin{equation}
\label{eq:state-space1}
\begin{split}
  & {\rm{State}}\;{\rm{transition}}:\;{{\bf{x}}_k} = {\bf{f}}({{\bf{x}}_{k - 1}},\;{{\bf{u}}_{k - 1}}) + {{\boldsymbol{\upomega }}_k},  \cr 
  & {\rm{Measurement:}}\;\;\;\;{{\bf{y}}_k} = {\bf{g}}({{\bf{x}}_k}) + {{\bf{v}}_k}, \cr
\end{split}
\end{equation}
where ${\bf{x}}_k$ is the vector of state variables at the $k^{\mathrm{th}}$ time step, ${\bf{u}}_k$ is the vector of measured exogenous inputs and often control inputs, ${{\boldsymbol{\upomega }}_k}$ is the vector of process noise variables for the state, ${\bf{y}}_k$ is the vector of system observations (or measurements), ${\bf{v}}_k$ is the vector of measurement noise variables, and ${\bf{f}}(\cdot)$ and ${\bf{g}}(\cdot)$ are the state transition and measurement functions, respectively. The state transition equation in Eq. (\ref{eq:state-space1}) models the time evolution of the state as a dynamic system, perfectly defined by ${\bf{f}}(\cdot)$, perturbed by a process noise ${{\boldsymbol{\upomega}}_k}$; it can be rewritten as the transition probability density $p({{\bf{x}}_k}\vert{{\bf{x}}_{k - 1}})$. The measurement equation depicts the dependence of the current measurement ${\bf{y}}_k$ on the current state ${\bf{x}}_k$; it can be rewritten as the conditional density of the measurement ${\bf{y}}_k$ given the state ${\bf{x}}_k$, $p({{\bf{y}}_k}\vert{{\bf{x}}_{k}})$. At this point, the state-space model in Eq. (\ref{eq:state-space1}) can be graphically represented as a hidden Markov model shown in Fig. \ref{fig:Hidden_MKC}. 

The objective of the state estimation task in Eq. (\ref{eq:state-space1}) and Fig. \ref{fig:Hidden_MKC} is to estimate, in a recursive way, the hidden state at the current time step, ${\bf{x}}_k$, from the current and past noisy measurements \citep{sarkka2013bayesian}. Mathematically, we want to compute the marginal conditional distribution of ${\bf{x}}_k$ given on all available measurements up to the current step ($k$), ${{\bf{y}}_{1:k}}$. This marginal conditional of the digital state can be denoted as $p({{\bf{x}}_k}\vert{{\bf{y}}_{1:k}})$ and be computed using Bayes' theorem as follows \citep{leser2020digital,vega2022diagnosis}
\begin{equation}
\label{eq:Bayesian}
\begin{split}
  & p({{\bf{x}}_k}\vert{{\bf{y}}_{1:k}}) = {{p({{\bf{y}}_k}\vert{{\bf{x}}_k})p({{\bf{x}}_k}\vert{{\bf{y}}_{1:k-1}})} \over {\int {p({{\bf{y}}_k}\vert{{\bf{x}}_k})p({{\bf{x}}_k}\vert{{\bf{y}}_{1:k-1}}){d}{{\bf{x}}_k}} }},  \cr 
  &\propto p({{\bf{y}}_k}\vert{{\bf{x}}_k})p({{\bf{x}}_k}\vert{{\bf{y}}_{1:k-1}}),
\end{split}
\end{equation}
where ${{\bf{y}}_{1:k}}$ are observations from $t_1$ to $t_k$, ``$\propto$" stands for ``proportional to", and as described earlier, $p(\cdot \vert \cdot)$ is a conditional probability density function.

The above equation is analytically intractable. Recursive Bayesian filtering approximates the marginal conditional by recursively executing two steps: a transition step and an update step. Alternative, non-filtering approaches such as Markov chain Monte Carlo and importance sampling approximate the full conditional, $p({{\bf{x}}_{1:k}}\vert{{\bf{y}}_{1:k}})$. These alternative approaches are not as attractive in the context of digital twin because (1) they could be computationally much more costly than recursive Bayesian filtering, especially when $k$ becomes large, and (2) approximating the full conditional is unnecessary for most digital twin applications. 

Table \ref{tab:filters} compares four (recursive) Bayesian filters regarding system nonlinearity, state distribution constraints, and computational efficiency \citep{sarkka2013bayesian}. If the state-space model in Eq. (\ref{eq:state-space1}) is perfectly linear and Gaussian (i.e., ${\bf{f}}(\cdot)$ and ${\bf{g}}(\cdot)$ are linear functions and ${{\boldsymbol{\upomega }}_k}$ and ${{\bf{v}}_k}$ follow Gaussian distributions), the Kalman filter, one of the simplest and most efficient Bayesian filters, finds the exact solution to the recursive filtering problem, i.e., the solution to the conditional marginal is exact. Suppose ${\bf{f}}(\cdot)$ or ${\bf{g}}(\cdot)$ is nonlinear and or ${{\bf{v}}_k}$ is non-Gaussian. In that case, the extended Kalman filter provides a first-order approximation to the marginal posterior by (1) linearizing the state-space model, and (2) assuming Gaussian noise \citep{gustafsson2010particle}. This approximation may be accurate for state-space models whose ${\bf{f}}(\cdot)$ or ${\bf{g}}(\cdot)$ has mild nonlinearity and where true marginal posteriors are unimodal and symmetric. For state-space models with higher function nonlinearity, the unscented Kalman filter propagates a small set of points (called sigma points), sampled based on the prior mean and covariance of ${\bf{x}}_k$, through the nonlinear ${\bf{f}}(\cdot)$ and ${\bf{g}}(\cdot)$ functions, from which a posterior mean and covariance of ${\bf{x}}_k$ are estimated while assuming Gaussian. It can accommodate quadratic (second-order) terms in general and even cubic (third-order) terms in some cases when ${{\boldsymbol{\upomega}}_k}$ and ${{\bf{v}}_k}$ are Gaussian. 

\begin{table*}[!ht]
    \centering
    \caption{A comparison of four different recursive Bayesian filtering methods}
    \begin{tabular}{ p{2.5cm}|p{2.5cm}|p{3cm}|p{3cm}|p{2.5cm}  }
     \hline \hline
     \textbf{Quantity of interest} & \textbf{Kalman filter} & \textbf{Extended Kalman filter} & \textbf{Unscented Kalman filter}& \textbf{Particle filter}\\
     \hline
     Ability to handle nonlinear functions (${\bf{f}}(\cdot)$ and ${\bf{g}}(\cdot)$)  & Linear & Mildly nonlinear (linearized with first- or second-order Taylor expansion) & Nonlinear (can handle quadratic and even cubic terms) & Nonlinear\\
    \hline
     Process noise and measurement noise  (${{\boldsymbol{\upomega }}_k}$ and ${{\bf{v}}_k}$)  & Gaussian & Gaussian & Gaussian & Gaussian and non-Gaussian (no limitation)\\
    \hline
     Posterior state distribution  ($p({{\bf{x}}_k}\vert{{\bf{y}}_{1:k}})$)  & Gaussian & Unimodal (only one peak) and symmetric  & Unimodal (only one peak) & Unimodal and multimodal (no limitation)\\
    \hline
     Computational efficiency  & High & Medium & Medium-low & Low\\
     \hline \hline
    \end{tabular}
    \label{tab:filters}
\end{table*}

Particle filters differ drastically from the family of Kalman filters in that it does not assume state and noise are Gaussian and characterize uncertainties using Gaussian mean and variance. Instead, particle filters draw a large set of weighted samples (or particles) from the marginal posterior (or a distribution proportional to the marginal posterior) at each time step and use these particles as an approximation to the posterior \citep{sarkka2013bayesian, gustafsson2010particle}. An example of pseudocode for a generic particle filter algorithm is given in Sec. \ref{Appendix A} (see Figure \ref{fig:PF_algorithm}). As shown in the pseudocode, particle filtering consists of three key steps: state transition, weight evaluation and normalization, and resampling. An easy-to-understand explanation of a particle filter can be found in an animated video on YouTube \citep{Andreas22Youtb}. This video gives a toy example of positioning an aircraft based on measurements of the distance to the ground and flight elevation. The MATLAB source code for this illustration can be accessed openly on GitHub \citep{Andreas22Git}. As a result of using the particle approximation, particle filters work well on any non-Gaussian state (${{\bf{x}}_k}$) and noise ($\boldsymbol{\upomega}_k$ and ${\bf{v}}_k$) distributions and highly nonlinear transition and measurement functions (${\bf{f}}(\cdot)$ and ${\bf{g}}(\cdot)$). These attractive advantages come with a price: higher computational costs. For example, the unscented Kalman filter,  the most computationally complex member in the Kalman family, needs to propagate a set of 7 sigma points to estimate the SOC, capacity, and internal resistance position velocity of a lithium-ion battery (capacity and internal resistance are two state-of-health parameters that affect the energy and power capability of a battery). However, a particle filter may propagate hundreds of weighted particles at each step to achieve good convergence speed and estimation accuracy. Also, particle filters do not scale easily to high-dimensional state spaces, where we may have no particles near the true state, also known as particle deprivation, and the numbers of particles required to achieve reasonable estimation become prohibitively high. Therefore, applications of particle filters for state estimation in the digital twin context have been mostly limited to a small number of state dimensions (typically $ < $ 5), consistent with what we found in \cite{li2020digital,wang2019digital,kapteyn2021probabilistic}. 

It is worth mentioning that approaches have been developed that use cheap-to-evaluate surrogate models to replace the state transition function (${\bf{f}}(\cdot)$), measurement function (${\bf{g}}(\cdot)$), or both when these functions are computationally expensive. These approaches alleviate, to some degree, the computational burden imposed by particle filers by allowing a more efficient propagation of thousands or even millions of weighted particles with a lower computational cost. These efforts make particle filter a popular technique for state estimation in many digital twin applications as reported in the literature \cite{li2017dynamic, kapteyn2021probabilistic,ye2020digital}.

\begin{figure}[!h]
  \centering
    {\includegraphics[scale=0.56]{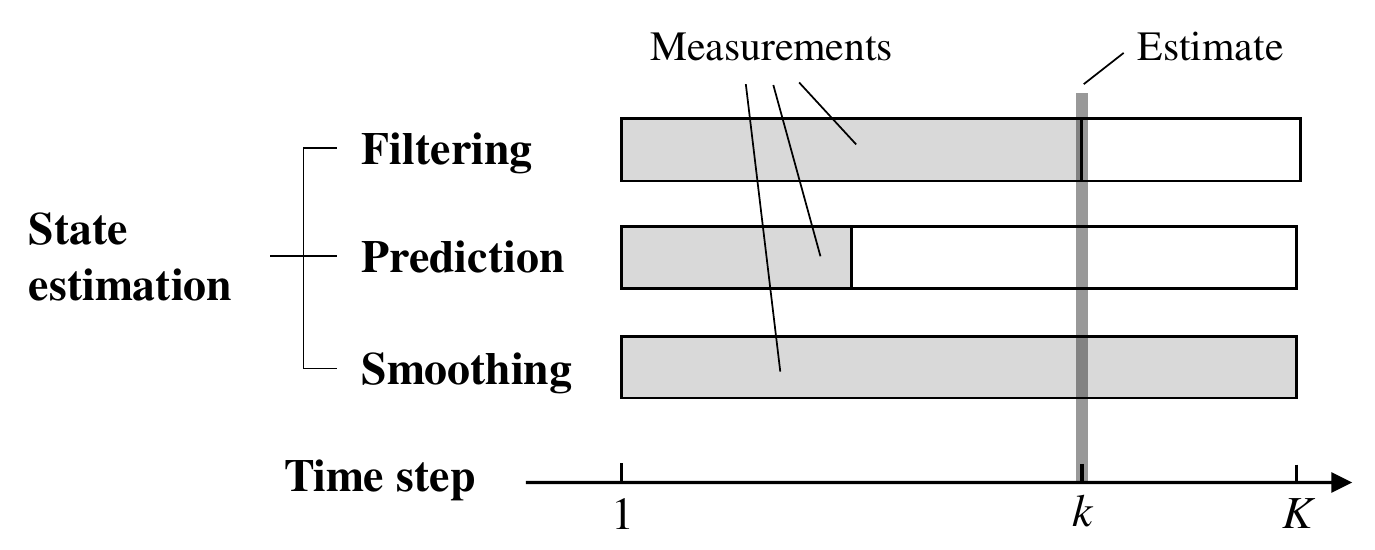}}
  \caption{A schematic distinguishing among three different state estimation problems. The differences lie in the period of the available measurements relative to the state estimation time.}
  \label{fig:state_estimate}
\end{figure}

Two other state estimation problems relevant to digital twin are called smoothing and prediction. These two problems aim to estimate past (smoothing) and future (prediction) states, respectively. Figure \ref{fig:state_estimate} illustrates the time requirements for all three state estimation methods. Each of the three methods play a different role in digital twins because they focus on different time frames. Smoothing methods operate on previously collected data and generally reduce the size of the data in the process. For example, smoothing methods can be used to understand a physical system's state history over time after the data has been collected. On the contrary, prediction methods are concerned with estimating future states using historical data. Shown in Fig. \ref{fig:state_estimate}, the \emph{prediction} state estimation method is estimating the state of the system at time $k$ using historical data. While smoothing was used to understand historical system state, prediction is typically used for understanding future system state. Prediction in digital twin is typically used for planning and control; two use cases which require some degree of understanding about the trajectory of system states.

In digital twin applications, state estimation results are used to inform decision making algorithms. The timeliness required for effective decision making using digital twin influences the choice of state estimation methodology. For example, fault diagnosis of machinery requires that the machine's health status be estimated at the present time. In this case, the ideal state estimation methodology to use in a fault diagnosis digital twin would be filtering, as it provides state estimates at the present time. On the other hand, health forecasting and remaining useful life (RUL) prediction require state estimates of system health at future times, and are thus only enabled by prediction methods which predict state at future time steps (See Fig. \ref{fig:state_estimate}) \citep{wang2019digital,tian2011condition,bangalore2018analysis}. Another example is that a structural digital twin of a UAV with an updated damage state can be used to predict (by simulation) the probability of failure for an upcoming flight maneuver and thus informing flight path planning and control \citep{kapteyn2021probabilistic}. The idea is to allow a structurally healthy UAV to perform aggressive maneuvers to reduce flight distance and time and constrain a structurally damaged UAV to act more conservatively to incur minimal damage in future maneuvers and prevent unexpected structural failure. The last example is that the state-of-charge and state-of-health estimate of a lithium-ion battery can be used to optimize the current profile (i.e., charging C-rate vs. SOC) during fast charging \citep{li2020digital,li2021electrochemical,wei2021deep}.

\subsubsection{Dynamic Bayesian network}
A Bayesian network (BN), which is also called a Bayes net, is a probabilistic graphical model that represents a set of random variables and their probabilistic relationships as a directed acyclic graph \citep{murphy2002dynamic,friedman1997bayesian}. A dynamic Bayesian network (DBN) extends a standard BN by considering the time evolution of variables and is used to model dynamic systems. It is a generalization of the hidden Markov model given in Fig. \ref{fig:Hidden_MKC} (see Sec. \ref{sec:probabilistic_updating} (b)). The difference between a hidden Markov model (HMM) and a DBN is that hidden Markov model uses a single hidden state variable to represent the entire state space whereas the DBN represents the hidden state as a set of random variables connected in a graph \citep{murphy2002dynamic}. A DBN allows for the modeling of nonlinear dynamic systems with arbitrary nonlinearities and distributions.

\begin{figure}[!h]
  \centering
    {\includegraphics[scale=0.95]{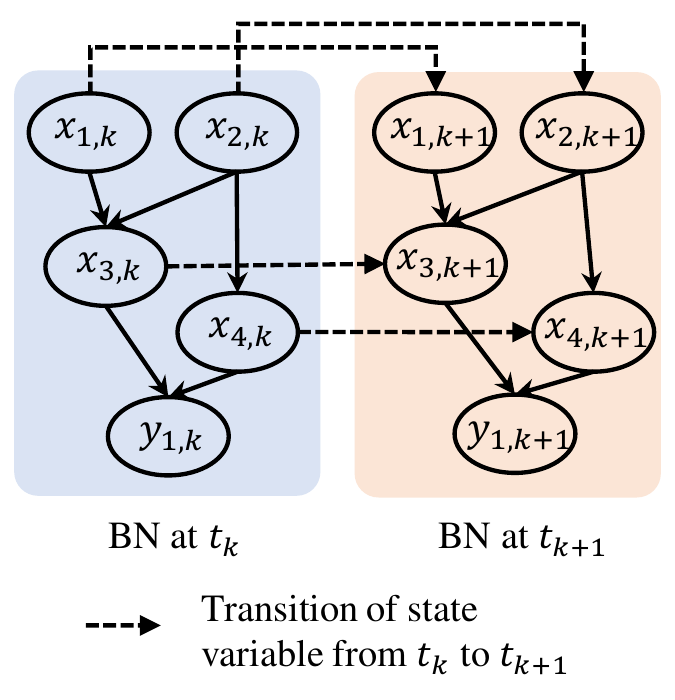}}
  \caption{Illustration of a DBN, which is a generalization of the hidden Markov model given in Fig. \ref{fig:Hidden_MKC}.}
  \label{fig:DBN}
\end{figure}

Fig. \ref{fig:DBN} presents an illustrative example of a DBN with five nodes. At each time instant $t_k$, the DBN has a static BN. The static BNs at two adjacent time instants $t_k$ and $t_{k+1}$ are connected through a transition BN. As a directed acyclic graph, a DBN consists of two essential elements, namely vertices (also called nodes) and edges. In a DBN, the vertices or nodes are random variables, such as ${x_{1,k}}, {x_{2,k}}, {x_{3,k}}, {x_{4,k}}, \textrm{and } {y_{1,k}}$ in Fig. \ref{fig:DBN}. The edges directed from one node to another, such as ${x_{3,k}} \rightarrow {y_{1,k}}$ and ${x_{1,k}}, {x_{2,k}} \rightarrow {x_{3,k}}$, represent the probabilistic causality relationship between the nodes. 

\emph{Nodes} in a DBN could represent discrete random variables or continuous random variables. For instance, as part of the U.S. Air Force funded digital twin effort, \cite{li2017dynamic} represented bolt looseness and a damage state (i.e., elastic/plastic zone) as discrete random variables and a crack length as a continuous random variable in their DBN-based digital twin that modeled the fatigue crack growth on an aircraft wing. Moreover, the nodes that an edge is directed from and directed to are called the parent node and the child node, respectively. For example, for edge ${x_{1,k}}, {x_{2,k}} \rightarrow {x_{3,k}}$ in Fig. \ref{fig:DBN}, ${x_{1,k}}, {x_{2,k}}$ are parent nodes and ${x_{3,k}}$ is a child node. Nodes without parent nodes are called root nodes (e.g., ${x_{1,k}}, {x_{2,k}}$ in Fig. \ref{fig:DBN}). 

\emph{Edges} connecting parent nodes with the corresponding child node(s) in a DBN, such as ${x_{3,k}} \rightarrow {y_{1,k}}$ and ${x_{4,k}} \rightarrow {y_{1,k}}$ in Fig. \ref{fig:DBN}, represent the causality relationship(s) between the parent node(s) and the child node(s). The probabilistic causal relationships between a group of parent nodes connected to a common child node are modeled as a conditional probability table (CPT) for discrete nodes and a conditional probability density (CPD) function for continuous nodes. The CPT contains the conditional probability mass function of a child node covering all possible realizations of its parent nodes. For example, the probability of having a bolt looseness or not for a given realization of load condition of an aircraft is described as a CPT in \cite{li2017dynamic}. The CPD could be a simple, analytical probability density function or a numerical probabilistic predictive model, such as a Gaussian process regression model \citep{karve2020digital}, a data-driven stochastic process model \citep{yu2021digital}, or a reduced-order model with noise \citep{kapteyn2021probabilistic}. For instance, if Gaussian process regression is employed to construct the CPD of $x_{3,k}$ given in Fig. \ref{fig:DBN}, a Gaussian process regression model ${x_{3,k}} = \hat G({x_{1,k}},\;{x_{2,k}})$ is first built according to the input-output dependence depicted by the edges ${x_{1,k}}, {x_{2,k}} \rightarrow {x_{3,k}}$. After that, we have the CPD of $x_{3,k}$ as
\begin{equation}
\label{eq:GP_CPD}
p({x_{3,k}}\vert{x_{1,k}},{x_{2,k}}) = \phi \left( {{{{x_{3,k}} - {\mu _{\hat G}}({x_{1,k}},{x_{2,k}})} \over {{\sigma _{\hat G}}({x_{1,k}},{x_{2,k}})}}} \right),
\end{equation}
in which ${{\mu _{\hat G}}({x_{1,k}},{x_{2,k}})}$ and ${{\sigma _{\hat G}}({x_{1,k}},{x_{2,k}})}$ are respectively the mean and standard deviation of the Gaussian process regressor-predicted $x_{3,k}$ for given values of the parent nodes $x_{1,k}$ and $x_{2,k}$, and $\phi \left(  \cdot  \right)$ is the probability density function of a standard normal random variable.

The physic-based models described in Sec. \ref{sec:physics_based_modeling}, such as FEA models, can also be used to construct CPDs by accounting for uncertainty sources in the models. Since digital twins require real-time model updating and each updating run may need to evaluate CPTs or CPDs thousands of times, high-fidelity physics-based models are usually replaced with computationally cheaper data-driven surrogate models (see Sec. \ref{sec:data_driven_modeling}) in the practical implementation of DBNs. In addition to the \emph{nodes} and \emph{edges}, as mentioned above, a transition BN is used to connect two BNs at two adjacent time instants. This is essential to making the BN \emph{dynamic}. Taking the DBN given in Fig. \ref{fig:DBN} as an example, the transition of each state variable between two adjacent time instants $t_k$ and $t_{k+1}$ is given by
\begin{equation}
\label{eq:Transient BN}
p({x_{i,k + 1}}\vert{x_{i,k}}), \forall i = 1,\;2,\;3,\;4,
\end{equation}
where $p({x_{i,k+ 1}}\vert{x_{i,k}})$ is the transition probability of state variable $x_{i}$ from $t_k$ to $t_{k+1}$, which is defined by a CPT or CPD and shown as a dashed edge in Fig. \ref{fig:DBN} (i.e., ${x_{i,k}} \rightarrow {x_{i,k+1}}, i=1, 2, 3, 4$).

An important property of BNs and DBNs is the \emph{conditional independence} property, which means that the nodes are independent from each other if they are not connected by directed edges, given the values of their parent nodes. For instance, for given values of $x_{1,k}$ and $x_{2,k}$ in Fig. \ref{fig:DBN}, $x_{3,k}$ and $x_{4,k}$ are independent from each other. Thanks to this property and based on the chain rule of conditional probabilities, the likelihood function of a DBN can be expressed as a product of individual CPTs and/or CPDs conditioned on the parent nodes (see an example given in Appendix B). The decomposition of the joint probability density into a product of individual CPTs and CPDs as illustrated in Eq. (\ref{eq:DBN}) in Appendix B really facilitates model updating and the construction of complex DBNs for digital twins. One just needs to properly define the CPTs and CPDs for individual nodes and connect them using a directed acyclic graph. After that, the inference can be performed using generic Bayesian inference methods including the commonly used Markov chain Monte Carlo simulation \citep{brooks1998markov} and sequential Monte Carlo methods such as particle filtering as described in Sec. \ref{sec:probabilistic_updating} (b), and given in Appendix A. As has been pointed out by \cite{murphy2002dynamic}, the flexibility of DBNs in incorporating various CPTs and CPDs using directed acyclic graphs makes DBNs advantageous over various hidden Markov models including classical hidden Markov models and hierarchical hidden Markov models. The fact that the CPDs in a DBN could be any type of probabilistic model with any nonlinearities and distribution types makes DBNs a flexible tool capable of fusing heterogeneous data sources (e.g., images, text, audio) for model updating (a.k.a. information fusion). \cite{kapteyn2021probabilistic} suggested DBNs as the foundation of a unifying mathematical formulation for digital twin at scale, by integrating Bayesian statistics, dynamical systems, and control theory. They also demonstrated the DBN concept using the UAV digital twin example as given in Fig. \ref{fig:UAV}. 

The directed acyclic graph in a BN or DBN is usually constructed based on physical knowledge of the causality relationships between different nodes, such as the logical relationships defined by the ontology and SysML as been described in Sec. \ref{sec:sys_modeling}. It is possible that the logical or causal relationship is unknown. In that case, the directed acyclic graph/causal relationships could be learned from data using BN learning methods, such as minimal description length, Bayesian–Dirichlet equivalence, and mutual information test \citep{scutari2009learning,murphy2002dynamic,hu2018bayesian}. In BN learning, the most probable topology of the graph is identified by minimizing or maximizing a score function, such as the aforementioned minimal description length, Bayesian–Dirichlet equivalence, or mutual information. A tutorial on BN learning can be found in \citep{heckerman2008tutorial}. Also, there are several open-sources toolkits available for the inference, learning, and construction of DBNs, such as the Graphical Models Toolkit \citep{gmtdnb2022} and the Bnlearn Python package \citep{bnlearn2022}.

\subsubsection{State and parameter estimation}
\label{sec:state_parameter_estimation}
When a digital twin involves multiple types of physics, the time scales of different digital state variables could be quite different. For example, the capacity of a lithium-ion battery cell typically fades slowly and only shows noticeable changes in months or years. In contrast, the cell’s SOC typically changes quickly with time and may go from 0$\%$ (fully depleted) to 100$\%$ (fully charged) within minutes in a fast charging cycle. As will be discussed in the battery case study in Sec. 3 of Part 2 of the review paper, SOC and capacity are two key performance measures that together assist in managing the health of lithium-ion batteries. The state estimation literature usually refers to estimating digital state variables that change very slowly or do not change with time as \emph{parameter estimation}. Estimating digital state variables that change rapidly with time is called \emph{state estimation}. Note that both the hidden state ${\bf{x}}$ and parameters ${\boldsymbol{\uptheta}}$ are digital state variables and part of the digital state.

When the state ${\bf{x}}$ is known and noise-free, then state estimation will not be needed, and only parameter estimation is needed. A straightforward way to estimate ${\boldsymbol{\uptheta}}$ is to solve a least squares problem, formulated based on the measurement function $g(\cdot)$ for a one-dimensional known state $x$ and measurable response $y$ as follows:
\begin{equation}
\label{eq:state-space_para}
\mathop {\min }\limits_{\boldsymbol{\uptheta }} \;\sum\limits_{k = 1}^K {{{({y_k} - g({x_k},\;{\boldsymbol{\uptheta }}))}^2}} ,
\end{equation}

\noindent where $K$ is the total number of measurements available for estimating ${\boldsymbol{\uptheta}}$ and $g(\cdot)$ is the measurement function that predicts $y$ at state $x$, given the vector of model parameters ${\boldsymbol{\uptheta}}$. Here, $g(\cdot)$ can be a physics-based model as discussed in Sec. \ref{sec:physics_based_modeling} or a data-driven model as discussed in Sec. \ref{sec:data_driven_modeling}. Three examples of least squares parameter estimation for a physics-based model are reported in (1) \cite{wang2019digital} for estimating two fault-sensitive parameters of a rotor dynamics model for unbalance diagnostics of a rotor system, (2) \cite{ramadesigan2011parameter} for estimating five degradation-sensitive parameters of an electrochemical model for every charge/discharge cycle for battery capacity fade prediction, and (3) \cite{peng2020digital} for estimating five degradation-sensitive parameters of an equivalent circuit model for condition monitoring of a buck DC-DC converter. Least squares estimation can also be used to estimate the parameters of data-driven models, such as estimating the weights and biases of an artificial neural network from a training dataset $\{(x_1, y_1), …, (x_K, y_K)\}$.

If $g(\cdot)$ is linear, Eq. (\ref{eq:state-space_para}) can be solved using linear least squares. When measurement data comes in a stream, the Kalman filter discussed in Sec. \ref{sec:probabilistic_updating} (b) and recursive least squares can be used to ensure a fixed computational cost at each time step \cite{simon2006optimal}. They produce optimal estimates, and if the assumption of Gaussian measurement noise holds, they are equivalent. 


If $g(\cdot)$ is nonlinear, numerical optimization algorithms, such as Gauss-Newton, gradient descent, and Levenberg–Marquardt, and population-based metaheuristic optimization algorithms, such as genetic algorithms and particle swarm optimization, can be used to solve Eq. (\ref{eq:state-space_para}). However, a potential issue is that the resulting estimate of ${\boldsymbol{\uptheta}}$ is deterministic. This estimate does not capture prior knowledge on  ${\boldsymbol{\uptheta}}$ or uncertainty in parameter estimation (e.g., due to low identifiability and measurement noise). This issue can be mitigated by deriving an approximate posterior using a Markov chain Monte Carlo method combined with a prior distribution of ${\boldsymbol{\uptheta}}$ under a Bayesian calibration framework \citep{ramadesigan2011parameter}. Although sequential Bayesian calibration (e.g., a sequential version of a well-known Bayesian calibration framework referred to as the Kennedy and O'Hagan (KOH) framework \citep{kennedy2001bayesian}) has been shown to track time-varying parameters, this approach may be computationally intensive and require longer run times than the response time needed to track sudden changes \citep{ward2021continuous}. A detailed discussion on the applicability of Bayesian calibration to UQ of dynamic system models is given in Sec. 2.1.2 in Part 2 of the review paper. Sequential, recursive filtering alternatives to these iterative optimizers are nonlinear extensions of the Kalman filter (e.g., the extended Kalman filter and unscented Kalman filter) and particle filters, which are more attractive in computational efficiency. It is not uncommon to see these recursive filtering alternatives succeed in training nonlinear models like neural networks with fixed optimal parameters \citep{puskorius1994neurocontrol,aitchison2020bayesian}, with a possible extension to time-varying parameters \citep{wagner2021kalman}. A new state-space representation can be defined for parameter estimation in the presence of a known and clean state. This new state-space model takes the following form: 
\begin{equation}
\label{eq:state-space2}
\begin{split}
  & {\rm{State}}\;{\rm{transition}}:\;{{\boldsymbol{\uptheta }}_k} = {{\boldsymbol{\uptheta }}_{k - 1}} + {{\bf{r}}_k},  \cr 
  & {\rm{Measurement}}:\;{{\bf{y}}_k} = {\bf{g}}({{\bf{x}}_k},\;{{\boldsymbol{\uptheta }}_k}) + {{\bf{v}}_k}, \cr
\end{split}
\end{equation}

Now suppose that $g$ is highly nonlinear, ${\boldsymbol{\uptheta}}$ is high-dimensional (e.g., consisting of $>$ 5 parameters), and the computational time and cost are not a concern. In that case, population-based optimization algorithms, such as genetic algorithms and particle swarm optimization, could be better alternatives due to the simple implementation and applicability to a wide variety of $g$ function forms and nonlinearities \citep{wang2019digital, peng2020digital}.

When the state is unknown or given with noise, state estimation and parameter estimation are needed. In these cases, state-space models have unknown parameters ${\boldsymbol{\uptheta}}$ and hidden states ${\bf{x}}$ that need to be estimated. As mentioned in Sec. \ref{sec:probabilistic_updating} (a), ${\boldsymbol{\uptheta}}$ is part of the digital state, and these parameters may first be calibrated offline but must be updated online. With both a hidden state and unknown parameters, the state-space model in Eq. (\ref{eq:state-space1}) becomes
\begin{equation}
\label{eq:state-space3}
\begin{split}
  & {\rm{State}}\;{\rm{transition}}:\;{{\bf{x}}_k} = {\bf{f}}({{\bf{x}}_{k - 1}},\;{{\boldsymbol{\uptheta }}_{k - 1}},\;{{\bf{u}}_{k - 1}}) + {{\boldsymbol{\upomega }}_k}, \cr
  & \; \; \;\; \; \;\; \; \;\; \; \;\; \; \;\; \; \; \; \;\; \; \;\; \; \; \;\; \;{{\boldsymbol{\uptheta }}_k} = {{\boldsymbol{\uptheta }}_{k - 1}} + {{\bf{r}}_k},  \cr 
  & {\rm{Measurement}}:\;{{\bf{y}}_k} = {\bf{g}}({{\bf{x}}_k},\;{{\boldsymbol{\uptheta }}_k}) + {{\bf{v}}_k}, \cr
\end{split}
\end{equation}

\noindent where ${\boldsymbol{\uptheta}}$ is the vector of unknown model parameters which changes slowly or do not change over time, ${{\bf{r}}_k}$ is the vector of process noise variables for the parameters at the $k^{\mathrm{th}}$ time step.  

\begin{figure}[!h]
  \centering
    {\includegraphics[scale=0.5]{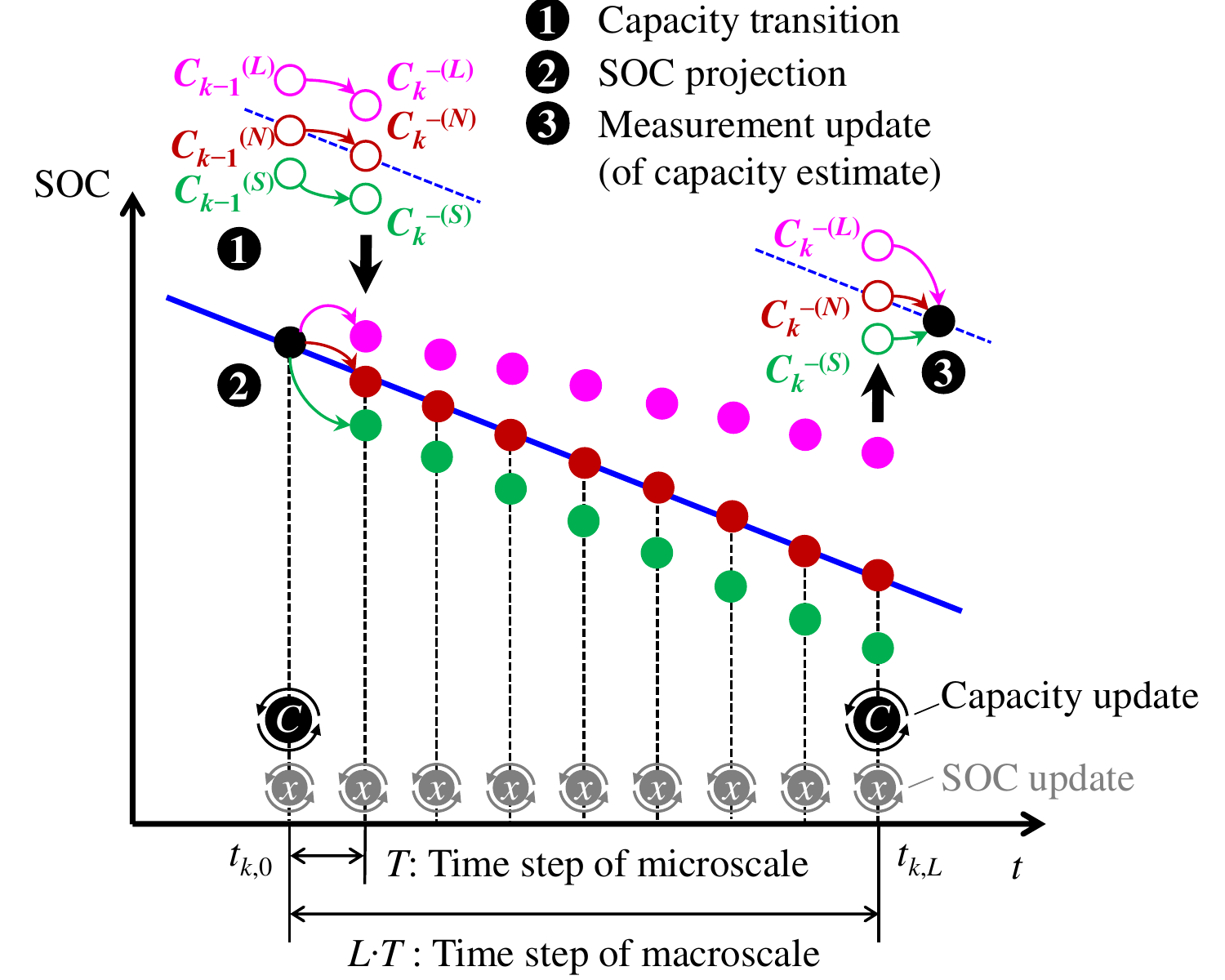}}
  \caption{Illustration of multiscale extended Kalman filtering for battery SOC and capacity estimation (adapted from \cite{hu2012multiscale}).}
  \label{fig:multiscale_KF}
\end{figure}

Similar to when the state is known and clean, iterative optimization algorithms can be used to jointly estimate ${\bf{x}}$ and ${\boldsymbol{\uptheta}}$. Sequential approaches such as recursive Bayesian filtering become favorable alternatives for online, close to real-time state estimation with limited computational power (e.g., in an embedded application). A straightforward sequential approach is to combine the state and parameters into an augmented state, ${\bf{z}} = [{\bf{x}}; {\boldsymbol{\uptheta}}]$, and estimate ${\bf{z}}$ using the extended Kalman filter or unscented Kalman filter. This approach is referred to as joint (extended or unscented) Kalman filtering \cite{cox1964estimation}, which was attempted more than half a century ago. As a new measurement from a physical system becomes available, this joint version of the extended Kalman filter simultaneously updates the state and parameter estimates. A drawback reported for this approach is divergence and bias in the parameter estimation \cite{ljung1979asymptotic}. An alternative is the dual extended or unscented Kalman filter that separately considers state and parameter estimation \citep{wan2001dual}. Instead of using fixed state transition and measurement functions such as in Eq. (\#), these approaches alternate between estimating the state given the most recent parameter estimate and estimating the model parameters given the most recent state estimate. Two notable applications of the dual extended Kalman filter are battery state (SOC) and parameter (capacity) estimation \cite{plett2004extended} and vehicle state (e.g., longitudinal and lateral velocities) and parameter (e.g., vehicle mass and moments of inertia) estimation \cite{wenzel2006dual}. 
Compared to the hidden state ${\bf{x}}$, the parameters ${\boldsymbol{\uptheta}}$ tend to vary more slowly with time. For example, the SOC and capacity of a lithium-ion battery cell change on two largely different time scales, as mentioned earlier. The different time scales of these digital state variables present unique challenges in estimating these variables. More recently, a multiscale filtering framework (see Fig. \ref{fig:multiscale_KF}) was proposed to estimate the SOC and capacity of a lithium-ion battery cell \citep{hu2012multiscale}. Multiscale filtering utilizes the temporal nature of SOC (fast varying) and capacity (slowly varying) to improve the efficiency and accuracy in estimating both the state and parameter over the joint and dual versions of the extended Kalman filtering.

\subsection{ML model updating}
\label{sec:ML_updating}
ML models are an increasingly critical component of digital twins, and they are often deployed in an environment with periodic operational changes in the physical world, such as sensors addition/removal, manufacturing process updates (e.g., add/remove a step in a manufacturing process), changes in measurement device settings/performance, etc. These significant changes might lead to test data that is sufficiently far from the distribution of the data on which the ML model is trained. Such phenomenon is typically referred to as ``out-of-distribution" in the ML field~\citep{lee2018simple,zhang2022explainable}. An ML model is often trained with a closed-world assumption (i.e., the training data and the test data are assumed to follow the same distribution). As a result, if out-of-distribution test data is not incorporated in model training timely and properly, ML model performance will drift significantly. 

It is crucial to detect the out-of-distribution situation and update the ML model to account for the operational changes associated with the physical asset. ML model updating will prevent severe performance degradation of a trained and deployed ML model. Updating digital models based on the continuous monitoring of the physical counterpart is an important characteristic of digital twins ~\citep{chakraborty2021role}. Besides, the timely update of the ML model is also the key to ensuring a consistent and acceptable predictive performance in characterizing the trend of quantities of interest associated with physical assets. One of the most commonly adopted paradigms is to utilize a performance-based indicator to determine when model drift onsets and subsequently trigger an update of the deployed ML model. ML model updating can be implemented in the cloud or on the edge. In the ML community, model drift (also referred to as model decay) is developed as an umbrella term to characterize the degradation of a model's predictive power due to unforeseeable changes in the environment. In practice, model drift often manifests itself in two principal forms:

\begin{figure*}[!ht]
    \centering
    \includegraphics[scale=0.5]{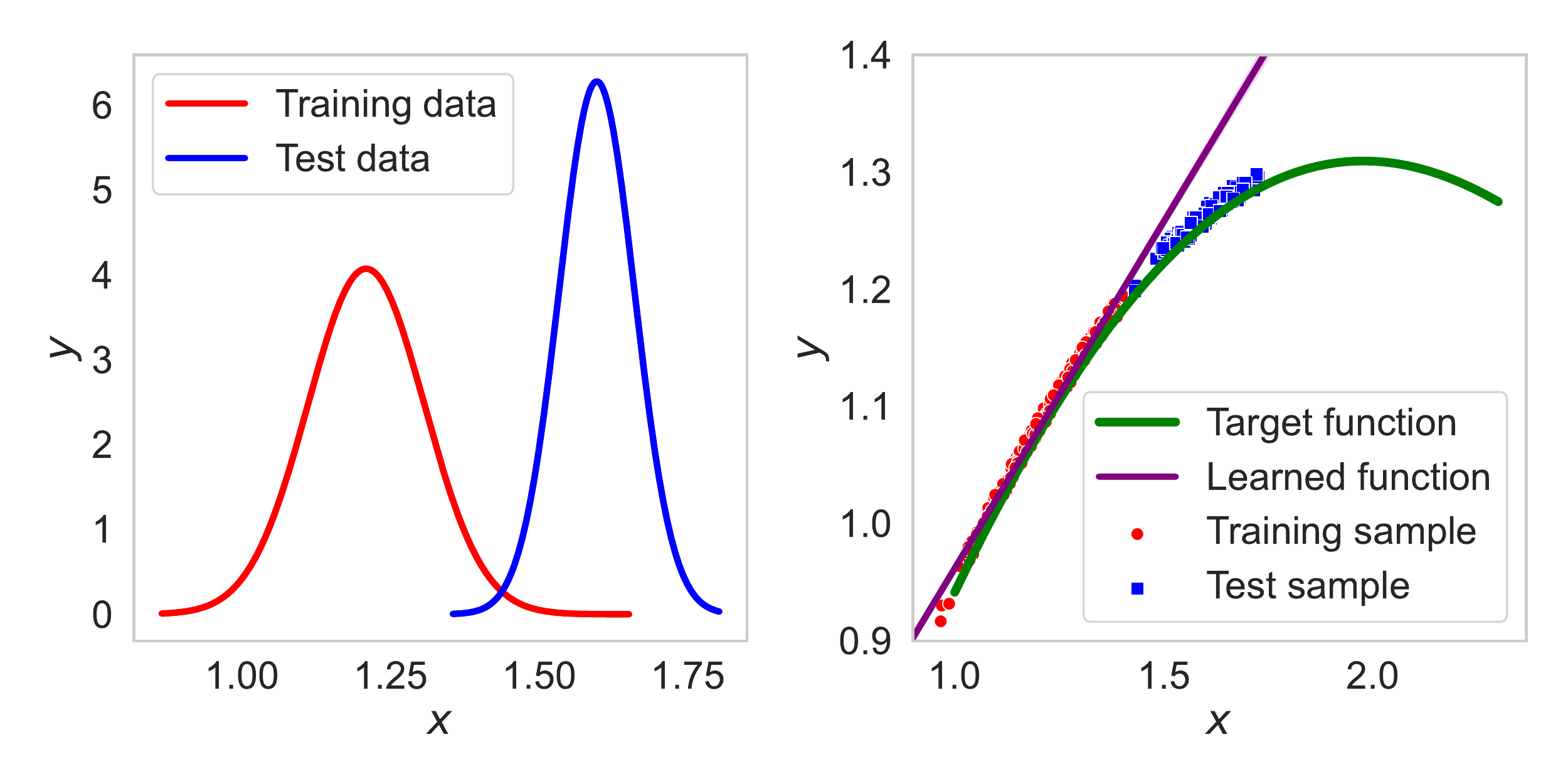}
    \caption{Demonstration of data drift. Adopted from~\cite{zhang2022towards}}
    \label{fig:data_drift}
\end{figure*}

\begin{enumerate}
\item {\bf{Data drift}} (also referred to as covariate drift): As mentioned before, the training of ML models is established upon the assumption that the training data distribution is representative of the test data distribution. However, this assumption can easily get violated in many real-world applications~\citep{sugiyama2008direct}. The term `` data drift" is used to describe the situation that statistical properties of the production data differ substantially from those of the training data. and it can be mathematically represented as $p^{\textrm{train}} \left( \mathcal{X} \right) \ne p^{\textrm{test}} \left( \mathcal{X} \right)$.

    Fig.~\ref{fig:data_drift} showcases an instance of data drift, where the relationship between $\mathcal{X}$ and $\mathcal{Y}$ remains the same (see the target function in Fig.~\ref{fig:data_drift}). Unfortunately, $\mathcal{X}$ follows two different distributions in the training data and test data. As a result of data drift, the function learned from the training data leads to erroneous predictions on the test data.

    \item {\bf{Concept drift}}: This form of model drift characterizes the phenomenon in which the underlying relationship between $\mathcal{X}$ and $\mathcal{Y}$ changes in non-stationary environments~\citep{lu2018learning}. In essence, the quantity we are trying to predict evolves over time. Mathematically, concept drift is represented as $p^{\textrm{train}} \left( \mathcal{Y} \vert \mathcal{X} \right) \ne p^{\textrm{test}} \left( \mathcal{Y} \vert \mathcal{X} \right)$. 
    
     Fig. \ref{fig:concept_drift} demonstrates an example of concept shift using the same regression problem. Obviously, the underlying functional relationship between $\mathcal{X}$ and $\mathcal{Y}$ has substantially changed from the training data to test data. 
     
    \begin{figure}[!ht]
    \centering
    \includegraphics[scale=0.4]{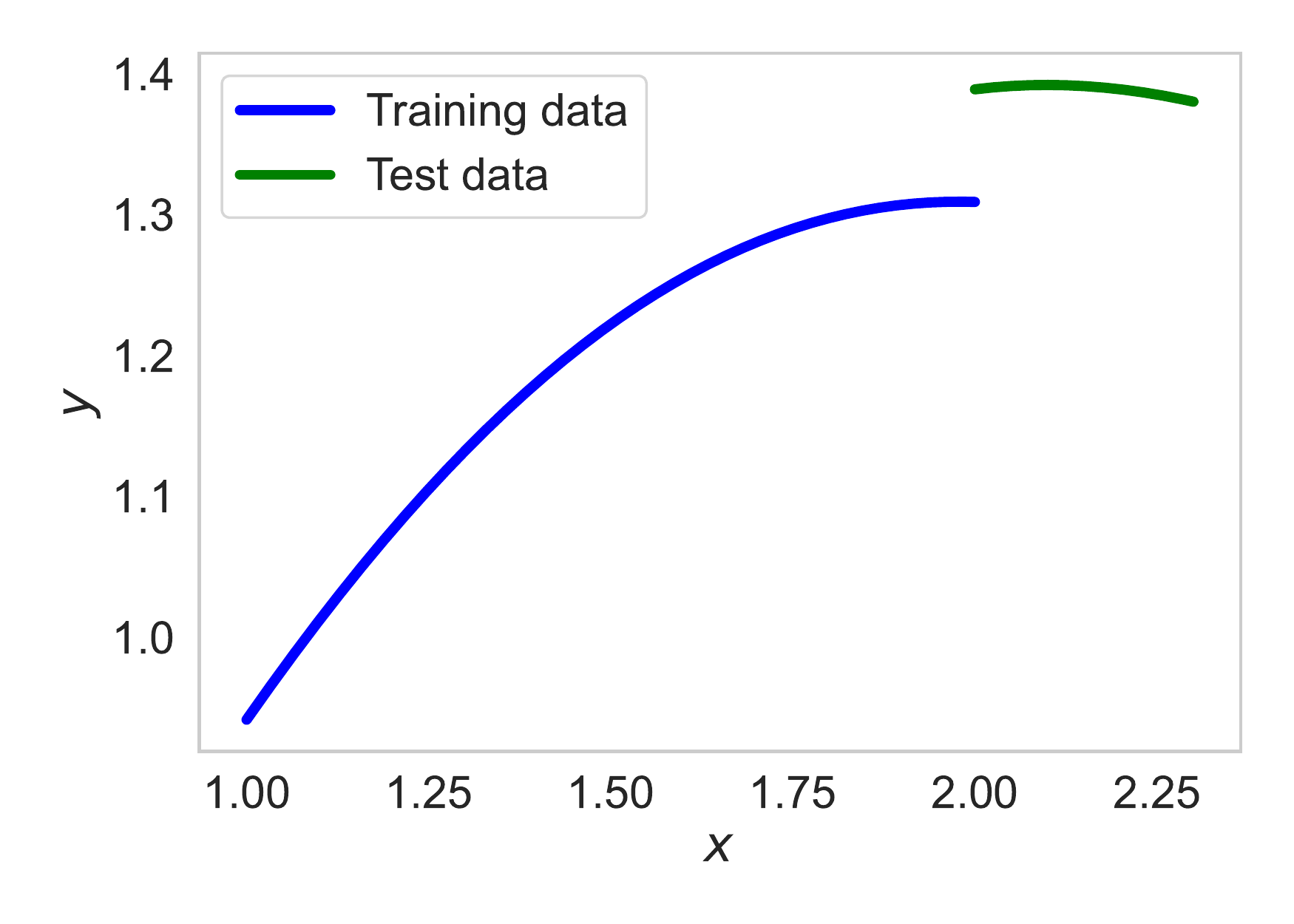}
    \caption{Demonstration of concept drift}
    \label{fig:concept_drift}
    \end{figure}
\end{enumerate}

In either case, model drift poses significant robustness and reliability challenges in practical ML applications. Such drift has led to substantial ML model performance degradation~\citep{tsymbal2004problem}. Safeguarding an ML model in digital twins from this issue entails a close monitoring of ML performance variation using a drift indicator and proactive update of the ML model on an as-needed basis. Here, a drift indicator is typically a quantitative application-specific metric. For example, one approach is to monitor an ML model's test data for drift from the training data with a statistic metric (e.g., Kullback–Leibler divergence or Jensen-Shannon divergence). A second approach is to monitor the variation of model predictive quality over a time window, allowing trending model quality. A third approach is to calculate a mean error between the ML predictions and actual measurements in the recent past and tune the last fully-connected layer or the last few layers, in the case of a neural network. In this direction, state-of-the-art studies have developed several drift-based approaches to determining when to update a deployed ML model. For example, ~\citet{chakraborty2021machine} investigated ML model updating in the digital twin of a multi-time scale dynamical system, where physical data was periodically used to update the ML model in learning the time-evolution of system parameters.~\citet{davis2017calibration} studied the deterioration of ML model performance in predictive analytics for the development of acute kidney injury with ten years of patient data and highlighted the importance of routine calibration of ML models when they were incorporated into clinical decision support systems. 

Q-learning is a reinforcement learning algorithm that identifies the best action given the current state of a physical system. ~\citet{sun2020adaptive} took advantage of Q-learning to sensitively quantify the deviation between the measurable response (e.g., CPU frequency of industrial devices and computer clusters or energy consumption) of a physical device (e.g., a sensor or a monitor) and its mapped value in the digital twin of an IIoT system and updated the ML model in the IIoT system to achieve a better trade-off between local update and global parameter aggregation. ~\citet{wu2019real} developed error- and event-triggered mechanisms to update a recurrent neural network model using the most recent process data for model predictive control (MPC) in a chemical process subject to time-varying disturbances. 

It is worth noting that ML model updating can be formulated as a parameter estimation problem, expressed as a discrete-time state-space model. In this state-space model, the ML model parameters to be updated (often a small subset of all ML model parameters) become ${\boldsymbol{\uptheta}}$ that changes by a vector of process noises (state transition equation), and the ML model becomes the measurement function that maps the ML model input and parameters to the measurement $\bf{y}$ (measurement equation). As new measurements from the physical system become available, Bayesian filters ~\citep{cox1964estimation,gustafsson2010particle}, such as the extended Kalman filter and particle filters, can be used to solve this state-state model for online updating the ML model parameters. See a more detailed discussion in Sec. \ref{sec:state_parameter_estimation}.

Accurate ML model prediction plays a cornerstone role in digital twins. ML model updating is just one of the mainstream ways to ensure satisfactory generalization performance of ML models in the open world. In addition to regular ML model updating, if the physics characterizing the underlying physical processes are known, incorporating these physics in the ML model can also be pursued as another viable path to improve ML models' generalizability. See a more detailed discussion of physics-informed ML in Sec. \ref{sec:PI_ML}.


\subsection{Fault diagnostics and failure prognostics}
\label{sec:diagnostics_and_prognostics}

\subsubsection{Overview of PHM}
\label{sec:overview_of_PHM}
Over the past two decades, sensor data collection, fault diagnostics, failure prognostics, and maintenance decision making have drawn significant attention, collectively forming a research field called prognostics and health management (PHM). This research field has been largely promoted by (1) various workshop committees on structural health monitoring (SHM), a research field largely parallel to and synergistic with PHM, (2) the PHM Society, and (3) the rise of journals like {\em Structural Health Monitoring: An International Journal}. While fault diagnostics and failure prognostics form an essential, integral part of PHM, the system health management goes beyond the predictions of the RUL and provides decision support for optimal maintenance and logistics decisions by integrating also the considerations on the available maintenance resources, the operating requirements, and the economic, environmental and operational impact of different faults. In general, health management can be considered as the process of taking optimal maintenance decisions based on the outputs from diagnostics and prognostics models, while taking the relevant operational, resource and monetary constraints into consideration. In short, the goal of PHM can be considered as optimally managing the health of components and systems by minimizing the operational and economic impact of failures, and proactively controlling the direct and indirect maintenance costs \cite{lee2014prognostics}.

The concept of PHM has been exploited in different application domains, leading to a diverse set of terms, algorithms, and tools. For example, machinery health monitoring is an engineering discipline dedicated to monitoring the health of machine components, including bearings, gears, shafts, motors, and pumps, and tracking and predicting their health evolution over time. Vibration-based fault diagnostics is a mature subdiscipline of machinery health monitoring that has a much longer history than PHM, with many algorithms and tools available for processing vibration signals \citep{antoni2006spectral} and diagnosing bearing faults \citep{randall2011rolling} and many applications in industrial sectors \citep{randall2021vibration}. Another example is that research on damage diagnostics and prognostics of civil infrastructures has, over time, formed an engineering discipline called SHM. As described earlier, SHM and PHM are two largely parallel yet synergistic research fields, with SHM having a much longer history than PHM. SHM primarily focuses on monitoring and detecting structural damage that alters the material and geometric properties of a structure \citep{farrar2007introduction}. In SHM, ``damage assessment" is a more appropriate and widely used term than fault diagnostics. It deals with detecting, localizing, and identifying the types of, and quantifying the severity of, damage as just defined. Given that structural failure is rare and difficult to introduce, prognostics plays less of an important role than damage assessment. A third example is when PHM is applied to lithium-ion batteries, creating an active discipline of battery health management. In this discipline, a better alternative to the term ``fault diagnostics" is ``degradation diagnostics", which identifies and quantifies degradation modes and mechanisms driving cell-level degradation \citep{birkl2017degradation}. Metrics used to measure cell-level degradation are called the state of health parameters, as discussed in \ref{sec:probabilistic_updating} (a). Two representative metrics in this regard are capacity and internal resistance, which together determine the energy and power capability of a battery cell. 

A few representative review papers on failure prognostics and SHM are (1) \cite{heng2009rotating, lee2014prognostics, lei2018machinery} for machinery health monitoring, with the first focusing on comparing traditional reliability engineering and prognostics, the second providing a balanced review on methodologies and industrial applications, and the third focusing on datasets and methodologies for health assessment and RUL prediction, and (2) \cite{rezvanizaniani2014review, waag2014critical, hu2020battery} for battery health management, with the first focusing on SOC and state of health monitoring as well as lifetime prognostics, the second on SOC and state of health monitoring, and the third on battery lifetime prognostics.

\subsubsection{Diagnostics vs. prognostics}
Before discussing diagnostics and prognostics in detail, let us first look at the key differences between fault diagnostics and failure prognostics. Fault diagnostics is focused on identifying the presence (fault detection), location (fault localization), type (fault identification) and possibly severity (fault severity assessment) of damage/defects representative of the system’s {\em current} health state. In contrast, failure prognostics is tasked with tracking the time evolution of fault properties and projecting their progression into the {\em future} to yield an updated estimate of the physical system’s health state or RUL, defined as how long the physical system can operate before it reaches a failure threshold or completely breaks down. Time evolution of damage or degradation is an essential factor that plays an almost dominant role in RUL prediction. In short, diagnostics gives us the current damage state, and prognostics projects that into the future to explore questions about criticality that could suggest actions such as maintenance, repair, etc.

\subsubsection{Fault diagnostics approaches}
\label{sec:fault_diagnostics}
Fault diagnostics is a broad topic that has been extensively studied in many application domains over the past decades. We intend to provide a brief survey of this topic in this subsection, focusing on ML-based fault diagnostics. Interested readers can gain more knowledge and insights by looking at review papers on fault diagnostics in respective domains, such as \cite{samuel2005review,jardine2006review,li2016recent,zhao2019deep} for machinery fault diagnostics, \cite{lynch2006summary, fan2011vibration} for sensors in SHM applications, \cite{farrar2007introduction,fan2011vibration} for structural damage detection and identification, and \cite{berecibar2016critical,li2019data} for battery state of health estimation. Furthermore, a recent special issue of the {\em Structural Health Monitoring} journal featured ML-based approaches ({\em Structural Health Monitoring} 20(4), 1353-2239, 2021.)

\begin{figure*}[!ht]
  \centering
    {\includegraphics[scale=0.6]{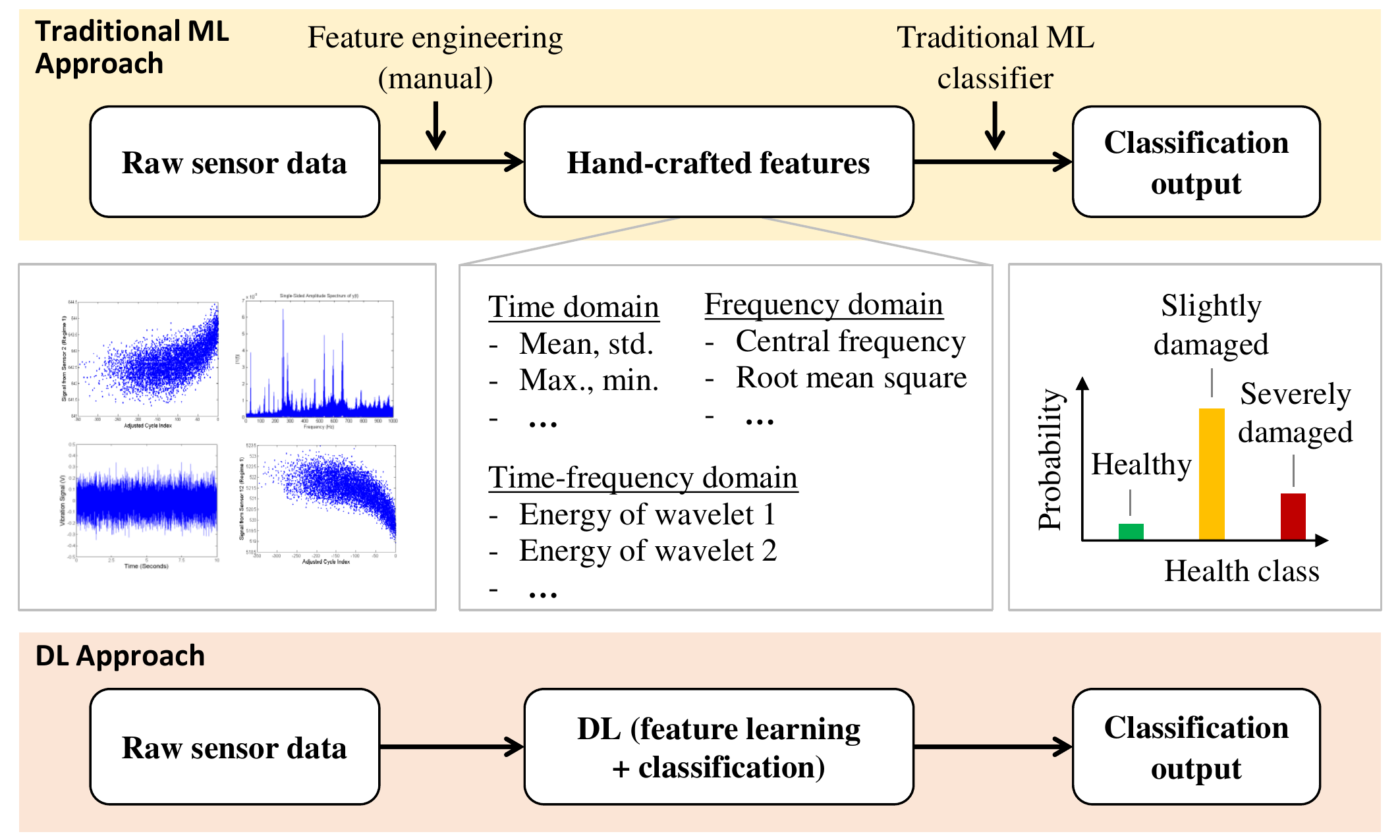}}
  \caption{A graphical comparison of traditional ML and emerging DL approaches to fault diagnostics}
  \label{fig:fault_diagnostics}
\end{figure*}

In ML-based fault diagnostics, ML models are either classification tools to classify the health state of a physical system (e.g., healthy, slightly damaged, or severely damaged of a roller bearing in a hydraulic motor \citep{shen2021physics}) or identify its fault type (e.g., unbalance, misalignment, mechanical looseness, rubbing, or oil whirl in a journal bearing rotor system \citep{oh2017scalable}) or predictive tools to quantify a fault/damage present in a physical system. Approaches have been developed using conventional ML and emerging DL methods. As indicated in Fig. \ref{fig:fault_diagnostics}, running a forward pass on conventional ML models typically involves two steps: feature extraction and classification/regression. Before training these ML models, their input features need to be properly defined. Low-dimensional features (typically less than 20) are first extracted from preprocessed sensor data and then used with other low-dimensional monitoring data (e.g., measurements of operating conditions) for fault classification or damage assessment. The commonly used feature extraction techniques include statistical feature extraction \citep{jegadeeshwaran2015fault}, principal component analysis \citep{misra2002multivariate,pei2021digital}, wavelet transform \citep{peng2004application}, and fast Fourier transform \citep{zhang2013fault}. For example, \cite{pei2021digital} used principal component analysis to extract important features from manufacturing process monitoring signals. Afterward, a support vector machine was used to map the extracted features to processing quality which is labeled as acceptable or abnormal according to the level of tolerance deviation. 

While successes using conventional ML methods for fault diagnostics can be found in some application domains, these success stories have been limited to cases where the domain knowledge required to extract hand-crafted features from the raw data and select appropriate features is available. Furthermore, extracting and selecting input features can be tedious and time-consuming. It is an iterative, manual process known as feature engineering, where features are added and removed, and ML models are redefined, trained, and evaluated. Feature engineering can account for more than 90\% of the total effort in building ML models. 

Emerging DL methods provide end-to-end fault diagnostics capability, eliminating the need for manual feature extraction, as indicated in Fig. \ref{fig:fault_diagnostics}. Unlike traditional ML algorithms, DL algorithms, such as those described in Sec.  \ref{sec:ML_models}, can automatically extract high-level, complex representations (or features) from large volumes of data. The DL models take preprocessed sensor data directly as input and produce the health class or fault/damage size as output. For instance, melt pool images and thermal images collected by cameras have been directly used as the inputs of DL models for defect detection in both additive and more conventional manufacturing processes \citep{liu2020digital,wang2020deep,franciosa2020deep}. 

In recent years, hundreds or even thousands of DL-based fault diagnostics approaches have been reported in the literature. Given large volumes of labeled data, DL approaches show better accuracy in fault diagnostics than their conventional ML counterparts in most applications. A review of DL methods for fault diagnostics can be found in \cite{zhang2020deep}. As expected, the effectiveness of ML models for fault diagnostics is significantly affected by the quantity and quality of the data used to train the model. In particular, data from faulty states are much more difficult to collect than data from the healthy state. This data collection issue could result in data imbalance and poor accuracy in fault diagnostics. Three strategies have been investigated to address this data challenge, and they all build on known physics.

\begin{itemize}
\item \emph{Synthetic data generation}: As reviewed in Sec. \ref{sec:PI_ML}, a variety of physics-informed ML approaches can be used in physical system modeling to improve generalizability, to increase interpretability, and to alleviate the data-size requirement over purely data-driven ML approaches. Similarly, physics-informed ML can also improve fault diagnostic performance. If we consider that data under faulty states are very difficult to collect, `` synthetic” faulty data can be generated from high-fidelity physics-based simulations and used to augment the original training dataset. The synthetic data can be generated by changing health-relevant parameters of the physics-based model (e.g., reducing material stiffness for selected regions where structural damage is intended to be injected \citep{kapteyn2021probabilistic}). Combining the synthetic data with available real data to train and validate ML models is essentially a strategy for physics-informed ML, illustrated as Approaches 2, 3, and 5 in Fig. \ref{fig:hybrid_model}. When high-fidelity physics-based models are not available, an alternative solution to synthetic data generation is to use a DL-based generative model called a generative adversarial network (GAN). After being trained on a real dataset, a GAN can generate synthetic healthy/faulty data, which share a similar distribution as the real training dataset \citep{shao2019generative}.  
    \item \emph{Physics-informed architecture or loss function design}: In many cases, physical domain knowledge can be leveraged to customize mathematical operations or node connections in ways that mimic essential physics and make black-box ML models more interpretable. The design of the model architecture can be designed to constrain the output of the model to better reflect the governing physics. For instance, \cite{sadoughi2019physics} proposed a physics-based (or better termed as physics-informed) CNN for fault classification of rolling element bearings. The uniqueness of their physics-informed CNN is that the kernel functions in its first convolution layer are designed based on known physical knowledge about bearings and their fault characteristics. \cite{kim2022health} developed a CNN-based classifier for gearbox fault diagnostics based on a health-adaptive time-scale representation that was built using multiscale convolutional filters. These convolutional filters were specially designed to incorporate known physics of faults, more specifically, the fault-related time and frequency characteristics in the vibration signals from a faulty gearbox. Other physics-informed ML approaches to fault diagnostics utilize frequency components of interest to customize the loss function used in ML model training, as reported in \cite{shen2021physics, russell2022physics}. These physics-informed ML approaches can help address the data challenge in ML-based fault diagnostics, as they can improve the generalizability of ML models to test samples unseen during model training, including those that may outside of the training distribution, as more extensively discussed in \ref{sec:PI_ML}. A a result, compared to  a purely data-driven ML model, a physics-informed ML model may be trained on a smaller set of faulty data to achieve a similar generalization performance.
    \item \emph{Fault diagnostics based on physics-based modeling}: For physical systems in the SHM domain, such as civil infrastructure systems and large aircraft structures, or in safe-critical industrial applications, it is unlikely one would be able to collect faulty data since these physical systems are generally well-maintained, and failures are extraordinarily rare. Thus, the amount of available faulty data is very limited. Even if it is possible to collect a reasonable quantity of faulty data and build a training dataset, this dataset may not be representative of other structures operating under different loading and environmental conditions. As a result, it is challenging to build ML models readily deployable for structures or machines that these models have not seen. One way to tackle this challenge is model-based fault diagnostics taking advantage of physics-based models. This approach classifies the health state of a physical system or quantifies its fault/damage by studying model predictions under healthy and faulty conditions \citep{park2016model}, comparing model prediction with sensor data, or using sensor data to update a physics-based model (see Bayesian model updating in Sec. \ref{sec:probabilistic_updating}), without resorting to building data-driven ML models. For example, \cite{park2016model} proposed a model-based fault diagnostics method to detect faults in planetary gears. A lumped parametric model was developed to simulate transmission error signals that contained gear fault-related information. The development of this physics-based model allowed for a simulation-based investigation into how to extract fault-sensitive features from these transmission error signals, informing the implementation based on actual sensor data in practice.  
\end{itemize}

\subsubsection{Failure prognostics approaches}
\label{sec:failure_prognostics}
Failure prognostics approaches can be broadly divided into \emph{physics-based}, \emph{data-driven}, and \emph{hybrid approaches}. An ideal example of a physics-based approach is using a physics-based model, as described in Sec. \ref{sec:physics_based_modeling}, depicting the physical processes, in combination with a degradation model, depicting the time evolution of damage or degradation parameters used in the physics-based model. Such a physics-based approach has been attempted in \cite{ramadesigan2011parameter}, as described in Sec. \ref{sec:state_parameter_estimation}. These ideal physics-based approaches can capture the physical processes and failure mechanisms, therefore offering good interpretability and generalisability. However, their applicability to prognostics is limited by difficulties in calibrating large numbers of parameters in the physics-based models due to, for example, poor model identifiability, as mentioned in Sec. \ref{sec:physics_based_modeling}  and \ref {sec:probabilistic_updating} (a), an insufficient understanding of the degradation mechanisms and their long-term evolution, and high computational costs (not suitable for real-time decision making). These limitations motivated researchers to look for models that are not strictly physics-based but rather semi-empirical or empirical. Examples of semi-empirical models include the Paris’ law equation describing fatigue crack growth widely used for fatigue damage prognostics \citep{sankararaman2011uncertainty} and the square-root-of-time dependence capturing battery capacity fade due to the solid electrolyte interphase growth for battery lifetime prediction \citep{bloom2001accelerated, smith2011high}. Empirical models do not have any physical meaning but show good agreement with data. Three representative examples of empirical models are (1) the statistical degradation models described in Sec. \ref{sec:data_driven_modeling} (b) for general-purpose prognostics (i.e., health forecasting of a general physical system), (2) exponential degradation models, originally developed for general-purpose prognostics \citep{gebraeel2005residual} and further developed and applied to bearing prognostics \citep{li2015improved} and battery prognostics \citep{he2011prognostics}, and (3) power-law models for battery prognostics \citep{lui2021physics, gasper2021challenging}, one of which reduces to the traditional model when the temporal power exponent equals 0.5.

\begin{figure*}[!ht]
    \centering
    \includegraphics[scale=0.55]{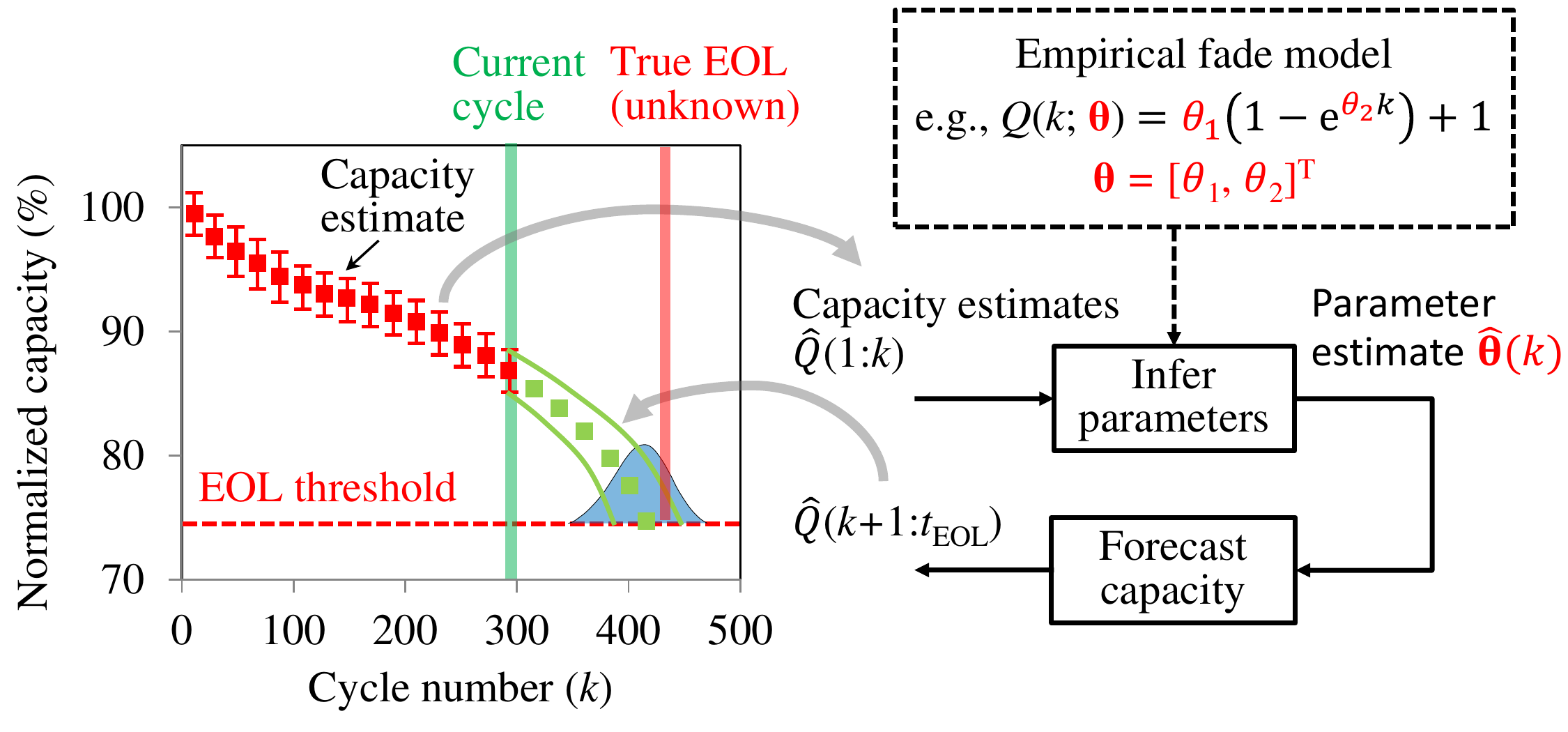}
    \caption{A schematic illustrating model-based prognostics for battery capacity forecasting and RUL prediction. $Q$ denotes the capacity of a battery cell normalized by its initial capacity, and $k$ denotes the cycle number. EOL in the figure stands for end of life.}
    \label{fig:battery_degradation}
\end{figure*}

Physics-based approaches, plus prognostics approaches using semi-empirical and empirical mathematical models, form a broader category called \emph{model-based prognostics}. Figure \ref{fig:battery_degradation} shows a diagram illustrating a model-based approach using an empirical exponential model for battery capacity forecasting and RUL prediction. This battery prognostics problem can be formulated as a state-space model for parameter estimation, as discussed in Sec. \ref{sec:state_parameter_estimation}, and solved using particle filtering, as detailed in Sec. \ref{sec:probabilistic_updating} (b). The model-based approach outlined in Fig. \ref{fig:battery_degradation} is used for online estimate of the RUL of a battery cell. In a digital twin, the RUL prediction could then be used in other process and control optimization algorithms to, for example, determine the optimal time to remove the cell from its first-life application. This application of a model-based prognostics method within a digital twin framework is extensively investigated as a case study in Part 2 of this review on digital twin.

\begin{figure*}[!ht]
  \centering
    {\includegraphics[scale=0.7]{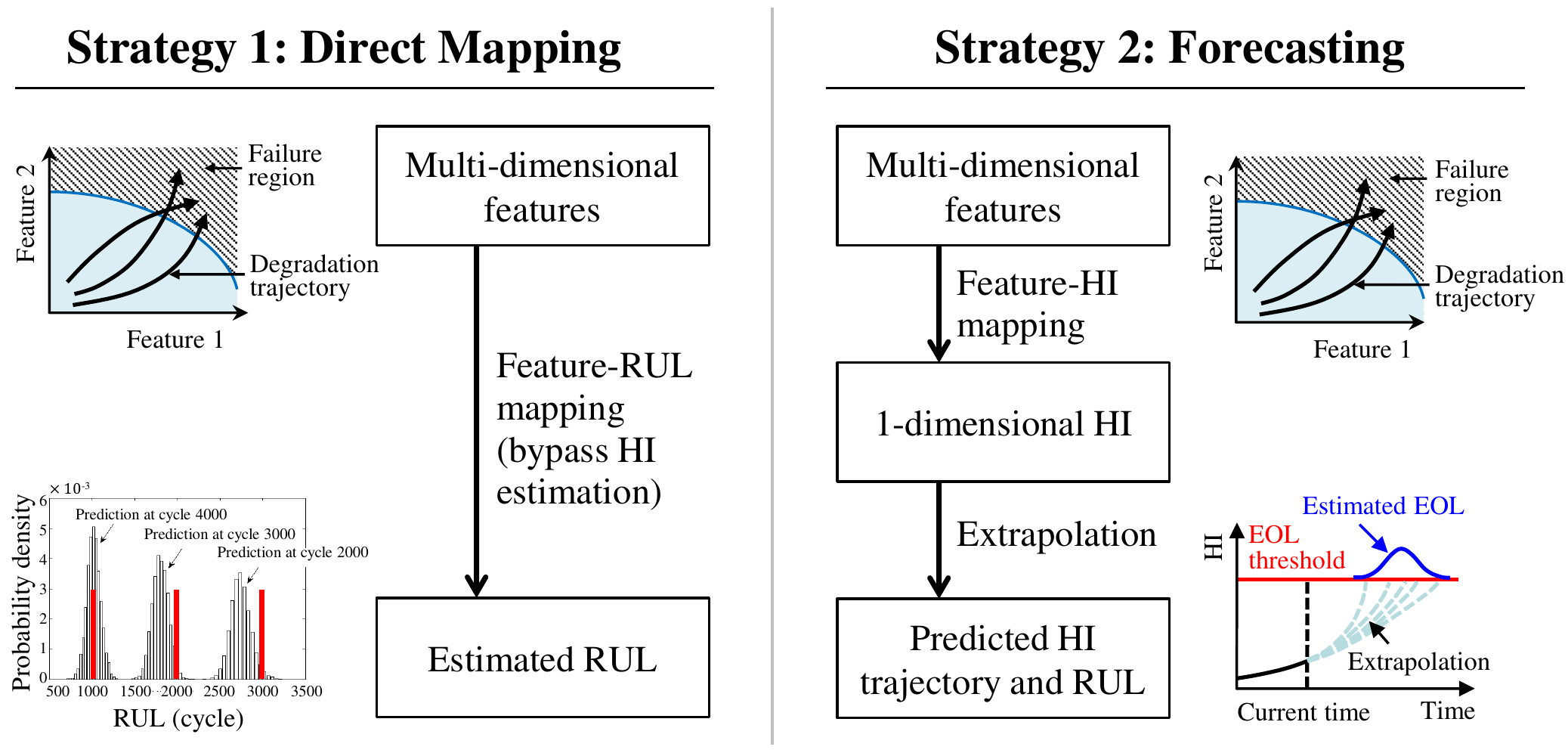}}
  \caption{A graphical comparison of two distinct strategies for data-driven prognostics. HI and EOL stand for health indicator and end of life, respectively.}
  \label{fig:Data-driven prognostics}
\end{figure*}

Suppose a good degradation model is unavailable, or degradation parameters are hard to estimate from observational data. In that case, data-driven approaches may be a better alternative to mode-based approaches, assuming medium to large volumes of degradation data are available. This is usually only possible in applications where life safety issues are not a concern. Data-driven approaches can be categorized further into direct mapping and time series forecasting approaches, as illustrated in Fig. \ref{fig:Data-driven prognostics}. Direct mapping approaches bypass health forecasting and directly map features (ML model input) to RUL (ML model output) using ML models such as standard feedforward neural networks \citep{huang2007residual, tian2012artificial}, deep CNN \citep{li2018remaining, li2019deep}, and LSTM recurrent neural networks \citep{yuan2016fault,huang2019bidirectional}. Some variants of direct mapping approaches do not output RUL and instead estimate a health indicator (e.g., an overall measure of engine/milling machine health \citep{malhotra2016multi} or bearing health \citep{guo2017recurrent}, battery capacity \citep{shen2020deep}), or a battery degradation parameter such as an electrode aging parameter \citep{tian2021electrode, thelen2021physics}). ML-based forecasting approaches predict how a health indicator or reliability metric (such as time to failure or survival probability) evolves beyond the current time as an intermediate step for RUL prediction. ML models used for forecasting include support vector machine \citep{pham2012machine}, relevance vector machine \citep{wang2013prognostics}, Gaussian process regression \citep{richardson2017gaussian}, feedforward neural networks \citep{fink2014predicting}, and LSTM recurrent neural networks \citep{nemani2021ensembles}, just to name a few. Applications of DL to PHM are well-reviewed in \cite{khan2018review,zhao2019deep,fink2020potential}, with the first and second papers focusing on machinery health monitoring, the second paper providing a comparative study on the tool wear prediction performance of ten ML and DL algorithms, and the third paper focusing on successes of DL in various application domains including PHM. 

Data-driven ML approaches can discover complex degradation patterns relevant to RUL prediction from large volumes of data, are simpler to implement than physics-based approaches, and are agnostic to the physical and degradation processes. However, large ML models can have many more parameters than physics-based models and require large training datasets to ensure low risks of overfitting. In reality, collecting large and representative run-to-failure datasets can be costly and time-consuming. Another well-known issue with ML models is that they have difficulties in generalizing to out-of-distribution data unseen during model training, as also discussed in Sec. \ref{sec:PI_ML} and \ref{sec:ML_updating}, and may produce predictions that do not make physical sense. In conclusion, the performance of purely ML models is limited by the quality and quantity of the available training data. Without consideration of physics, the accuracy numbers on a test dataset that falls within the training data distribution may not hold for other datasets outside the training distribution.

As discussed above, purely physics-based and data-driven approaches have complementary advantages that can be leveraged to tackle the limitations. For example, physics-based models offer opportunities to generate synthetic data that can help alleviate the data-size requirement of data-driven approaches, extending the applicability of these approaches to cases where run-to-failure trajectories are scarce. Using physics-based models to generate synthetic training data is illustrated as Approaches 2, 3, and 5 in Fig. \ref{fig:hybrid_model}. Another example is that data-driven ML models can learn complex bias functions from data to compensate for discrepancies in physics-based models. This strategy in combining physics-based and model-based approaches is illustrated in Fig. \ref{fig:hybrid_model} as Approaches 4 and 5, where the bias functions respectively capture the prediction inaccuracy of a physics-based model (Approach 4) and the prediction inaccuracy of an ML model trained with physics-based synthetic data (Approach 5). All the approaches described in Sec. \ref{sec:PI_ML} can be adapted for failure prognostics, with some successes recently reported in \cite{chao2022fusing,kohtz2022physics,yucesan2022hybrid}.

\subsection{Ontology-based reasoning}
\label{sec:Ontology_reasoning}
Ontology maps and knowledge graphs provide a straightforward way to represent empirical knowledge in a structured manner. The unified knowledge engineering and representation through descriptive logic lays a firm foundation for knowledge reasoning over knowledge graph in the semantic space. In practice, they are primarily implemented through ontology markup languages~\citep{chen2010development}, such as Web Ontology Language, XML Schema, Resource Description Framework Schema (RDFS).

The proper construction of an ontology map and knowledge graph facilitates knowledge inference along the semantic dimension~\citep{chen2020review}. In general, knowledge reasoning methods can be categorized into three classes.
\begin{enumerate}
    \item \textbf{Logic rules-based reasoning}. Rules-based reasoning manifests in several forms: (a) first-order predicate logic rules, such as First-Order Inductive Learner (FOIL)~\citep{schoenmackers2010learning}, that is to use the predicate as the basic unit for knowledge reasoning; (b) manually inject user-defined or simply learned rules in knowledge graph to generate new knowledge, such as the inference component in the Never-Ending Language Learning system (NELLs)~\citep{mitchell2018never}.
    (c) to incorporate frequently present patterns, constraints or paths to extract new knowledge. That is to reason with the Web Ontology Language. (d) identify path rules and inject them into knowledge graph to facilitate reasoning. For example,~\cite{lao2010relational} utilized a path ranking algorithm to identify the associations between edge types and the instances of edge types, and exploited such associations to predict missing edges in the graph.
    
    \item \textbf{Distributed representation-based reasoning}. This line of studies aim to project the basic elements in knowledge graph (i.e., entities, relations, attributes) into a continuous vector space through embedding-based approaches~\citep{nickel2011three}, including tensor decomposition, distance-based mapping, and semantic matching models. The distributed representation allows to discover latent rules and connections in the transformed space. 
    
    \item \textbf{Neural network-based knowledge reasoning}. In recent years, more and more studies have explored the representation learning ability in deep neural networks for knowledge reasoning. Specifically, the CNN-based learner enables to account for the fine-grained multi-source heterogeneous information encoded in entity description~\citep{xie2016representation}; the recurrent neural network is capable of reasoning about multi-hop relations in knowledge graph~\citep{neelakantan2015compositional}. These unique advantages have positioned neural network as an appealing paradigm for knowledge inference in the semantic space.
\end{enumerate}

Over the past few years, there has been an increasing investigation of ontology techniques for data management as well as knowledge representation and inference in digital twins. For example,~\citet{ladj2021knowledge} combined business rules with the knowledge deriving from industrial data analytics to build ontology-based knowledge model for incident detection in the machining process, and the knowledge base served as an inference platform to detect detrimental incidents.~\citet{lim2020digital} employed knowledge graph-based method to accommodate heterogeneous data sources, such as asset ontology, environment ontology, and system ontology. The comprehensive data integration in knowledge graph drives the inference of smart solutions for asset configuration, resource planing and component maintenance.~\citet{liu2021multi} developed multi-scale product quality knowledge model to characterize the relationship between product quality and the product quality factors explicitly.

\section{Virtual-to-physical (V2P) twinning enabling technologies}
\label{sec:V2P}

\begin{figure*}[!ht]
  \centering
    {\includegraphics[scale=0.6]{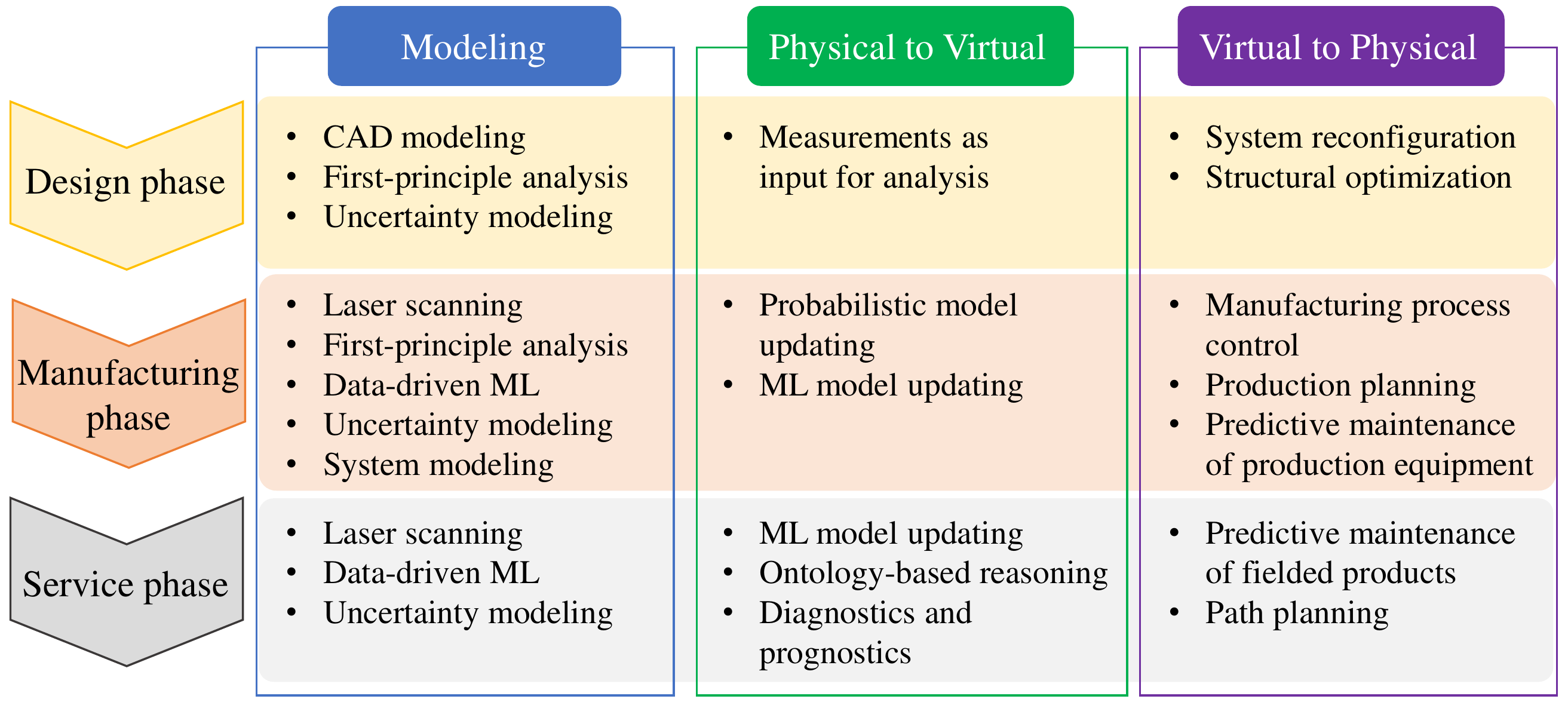}}
  \caption{Classification of modeling and twinning techniques according to phase of a physical system's life cycle}
  \label{fig:method_by_domain}
\end{figure*}

Figure \ref{fig:method_by_domain} summarizes the modeling and twinning methods by application in different phases of a physical system's life cycle. As illustrated in this figure, there are numerous ways of establishing V2P connection, such as system reconfiguration, process control, production planning, maintenance scheduling, and path planning. In this section, we mainly concentrate on two classical V2P twinning enabling technologies, namely model predictive control and predictive maintenance.

\subsection{Model predictive control} 
\label{sec:MPC}

Model predictive control (MPC) includes a wide range of advanced control methods, which in common use a model to predict the future behavior of the process and determine the optimal control within a set of constraints.  MPC is an essential component for smart manufacturing, which enables not only the optimal use of resources for high-quality products, and but also quick production responses to changes in market demands and supply chains \citep{Lu2016SM}. It is one of the most widely used methods to establish the V2P connection in a digital twin.

    \begin{figure*}[!h]
    \centering
    \includegraphics[scale=0.60]{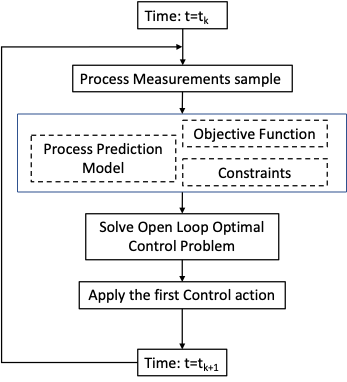}
    \caption{MPC structure \cite{Nikolaou2001MPC}}
    \label{fig:mpc_structure}
    \end{figure*}

Figure ~\ref{fig:mpc_structure} shows a general structure of MPC, including a process model, an objective function, process measurements, constraints and sampling points \citep{Nikolaou2001MPC}. 

Assuming an arbitrary process with a state space representation: 
\begin{equation}
\label{eq:MPC state-space1}
\begin{split}
  & {{\bf{x}}({k+1})} = {\bf{f}}({{\bf{x}}(k}),\;{{\bf{u}}(k})),  \cr 
  & {{\bf{y}}({k+1})} = {\bf{h}}({{\bf{x}}({k+1})}). \cr
\end{split}
\end{equation}

MPC minimizes a user-defined cost function $J$,
\begin{equation}
\label{eq:MPC optimization 1}
\begin{split}
    \mathop {\min \;}\limits_{\bf{u}} \sum\limits_{t = 0}^k {J({\bf{x}}(t),\;{\bf{u}}(t))} ,\cr
\end{split}
\end{equation}
in which $J({\bf{x}}(t),\;{\bf{u}}(t)):{\mathbb{R}^s} \times {\mathbb{R}^c} \to \mathbb{R}$ is the cost function at $t$.

For a trajectory following process control problem, the objective functions can be any norm of the tracking error between the reference vector $\bf{r}$ and the model output $\bf{y}$, as shown in Eq. (\ref{eq:MPC optimization 2}). The optimization problem is defined under hard state and input constraints:

\begin{equation}
\label{eq:MPC optimization 2}
\begin{split}
  & \min_u \;\;\;\; {\sum^{N_2}_{i=N_1}} \; \lVert { {\bf{r}}({{k}+i}\mid{k})-{\bf{y}}({{k}+i}\mid{k}})\rVert \cr
  & s.t. \;\;\;\;\;\;\;\;\;{\bf{u}}_{lb} \leq {\bf{u}}({k} + j \mid {k}) \leq {\bf{u}}_{ub} \cr
  & \;\;\;\;\;\;\;\;\;\;\;\;\;\;{\bf{y}}_{lb} \leq {\bf{y}}({k} + j \mid {k}) \leq {\bf{y}}_{ub} \cr
  & \;\;\;\forall \;\; i \in \{N_1, ..., N_2 \} \;\;and\;\; j\in \{0, ..., N_u \}, \cr
\end{split}
\end{equation}
where $N_1 \leq N_u \leq N_2$.

\begin{figure*}[!ht]
    \centering
    \includegraphics[scale=0.50]{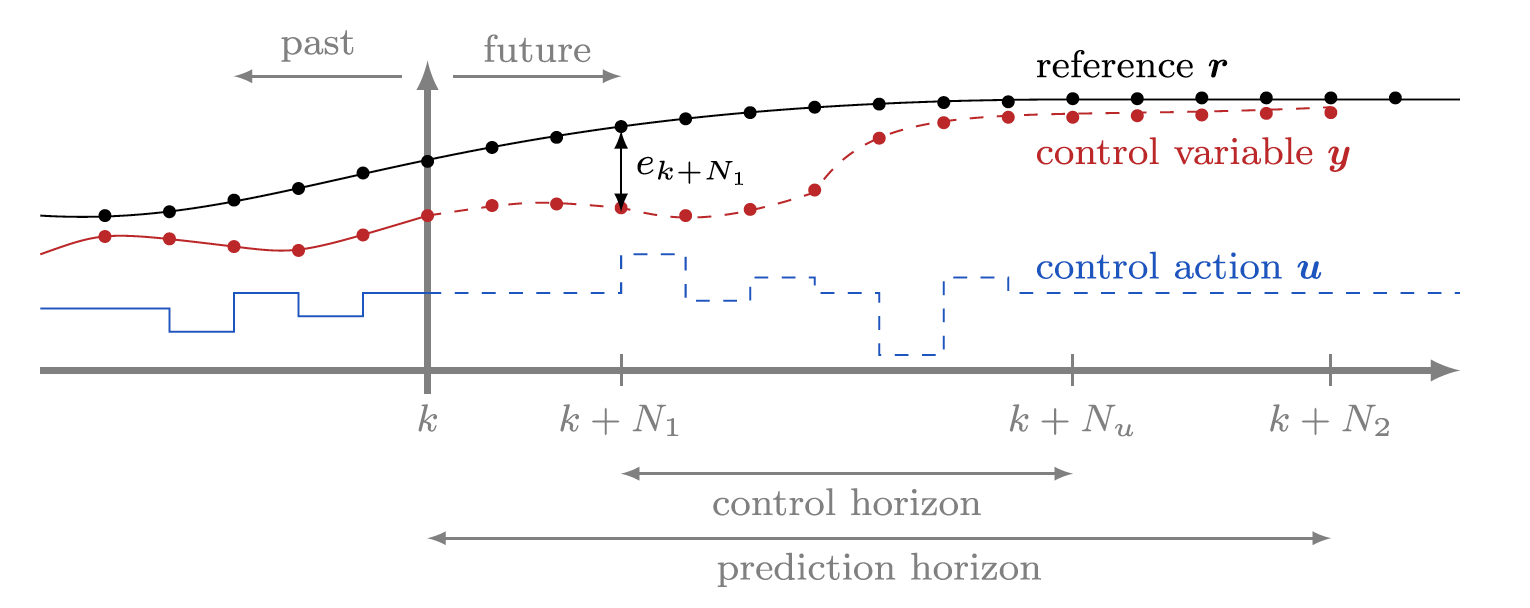}
    \caption{Iterative Horizon Receding Control of MPC \cite{Schwenzer2021MPC} }
    \label{fig:mpc_function}
    \end{figure*}

The iterative, finite-horizon optimization of a process can be further illustrated in Fig. \ref{fig:mpc_function}. At time $t_k$ the current process measurements are acquired and a cost minimizing control strategy is computed to minimize the error between the system output $y$ and the given reference $r$ for a horizon $N_2$. $N_2$ must be long enough to represent the effect of a change in the control commands $u$ on the control variable $y$. If there a pure delay exists, the sum of the objective function starts from $N_1$. Only the first value of the optimal control commands is implemented, then the process measurements are sampled again and the optimization is repeated, yielding new sets of control commands and new predicted state path. The prediction horizon keeps being shifted forward so MPC is also called receding horizon control. 

MPC has become the most widely implemented advanced process control technology and is offered by major automation suppliers, with applications in process industries such as refining, petrochemicals, paper and food processing \citep{RockwellMPC,SiemensMPC} . In the last decade, MPC is extended and also adopted in power plant control, build thermal management and discrete manufacturing \citep{PowerPlantMPC,MicrogridMPC,AutomotiveMPC}.  In the era of smart manufacturing, the process and manufacturing industries are embracing tremendous opportunities brought by the Internet-of-Things (IoT) and advanced data analytical technologies, which also enhance the application of MPC by expanding the objective function to more some general performance measure \citep{EconomicMPC}, and by combining machine learning techniques in process modeling \citep{DLMPC1}. Economic MPC employs cost function connected to the economics of the considered process which makes economic MPC well suited as a tool to achieve the goals of smart manufacturing. Machine learning-based MPC is reported to use recurrent neural network modeling approaches for a general class of nonlinear dynamic process systems to improve prediction accuracy and control performance \citep{RNNMPC}. There are also efforts reported to integrate a DL architecture with MPC for robot manipulation control \citep{DeepMPCLD}. Rather than developing a dynamics model from first principles, the future MPC will more likely to learn digital models using a novel deep architecture and learning algorithms directly from data. For example, some of the data-driven models summarized in Sec. \ref{sec:data_driven_modeling} can be used as predictive models for MPC in digital twins.

\subsection{Predictive maintenance}
\label{sec:predictive maintenance}
A recent industry-wide shift called \emph{Industry 4.0} capitalizes on increased interconnectivity and automation to make machines on a production floor smarter. Benefits of this digital transformation to Industry 4.0 include increased productivity, improved workflow efficiency, and improved product quality. A key enabling step of Industry 4.0 is predictive maintenance, mostly discussed in the manufacturing domain. Predictive maintenance is a proactive approach to equipment maintenance that centers on (1) identifying signatures in sensor data (continuous monitoring or periodic inspections) indicative of changes in equipment health (see fault diagnostics discussed in Sec. \ref{sec:fault_diagnostics}), (2) predicting when a machine, component or part might fail (see failure prognostics discussed in Sec. \ref{sec:failure_prognostics}), and (3) scheduling maintenance work during planned downtime just before equipment failure, sometimes referred to as “just-in-time maintenance” \citep{lee2013recent,
lee2013predictive}.

Now let us answer a simple question: why is predictive maintenance needed, for example, for a rotating machine (e.g., motor, pump, and fan)? The answer is quite simple. The unexpected failure of this rotating machine often incurs high maintenance and downtime costs, reduces customer satisfaction with a produced good, and may even cause human injuries and fatalities. These consequences also impact the machinery manufacturer by tarnishing their reputation and potentially putting them at a competitive disadvantage. Therefore, it will be value-added to develop, implement, and deploy predictive maintenance solutions, especially justifiable for high-value equipment whose failure cost is high. Note that predictive maintenance differs from the more traditional \emph{reactive maintenance}, which performs maintenance work after an equipment failure and may lead to costly unexpected downtime, and \emph{preventive maintenance}, which performs maintenance work at a fixed time interval, typically more often than necessary, and may cause a waster of time and resources due to unnecessary maintenance.

A typical ML pipeline for building and deploying a predictive maintenance solution for a rotating machine is shown in Fig. \ref{fig:Predictive maintenance} and consists of four main steps. 
\begin{enumerate}
\item \textbf{Acquire sensor data (P2V):} This first step identifies critical machine assets for condition monitoring and installs sensors on these assets to measure signals that are sensitive to machine health. For example, sensor signals sensitive to bearing health include acoustics, vibration, oil wear particle count, and temperature, ranked from the earliest to latest in detecting a bearing fault. Note that in SHM, the placement of a sensor network needs to be optimized to maximize the amount of useful information for structural damage assessment, as discussed in detail in Sec. 2.2.1 of Part 2 of the review paper.


\begin{figure*}[!ht]
  \centering
    {\includegraphics[scale=0.70]{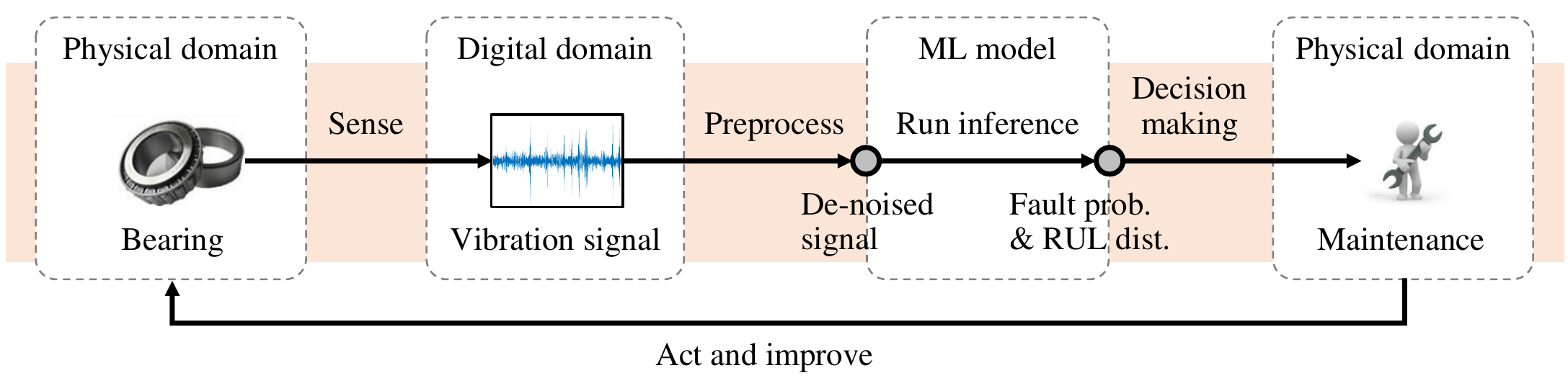}}
  \caption{An overview of an ML pipeline for predictive maintenance}
  \label{fig:Predictive maintenance}
\end{figure*}

\item \textbf{Preprocess data (P2V):} Field data is rarely clean; it contains outliers and noise that need to be removed to avoid corrupting ML model performance. An example in bearing health monitoring is that vibration signal processing is almost always needed to denoise vibration signals by, for example, designing frequency-domain filters used to remove background noise irrelevant to machine faults \citep{antoni2006spectral}. Another example is that outliers in battery voltage and current measurements must be removed before they are fed into ML models for capacity estimation or RUL prediction enabling battery health management. 
\item \textbf{Train ML models (P2V):} This step trains ML models to classify the machine condition as healthy/faulty (green/red) or healthy/lightly damaged/heavily damaged (green/yellow/red) and estimate the RUL of the machine. For both fault classification and RUL prediction, the predictive uncertainty of the ML models needs to be estimated, accurately reflecting the models’ confidence in making a health class/RUL prediction (see Sec. 2.1.1 in Part 2 of the review paper for a detailed discussion on ML model uncertainty). A probabilistic estimate of RUL can be in the form of a mean RUL and an associated confidence interval representing the predictive uncertainty. 
\item \textbf{Optimize maintenance decision making (V2P):}  This last step optimizes maintenance schedules based on the predictive information from Step 3. In practice, what is preferred is to repair or replace equipment during a scheduled downtime just before failure occurs, without affecting the chance of meeting production goals. Maintenance decisions need to be made from a risk perspective, given that ML model predictions are associated with varying degrees of uncertainty. A dedicated review of maintenance schedule optimization will be given in Sec. 2.2.3 (c) in Part 2 of the review paper.
\end{enumerate}

Digital twin can be an enabler of predictive maintenance. As discussed earlier in Sec. \ref{sec:fault_diagnostics} and \ref{sec:failure_prognostics}, one of the main issues with diagnostics/prognostics is the lack of faulty/run-to-failure data (i.e., data acquired from a physical system whose health state or RUL is known). This data challenge can be addressed by using a physics-based model in a digital twin, calibrated offline and updated online (Sec. \ref{sec:probabilistic_updating}), to generate synthetic faulty/run-to-failure data. This strategy for tackling the data challenge has already been discussed in Sec. \ref{sec:fault_diagnostics}.


An example where a digital twin can help with the online phase of predictive maintenance is that a physics-based model in a digital twin, considering both physical and degradation processes, takes sensor data as input, estimates digital state variables that significantly affect lifetime, and predicts RUL. Sometimes, ML models can assist physics-based models in estimating the digital state variables. This strategy is similar to the ML-assisted approach (Approach 6) shown in Fig. \ref{fig:hybrid_model}, except that ML models may not always be used in the digital twin. A lot of success stories have been reported in the manufacturing industry on this strategy, for example, Ansys Twin Builder and PTC thingworx collaboratively working on estimating the maximum temperature of the rotor, stator, and case of an electric motor based on sensor measurements to predict the motor’s RUL \citep{ansysptcmotor}. 

\begin{figure*}[!ht]
  \centering
    {\includegraphics[scale=0.65]{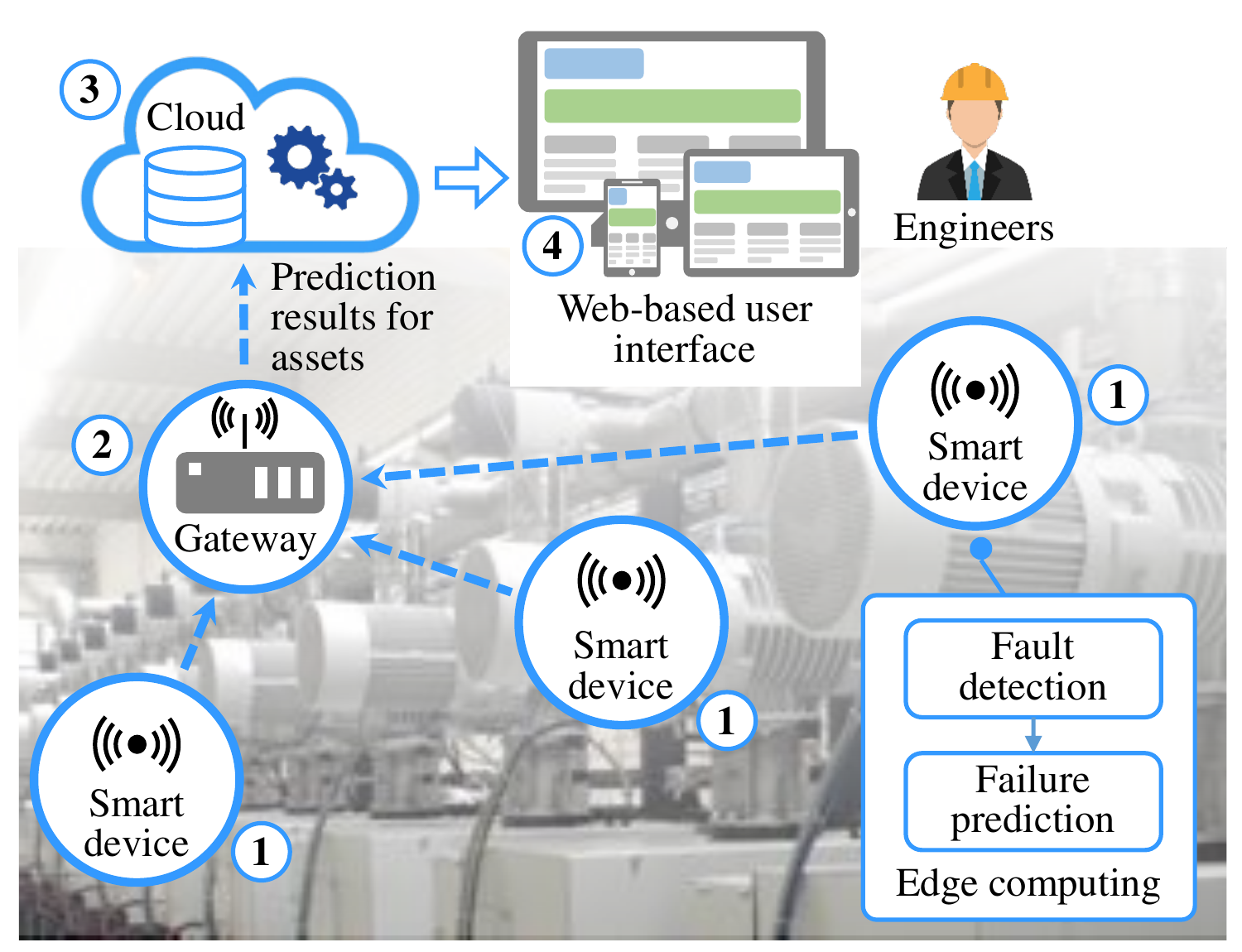}}
  \caption{An IIoT platform for predictive maintenance of industrial assets}
  \label{fig:Deep Learning_Predictive Maintenance}
\end{figure*}

Another industrial success story is the predictive maintenance alarm system in MATLAB developed by Baker Hughes for the positive displacement pumps on their trucks. The core of this system is the pump health monitoring software that analyzes real-time pressure, vibration, and timing signals and uses a trained neural network to predict pump failures. It is shown that the software can classify pump health into one of three health classes, namely “normal operation”, “monitor closely”, and “maintenance needed”, enabling better maintenance planning. However, it is unclear whether the software has predictive capabilities (i.e., whether it can predict the RUL of a pump). As the emergence of IIoT platforms facilitates the industry-scale adoption of digital twins, we can expect the number of industrial success stories of predictive maintenance to continue increasing rapidly. Figure \ref{fig:Deep Learning_Predictive Maintenance} illustrates an industry-grade IIoT platform for proactively predicting and preventing the failure of rotating machinery on a manufacturing floor. In this IIoT platform, a network of smart devices perform vibration analysis, fault detection, and failure prediction on the edge, and then send detection and prediction results to a web-based user interface that provides real-time analytics, dashboards, and alert capabilities to the maintenance and reliability engineers responsible for keeping rotating equipment in service. Maybe talk about the possibility of building centralized digital twins and distributed digital twins on the edge and sharing ML model information among clients without sharing data using federated learning in Sec. \ref{sec:federated_learning}.

\subsection{Real-time perception and decision making}\label{sec:real-time-ML}
After sensor measurements characterizing the state of a physical system are transmitted to the digital counterpart, simulation and optimization models enable the \textbf{P2V and V2P} connections within the digital twin. Specifically, simulation enables precisely modeling of the behavior of the physical system under uncertainty (P2V), and optimization drives control, maintenance, and decision making on the physical asset (V2P) given its current state. Ideally, the simulation and optimization algorithms in the digital model will be performed in real-time and will not have significant latency. However, this is not the case, as simulation and optimization are known to be time-consuming and computationally expensive, which substantially hampers their ability to provide timely results in digital twins which operate on physical systems with short time scales. To this end, the power of ML can be fully exploited to alleviate many of the inherent issues incurred by using classical simulation and optimization models. Various ML techniques can be used to facilitate real-time or near real-time (sometimes termed just-in-time) perception and decision making. Specifically, ML can be leveraged to infer the parameters of or replace the simulation model for accurately modeling the physical system~\citep{goodwin2022real}. For example,~\cite{zohdi2020machine} used ML to ascertain the parameters of a multi-submodel system for fast and accurate simulation of fire propagation in order to support continuous update of the digital twin of the physical counterpart in near real-time. In each case, the simulations are expedited by using pre-trained ML models which are very computationally inexpensive to evaluate. These examples are a few of many where ML can be used to speed up simulation operations in digital twins.

Regarding optimization, the expensive iterations in classical optimization algorithms can be replaced by an end-to-end ML-based optimization model proxy. In this method, the ML-based optimizer typically adopts a predict-then-repair paradigm. The basic idea is to first reformulate the optimization problem as a learning problem and then incorporate fast feasibility-restoration procedures to guarantee the feasibility of the produced solution~\citep{bengio2021machine,barry2022risk}. ML-based optimization proxies have been shown to surpass conventional optimization algorithms by a large margin in computational speed. For example, ~\cite{van2021machine} showcased the role of ML in approximating the optimal solutions of power flow problems and the great potential of ML-based optimization proxy in speeding up existing algorithms. ~\cite{chen2022learning} proposed an end-to-end DL model that can predict an optimal solution for Security-Constrained Economic Dispatch (SCED) in milliseconds to meet the need of power gird in real-time operations. Creating this end-to-end DL model overcomes the fundamental limitation of classical optimization techniques in computational efficiency.

Ultimately, machine learning will play a large role in enabling real-time simulation, optimization, and control of physical systems. Researchers are actively investigating methods to increase the speed of physics-based simulations and traditional optimization algorithms. These techniques will pave the way for digital twins which operate on the millisecond and sub-millisecond timescales.

\section{Perspectives on modeling and twinning in digital twins}
\label{sec:perspectives}

\subsection{Federated learning in digital twins}
\label{sec:federated_learning}
For data-driven digital twins, a single system or component may not have sufficient condition monitoring data to train a representative model. This is particularly due to the fact that industrial and infrastructure systems are operated under varying conditions. Thus, monitoring a system for a limited period of time may not cover all possible expected operating conditions. Moreover, since faults or critical system states are rare, one single system may not observe a sufficient number and type of faults to train the model. Therefore, data from similar systems can be leveraged to improve the performance and generalization of the developed model. 

Accumulating data from multiple similar systems can eventually result in large-scale datasets that enable representative data-driven digital twins to perform well under all relevant operating conditions. Generally, compiling such representative datasets could be achieved through sharing the data between different stakeholders, for example between different operators of similar systems. However, even without direct competition, companies are often reluctant to share data mainly due to the concern of disclosing proprietary company information. Federated learning could potentially overcome that concern enabling to take advantage of the benefits of a shared model without the need of data disclosure. Federated learning is a privacy preserving learning concept. It enables to benefit from the local data subsets (e.g., of several power plant operators) without the need of sharing the data \citep{McMahan2017, Li2020}. It is a decentralized machine learning technique that enables to train a neural network across multiple local datasets without exchanging or sharing data samples across the decentralized nodes. A global model is constructed by aggregating (typically averaging) the parameters (either gradients or weights) of the locally trained models. Hence, it can make predictions based on the experience of the entire fleet but without the stakeholders having access to each other’s data. Each individual system retains its own collected condition monitoring data on the edge, without sharing it with other systems. It gets a current version of the model, improves it based on its collected data and sends the updates to the model.  
Two different concepts of federated learning are typically distinguished: cross-device federated learning (also referred to as sample-based or horizontal federated learning) \citep{yang2019federated} and cross-silo federated learning (also referred to as feature-based or vertical federated learning) \citep{yang2019federated}. In the context of digital twins, in fact, particularly the horizontal federated learning concept is the most relevant. However, there are also contexts where vertical federated learning could be applied. 

Privacy preservation is an essential characteristic of federated learning. However, even though data is not directly shared, only sharing model updates during the training process can still potentially reveal sensitive information \citep{zhu2019deep}. Different privacy preserving approaches have been proposed that provide privacy guarantees and prevent potential indirect data leakage \citep{truex2019hybrid, chen2020training, yin2021comprehensive}. 
Federated learning has not yet been broadly applied for digital twins. Federated learning will then face similar challenges as other data-driven approaches, such as the lack of representative datasets, differences in operating conditions and system configurations between different entities of the digital twins. One of the potential directions to tackle these challenges is combining federated learning and transfer learning. 

\subsection{Domain adaptation in digital twins}
Data-driven digital twins can show a significant performance drop if they are applied under operating or environmental conditions that they have not been trained on. However, evolving operating conditions are very common in practical applications. Moreover, if a data-driven digital twin is applied across different units of a fleet, the performance may also drop significantly due to discrepancies either in system configurations or operating regimes of different units of a fleet. It is, however, unrealistic to assume that sufficient data can be collected for all units and for all operating conditions. Different operating and environmental conditions and different system configurations result in  the so-called domain shift or distributional shift, a change in the data distribution between the source (training) and the target (testing) dataset \citep{fink2020potential, li2022perspective}, which is also described in Sec.~\ref{sec:ML_updating} in the context of machine learning. Typically, no labels are available in the target dataset, so that fine-tuning of the trained model cannot be performed. Further to the two scenarios of domain shifts caused by differences in operating conditions or by differences between different units of a fleet, a third scenario, particularly relevant to digital twins is the synthetic to real gap existing between synthetically generated data stemming potentially from model-based simulators and the actual observation \citep{bousmalis2018using, wang2021integrating, li2021synthetic}. 

Domain adaptation, a subfield of transfer learning, aims at overcoming the differences between the domains so that the algorithms are able to generalize across domains and learn domain-invariant features. Domain adaptation solves the domain shift problem by finding a mapping from the source data distribution to the target distribution. Different approaches for domain adaptation have been proposed. One of the most obvious approaches to overcome the domain gap is distribution alignment approaches by statistics matching, including Adaptive Batch Normalization (AdaBN) and Automatic DomaIn Alignment Layers (AutoDIAL), using the statistics of source and target mini-batches at different layers to align the two distributions \citep{long2015learning, yang2016revisiting, carlucci2017autodial}. Some of the other commonly applied approaches have been relying on the principle of minimizing a divergence-based criterion between source and target distributions, including such criteria as Maximum Mean Discrepancy (MMD) or Wasserstein Discrepancy \citep{yan2017mind, lee2019sliced}. Due to the superior performance on many different tasks, adversarial domain adaptation has recently emerged as one of the most frequently applied approaches to confront domain shift \citep{ganin2014unsupervised, tzeng2017adversarial, ganin2016domain}. A domain discriminator is then applied to ensure that the extracted features become indistinguishable between the two domains, while the performance on the main classification task is maximized.

Domain adaptation has been recently broadly applied to many different tasks in prognostics and health management, partly also for digital twins \citep{ li2022perspective,liu2020domain}.
One of the additional challenges that is typically encountered in real applications is that the label space of the source (training) is often not identical with that of the target (testing) dataset. This is particularly relevant in the context of fault diagnostics, where different systems may have experienced different fault types within the (short) observation time period. Possible setups in real applications are that some classes are missing in the target dataset (Partial Domain Adaptation), some classes may also be missing the source datasets (OpenSet Domain Adaptation). Or we may be even facing only a partial overlap of the label space between source and target datasets, resulting in a combination of the two types (Partial \& OpenSet Domain Adaption) \citep{boris2021universal}. Some approaches have been recently proposed to tackle such challenges \citep{ li2020partial, wang2020missing, li2020deep2,rombach2022controlled}, on the one hand by overcoming misalignment in case of different label spaces and on the other hand by developing generative models for a controlled generation of the missing fault types. 
While domain adaptation has been flourishing for classification tasks, its application to regression has been rather limited \citep{boris2021universal}. This has been also reflected in the tasks tackled in the field of prognostics and health management, where the main application field has been fault diagnostics, with only few research works addressing prognostics tasks as well \citep{ wu2019weighted, da2020remaining}. Similar to classification tasks, domain adaptation for regression is also facing the challenge of label discrepancy between the source and the target label spaces \citep{boris2021universal}.

\subsection{Deep reinforcement learning in digital twins}
\label{sec:deep_reinforcement_learning}
Reinforcement learning is concerned with learning an optimal policy for sequential decision making problems through an agent's interactions with the environment over time by trial and error~\citep{mnih2015human}. Towards policy learning, reinforcement learning not only considers the immediate reward of an action, but also takes into account the accumulative reward resulting from the action in a discounted manner. In general, reinforcement learning is composed of several key components: state space, action space, policy to map from state space to action space, state transition probability, and reward function. The advantage of reinforcement learning in long-term cumulative reward modeling has 
put it in a good position to tackle complex sequential decision making problems. 

Deep neural networks have recently been used to approximate such functions as value functions (value function is used to predict the expected cumulative reward indicting how good a state-action pair is), policies (policy describes which action to take given a specific state), reward functions, and state transition function. For example, the well-known deep Q-network (DQN) developed by~\cite{mnih2015human} approximates the state-value function in a Q-Learning framework with a neural network. Similarly, deep deterministic policy gradient~\citep{silver2014deterministic} concurrently learned the action-value function and the policy.

For digital twins, deep reinforcement learning has only recently started to be applied. There are two main directions that have been pursued:  1) Establishing the P2V connection for model updating; 2) Building upon the operational digital twin, deep reinforcement learning has been used for control tasks and decision support, in particular due to its learning capabilities. As one example for the P2V connection, ~\cite{tian2022real} trained an agent with Lyapunov-based maximum entropy deep reinforcement learning to perform model calibration and infer the uncertain model parameters of the physics-based model, such that the response of physical model matched the observed data. The proposed framework provides a similar performance for the model calibration as model-based approaches, such as UKF, but provides real-time update capability. As an example for the the P2V connection, ~\cite{xia2021digital} investigated the application of deep reinforcement learning in industrial process control for smart manufacturing, where deep reinforcement learning was trained to learn an intelligent operations scheduler. Some other examples include a deep reinforcement learning algorithm trained to find the optimal off-loading policy in an IoT digital twin framework with a goal of reducing  energy consumption while enhancing data processing efficiency \cite{dai2020deep}. Moreover, \cite{khan2020requirements} provided an interesting perspective postulating that the digital twin as a technology for decision automation could be treated as a deep reinforcement learning model, where the physical system provides states and rewards information to the digital system via the P2V connection and digital system recommends actions to physical system to maximize the long-term reward.

Future research directions could make progress both directions: on the one hand integrating deep reinforcement learning more in the learning process of data-driven (and hybrid) digital twin, such for example for meta-learning tasks; on the other hand deep reinforcement learning is essential for interweaving more the digital twin and decision support, respectively control tasks enabling to harness more benefits from digital twin technology. One of such directions could be fore example, prescriptive operations, where deep reinforcement learning is used to prescribe an optimal course of actions with a potential objective to prolong the remaining useful lifetime \cite{TIAN2022108529}. 

\section{Conclusion}\label{sec9}
In this paper, we propose a five-dimensional definition for digital twin based on how data flows between the physical and virtual systems. The five dimensions are the: physical system, digital system, an updating engine (P2V), a prediction engine (V2P), and an optimization dimension (OPT). The five-dimensional digital twin provides a clear and complete picture on the different level of interactions between the physical system and its digital counterpart (i.e., data exchange, modeling, actions). Following the five-dimensional digital twin definition, we comprehensively review the underlying techniques that are commonly used in the state-of-the-art literature to enable each dimension of the digital twin and present our findings in a series of two papers. This paper covers the enabling techniques for three specific dimensions of digital twin: physical system modeling, P2V, and V2P. In the following paper, we will concentrate on articulating how to fully incorporate UQ and optimization in the three dimensions of digital twin described in this paper.  

Specific to this paper, we provide a comprehensive and rigorous examination on various means of modeling different aspects of a physical system (e.g., geometry, structure, system dynamics, relationship), the key P2V enabling techniques, and the actions and benefits (e.g., model predictive control, predictive maintenance) enabled by V2P twinning techniques across different phases of a physical system's life cycle. The accurate representation of a physical system together with the constant interactions enabled by P2V and V2P techniques form a closed loop to drive the implementation of digital twin in practice. In Part 2 of the review paper, we investigate how to ensure robust performance of a digital twin model by incorporating UQ and optimization techniques into the underlying enabling technologies.

\section*{Appendix A: A generic particle filter algorithm}
\label{Appendix A}

\begin{figure*}[!ht]
  \centering
    {\includegraphics[scale=0.65]{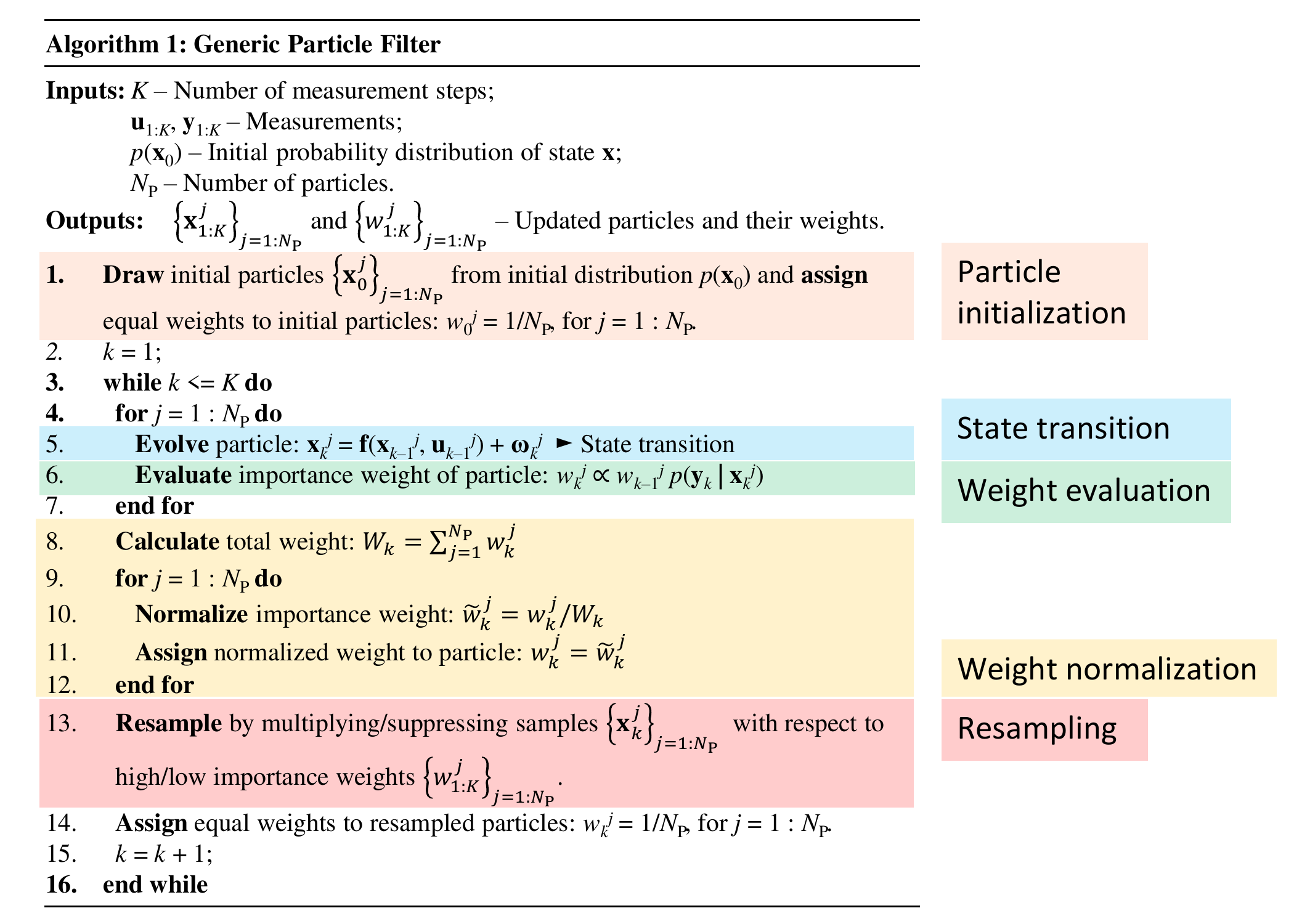}}
  \caption{Pseudocode of a generic particle filter algorithm}
  \label{fig:PF_algorithm}
\end{figure*}

Particle filtering is a widely used Bayesian filtering technique for state estimation in generic state-space models and DBNs. It is a key enabler in many digital twin applications where probabilistic model updating with state estimation plays an essential role. Let us briefly look at how particle filtering works. Particle filters represent a family of algorithms that recursively execute Bayesian filtering through sequential Monte Carlo simulation \citep{arulampalam2002tutorial}. In a particle filter, the marginal posterior of a state at the current time step $k$ is approximated using a large set of weighted particles. This approximation takes the following form
\begin{equation}
\label{eq:Expected K-L}
\begin{split}
p({{\bf{x}}_k}\vert{{\bf{y}}_{1:k}}) \approx \sum\limits_{j = 1}^{{N_p}} {w_k^j\delta ({{\bf{x}}_k} - {\bf{x}}_k^j)} ,
\end{split}
\end{equation}
where ${\bf{x}}_k^j$ is the jth particle of the state x at the kth time step, $w_k^j$ is its associated weight, NP is the total number of particles, and $\delta$ is the Dirac delta function. Algorithm 1 depicted in Fig. \ref{fig:PF_algorithm} gives the pseudo-code of a generic particle filter \citep{hu2018remaining,cappe2007overview}. Three key steps of the particle filtering procedure are described as follows:

\begin{itemize}
  \item State Transition: Each particle of the state transitions forward in time by one step (from the previous, $(k–1)^{th}$ step to the current, $k^{th}$ step according to the state transition equation in Eq. (\ref{eq:Expected K-L}) (Line 5). These particles are equally weighed ($1/N_P$), and they form the prior of ${{\bf{x}}_k}$ before the current measurement is used.
  \item Weight Evaluation and Normalization: The weight of each particle is updated using the likelihood of the current measurement given the state values of the particle, i.e., ${\bf{w}}_k^i \propto {\bf{w}}_{k - 1}^ip({{\bf{y}}_k}\vert{\bf{x}}_k^i)$ (Line 6). The likelihood is calculated by comparing the measurement distribution predicted by the measurement equation ${\bf{g}}$ in Eq. (\ref{eq:Expected K-L}) to the actual measurement ${{\bf{y}}_k}$. After that, the updated weights are normalized, and the normalized weights sum up to 1 (Lines 8-12).
  \item Resampling: Particles with negligible weights are replaced by new particles that are copies of particles with higher weights (Line 13). This step mitigates the issue of particle degeneracy, where the weights become overly concentrated on a very small subset of particles (in an extreme case, only all by one particle have close-to-zero weights). 
\end{itemize}

\section*{Appendix B: Decomposition of likelihood and Bayesian inference in a DBN}
\label{Appendix B}
Using the DBN given in Fig. \ref{fig:DBN} as an example, the posterior distribution $p({x_{1,k}},{x_{2,k}},{x_{3,k}} ,{x_{4,k}} \vert {y_{1,k}})$ of state variables ${x_{1,k}}$, ${x_{2,k}}$, ${x_{3,k}}$, and ${x_{4,k}}$ for given observation ${y_{1,k}}$ at time $t_k$ is given by 
\begin{equation}
\label{eq:DBN}
\begin{split}
  & p({x_{1,k}},{x_{2,k}},{x_{3,k}},{x_{4,k}} \vert {y_{1,k}})  \cr 
  &  \propto p({y_{1,k}} \vert {x_{3,k}},{x_{4,k}})p({x_{3,k}} \vert {x_{2,k}},{x_{1,k}}) \cr
  &\times p({x_{4,k}} \vert {x_{2,k}},{x_{1,k}})p'({x_{1,k}},{x_{2,k}},{x_{3,k}},{x_{4,k}}), \cr
\end{split}
\end{equation}
where $p({y_{1,k}} \vert {x_{3,k}},{x_{4,k}})$, $p({x_{3,k}} \vert {x_{2,k}},{x_{1,k}})$, and $p({x_{4,k}} \vert {x_{2,k}},{x_{1,k}})$ are the conditional probability tables or conditional probability distributions describing the probabilistic causality relationship between the parent nodes (e.g., $x_{2,k}$ and $x_{1,k}$) and a child node (e.g., $x_{3,k}$), $p'({x_{1,k}},{x_{2,k}},{x_{3,k}})$ is the prior distribution of state variables ${x_{1,k}},{x_{2,k}}$, ${x_{3,k}}$, and ${x_{4,k}}$ at $t_k$, which is obtained at each time step by recursively performing Bayesian inference and uncertainty propagation based on observations and state transition probabilities given in Eq. (\ref{eq:Transient BN}).

Theoretically, all the Bayesian inference methods discussed in Sec. \ref{state est and bayesian filters} can be employed to update the posterior of state variables based on the formulation given in Eq. (\ref{eq:DBN}). In the context of digital twins, particle filters such as the one presented in Appendix A are usually used in conjunction with DBNs to update digital states for two main reasons: (1) particle filters have fewer assumptions on the nonlinearity of the state transition and measurement functions and noise distributions than Kalman filters (see Table \ref{tab:filters}); and (2) using surrogate models to approximate conditional probability distributions (see Eq. (\ref{eq:GP_CPD})) allows for an efficient evaluation of likelihood functions, such that a large number of particles can be used for the inference. Using Eq. (\ref{eq:DBN}) as an example, in a particle filter implementation, we first generate equally weighted particles of state variables ${x_{1,k}},{x_{2,k}}$, ${x_{3,k}}$, and ${x_{4,k}}$ as their prior at $t_k$ according to the transition probabilities of state variables defined by the transition BN (i.e., Step 5 in Fig. \ref{fig:PF_algorithm}). After that, the likelihood of each particle is computed using the decomposed likelihood function $ p({y_{1,k}} \vert {x_{3,k}},{x_{4,k}})p({x_{3,k}} \vert {x_{2,k}},{x_{1,k}}) 
\times p({x_{4,k}} \vert {x_{2,k}},{x_{1,k}})$ given in Eq. (\ref{eq:DBN}). This corresponds to the computation of particle weights in Step 6 of Fig. \ref{fig:PF_algorithm}. The likelihood values of the particles are then normalized and the particles are subsequently re-sampled based on the normalized weights to obtain the posterior samples of the state variables at $t_k$. This process is implemented repeatedly over time to dynamically update the state variables. 

\backmatter

\section*{Acknowledgements}
Adam Thelen and Chao Hu would like to thank the financial support from the U.S. National Science Foundation under Grant No. ECCS-2015710. Xiaoge Zhang is supported by a grant from the Research Committee of The Hong Kong Polytechnic University under project code 1-BE6V and G-UAMR. Sankaran Mahadevan acknowledges the support of the National Institute of Science and Technology. Michael D. Todd and Zhen Hu received financial support from the U.S. Army Corps of Engineers through the U.S. Army Engineer Research and Development Center Research Cooperative Agreement W912HZ-17-2-0024.

\section*{Competing interests}
The authors have no relevant financial or non-financial conflicts of interest to disclose.

\section*{Replication of results}
The Python code and preprocessed dataset used for the battery case study are available for download on \cite{github_code}.

\section*{Authors' contributions}
All authors read and approved the final manuscript. Hu, C. and Hu, Z. devised the original concept of the review paper. Hu, Z, Thelen, A., and Zhang, X. were responsible for the literature review. Thelen, A. was responsible for geometric modeling. Hu, C., Thelen, A., and Hu, Z. were responsible for physics-based modeling. Hu, Z. was responsible for data-driven modeling. Hu, C. and Hu, Z. were responsible for physics-informed ML. Zhang, X. was responsible for system modeling. Hu, C., and Hu, Z., were responsible for probabilistic model updating. Zhang, X. was responsible for ML model updating. Hu, C. and Hu., Z were responsible for fault diagnostics, failure prognostics, and predictive maintenance. Lu, Y. was responsible for MPC. Fink, O. was responsible for federated learning and domain adaptation. Zhang, X., Hu, Z., and Fink, O. were responsible for deep reinforcement learning. Hu, C. was responsible for UQ of ML models. Hu, Z. was responsible for UQ of dynamic system models, optimization for sensor placement, and optimization for physical system modeling. Lu, Y. was responsible for the optimization of additive manufacturing processes. Zhang, X. and Hu, Z. were responsible for real-time mission planning. Thelen, A. and Hu, C. were responsible for the case study and predictive maintenance scheduling. Hu, C. was responsible for open-source software and data. Ghosh, S. was responsible for the industry demonstration. Hu, C., Todd, M., and Mahadevan, S. were responsible for perspectives. All authors participated in manuscript writing, review, editing, and comment.


\bibliography{sn-bibliography}

\end{document}